\documentclass[a4paper,fleqn,usenatbib]{mnras}

\usepackage{times}
\usepackage{aas_macros}

\usepackage{graphicx}
\usepackage{amstext}
\usepackage{enumerate}

\newcommand{\unit}[1]{\ensuremath{\, \mathrm{#1}}}

\providecommand*{\printsecond}[2]{#2}

\title[Asteroseismology of core helium burning stars]{The treatment of mixing in core helium burning models -- I. Implications for asteroseismology}

\author[Constantino et al.]{Thomas Constantino$^{1}$\thanks{E-mail: thomas.constantino@monash.edu}, Simon W. Campbell$^{1,2}$, J{\o}rgen Christensen-Dalsgaard$^{3}$, \newauthor John C. Lattanzio$^{1}$ and Dennis Stello$^{4,3}$\\
$^{1}$Monash Centre for Astrophysics (MoCA), School of Physics and Astronomy,  Monash University, Victoria, 3800, Australia\\
$^{2}$Max-Planck-Institut f\"{u}r Astrophysik, Karl-Schwarzschild-Stra{\ss}e 1, 85748 Garching bei M\"{u}nchen, Germany\\
$^{3}$Stellar Astrophysics Centre, Department of Physics and Astronomy, Aarhus University, Ny Munkegade 120, DK-8000 Aarhus C, Denmark\\
$^{4}$Institute for Astronomy (SIfA), School of Physics, University of Sydney, NSW 2006, Australia}

\begin{document}



\maketitle

\begin{abstract} 
The detection of mixed oscillation modes offers a unique insight into the internal structure of core helium burning (CHeB) stars.  The stellar structure during CHeB is very uncertain because the growth of the convective core, and/or the development of a semiconvection zone, is critically dependent on the treatment of convective boundaries.  In this study we calculate a suite of stellar structure models and their non-radial pulsations to investigate why the predicted asymptotic g-mode $\ell = 1$ period spacing $\Delta\Pi_1$ is systematically lower than is inferred from \textit{Kepler} field stars.  We find that only models with large convective cores, such as those calculated with our newly proposed ``maximal-overshoot'' scheme, can match the average $\Delta\Pi_1$ reported.  However, we also find another possible solution that is related to the method used to determine $\Delta\Pi_1$: mode trapping can raise the observationally inferred $\Delta\Pi_1$ well above its true value.  Even after accounting for these two proposed resolutions to the discrepancy in average $\Delta\Pi_1$, models still predict more CHeB stars with low $\Delta\Pi_1$ ($ \la 270$\,s) than are observed.  We establish two possible remedies for this: i) there may be a difficulty in determining $\Delta\Pi_1$ for early CHeB stars (when $\Delta\Pi_1$ is lowest) because of the effect that the sharp composition profile at the hydrogen burning shell has on the pulsations, or ii) the mass of the helium core at the flash is higher than predicted.  Our conclusions highlight the need for the reporting of selection effects in asteroseismic population studies in order to safely use this information to constrain stellar evolution theory.
\end{abstract}

\begin{keywords}
asteroseismology --- stars: evolution --- stars: horizontal-branch --- stars: interiors
\end{keywords}

\section{Introduction}

In low-mass stellar evolution the core helium burning (CHeB) phase is the third stage of nuclear burning -- after core and shell hydrogen burning.  In evolution calculations the core structure during this phase is highly uncertain, but it has been postulated for more than four decades that CHeB stars develop a zone of slow mixing, or ``semiconvection'', beyond the fully convective core \citep[e.g.][]{1969BAAS....1...98SH}.  How this develops in models, if at all, depends on the treatment of convective boundaries -- a major source of uncertainty generally in stellar models (see Section~\ref{sec:mixing_schemes}). Later evolution depends on the structure at the end of this phase, and this is typically where the results of different stellar evolution codes begin to diverge (see e.g. \citealt{1971Ap&SS..10..355C}; Fig. 15 in \citealt{2013ApJS..208....4P}).  

\subsection{Brief overview}
In this paper we make use of potent new constraints on the structure of core helium burning models -- mixed modes of oscillation detected from asteroseismology.  We present calculations of non-radial pulsations for CHeB models evolved with a variety of mixing prescriptions: models with and without convective overshoot as well as those with a semiconvection region.  We also introduce a new algorithm for core mixing during the CHeB phase that can better match the asymptotic g-mode period spacing inferred from asteroseismology with the use of otherwise standard physics.  Finally, we compute pulsation spectra at different stages of the core helium burning phase.  In order to make this paper more accessible to non-experts, we provide an extensive summary in Section~\ref{sec:conclusions}, which includes clear references to previous sections that contain more technical detail.

\subsection{The problem of convective boundaries}

Core helium burning stars, in their various (observational) flavours, may be known as subdwarf B (sdB), horizontal branch (HB), RR-Lyrae, red clump (RC), or secondary clump \citep{1999MNRAS.308..818G} stars.  Most of the variation between these types of CHeB stars is due to differences in the mass of the convective envelope and the hydrogen-exhausted core beneath it (where all of the hydrogen has been burned to helium).  The common thread between them is that they are all thought to contain a central helium-burning convection zone that is surrounded by a helium-rich region that is not convective.

In their convective cores, CHeB stars produce carbon via the triple-$\alpha$ reaction, and oxygen via $^{12} \text{C} (\alpha ,\gamma) ^{16}\text{O}$.  This burning produces a growing abundance discontinuity at the formal boundary of the convection zone if there is no convective overshoot to induce mixing beyond it.  This boundary is usually defined as the point of convective neutrality (i.e. where a displaced fluid element experiences no acceleration).  If the \citet{1906WisGo.195...41S} criterion is applied then this is where the radiative temperature gradient $\nabla_\text{rad}$ is equal to the adiabatic temperature gradient $\nabla_\text{ad}$.  Therefore the criterion for convective stability is
\begin{equation}
\nabla_\text{rad}<\nabla_\text{ad},
\end{equation}
where 
\begin{equation}
\nabla_\text{ad}=\left(\frac{\partial\ln{T}}{\partial\ln{p}}\right)_\text{ad}\hspace{0.05cm}, \hspace{0.25cm}\nabla_\text{rad}= \left(\frac{\text{d}\ln{T}}{\text{d}\ln{p}}\right)_\text{rad},
\end{equation}
and $\nabla_\text{rad}$ is the temperature gradient required for radiation to carry the total energy flux.  In low-mass CHeB models the location of this boundary is unstable.  The increasingly C- and O-rich mixture in the convection zone is more opaque than the He-rich material just beyond the boundary.  Because of this, any mixing from convective overshoot (which has a sound physical basis because the boundary is defined only as the point of zero acceleration but where convective elements still carry momentum; see \citealt{1971Ap&SS..10..340C} for a quantitative analysis) will increase the opacity and therefore $\nabla_\text{rad}$, and cause the convection zone to grow.  \citet{1969BAAS....1...98SH} first found that in CHeB models a partially-mixed or ``semiconvection'' region can then develop \citep[see also][]{1970QJRAS..11...12S,1970AcA....20..195P}.  While only marginally stable according to the Schwarzschild criterion ($\nabla_\text{rad} \approx \nabla_\text{ad}$), the semiconvective region is stable when the effect of the molecular weight gradient is considered.  This is accounted for in the \citet{1947ApJ...105..305L} criterion for convective stability
\begin{equation}
\nabla_\text{rad} < \nabla_\text{ad} +\frac{\varphi}{\delta}\nabla_\mu,
\end{equation} 
where
\begin{equation}
\nabla_\mu = \frac{\text{d}\ln{\mu}}{\text{d}\ln{p}}, \hspace{0.07cm} \varphi=\left( \frac{\partial \ln{\rho}}{\partial \ln{\mu}}\right)_{T,p}, \hspace{0.07cm} \delta = -\left(\frac{\partial \ln{\rho}}{\partial \ln{T}}\right)_{p,\mu}.
\end{equation}
Semiconvection regions are usually defined as stable (not convective) according to the Ledoux criterion but convective (or neutral) according to Schwarzschild.

In early studies, algorithms were developed that produce what we shall refer to as ``classical semiconvection'', where the composition is adjusted to produce $\nabla_\text{rad}/\nabla_\text{ad}=1$, which results in a smooth abundance profile \citep[e.g.,][]{1971PASAu...2...27S,1972ApJ...171..309R,1973ApJ...180..435F}.  Even if there is no explicit process for allowing semiconvection, a similar chemical profile and temperature gradient is produced from localized mixing episodes in evolution sequences that have instantaneous mixing in convection zones and an overshooting prescription that allows mixing beyond the Schwarzschild boundary \citep{1986ApJ...311..708L,1990A&A...240..305C}.  The most obvious difference is that the local mixing events leave behind numerous small composition discontinuities.  By the end of core helium burning, both of these schemes permit the partially mixed region to grow to such an extent that its enclosed mass is around double that of the convective core (Figure~\ref{figure_helium_ev}b).  Typically, models initially experience a period of rapid growth of the convective core.  The expansion rate of the convective core then slows as a result of the emergence and subsequent growth of a partially mixed region.  The semiconvection region in the 1.5\,$\text{M}_\odot$ model from \citet{1973ApJ...180..435F}, for example, appears when the central helium mass fraction has reduced to 0.75 (from 0.978 initially), at which time the convective core growth rate is halved.

The total mass of helium that burns during the CHeB phase differs from code to code.  The principal reason is mixing: specifically whether the criterion for convection is Schwarzschild or Ledoux, and whether convective overshoot or a scheme for semiconvection is applied.  Recently, \citet{2014A&A...569A..63G} also highlighted the significance of whether the location of a convective boundary is determined from inside or outside the convection zone.  The greater opacity of the products of helium burning means that in this phase, numerical subtleties such as these have a compounding effect on the evolution.

\subsection{Classical constraints from globular clusters}

Despite making a vast difference to the evolution of the interior, the core mixing has little immediate effect on the conditions at the surface.  By controlling the amount of helium that is burnt, the mixing scheme does, however, affect the CHeB and early-asymptotic giant branch (AGB) lifetimes.  Empirically, the lifetime of various phases of evolution can be inferred from star counts in globular clusters.  This is because they have large and (relatively) homogeneous stellar populations.  By using the so-called R-method on a sample of 15 globular clusters (i.e. determining $R_1$, the ratio of AGB to red giant branch stars), \citet{1983A&A...128...94B} found indications for the existence of a fully developed semiconvective zone.  

Late in the CHeB phase, models can also show the phenomenon of ``core breathing pulses'', the rapid growth in the mass of the convective core when the central helium abundance is very low \citep{1973ASSL...36..221S,1985ApJ...296..204C}.  Despite their emergence in stellar evolution calculations, \citet{1989ApJ...340..241C} and \citet{2001A&A...366..578C} contend that evidence from star counts in globular clusters discredits the existence of core breathing pulses, because they further prolong the HB lifetime and shorten the early-AGB.  This conflict between theoretical predictions and observations exposes the uncertainty of stellar models during the CHeB phase.

\subsection{Asteroseismology of CHeB stars}

The study of asteroseismology promises a unique chance to constrain CHeB models.  The long time series observations from the CoRoT and \textit{Kepler} missions have yielded unprecedented potential for red-giant asteroseismology.  Solar-like oscillations have now been detected in more than 13,000 giants in the \textit{Kepler} field \citep{2013ApJ...765L..41S}. 

\citet{2010ApJ...713L.176B} first detected mixed modes in the surface oscillations of red giants in the \textit{Kepler} field.  These propagate as acoustic modes in the convective envelope and as gravity modes in the radiative core \citep{1977A&A....58...41A}.  Crucially, the observed period spacing $\Delta P$ of the dipole ($\ell = 1$) modes is thought to provide a lower bound on the asymptotic `pure g-mode' spacing $\Delta\Pi_1$ \citep{2011Sci...332..205B}.  \citet{2011Natur.471..608B} showed that this period spacing can reliably distinguish CHeB stars from photometrically similar, but shell-hydrogen burning, red giant branch (RGB) stars.  This is possible because the mixed-mode period spacing is sensitive to the conditions in the core, which change substantially between the RGB and CHeB phases.  More recently, \citet{2012A&A...540A.143M} have developed a method to infer $\Delta\Pi_1$ from the relatively small fraction of mixed modes that are detectable.  The CHeB stars for which they reported $\Delta\Pi_1$ mostly have asteroseismic scaling-relation-determined masses of $0.8<M/\text{M}_\odot < 2.6$, while a handful have masses up to $M=3.4$\,$\text{M}_\odot$.  The metallicity ([M/H]) distribution of the stars  in the latest (and larger) core helium burning sample with $\Delta\Pi_1$ determinations from \citet{2014A&A...572L...5M} is strongly peaked around the solar value (determined from the stars also in the {\sc apokasc} catalogue; \citealt{2014ApJS..215...19P}).

Measurement of $\Delta \Pi_1$ is a particularly useful diagnostic because it depends only on the Brunt--V{\"a}is{\"a}l{\"a} frequency $N$, which is easily computed from a given stellar structure.  Specifically, $N$ is the frequency of oscillation that an adiabatically displaced mass element will undergo due to buoyancy forces.  In a convective region, displaced elements are buoyantly unstable, therefore such oscillations cannot occur, and gravity waves are damped.  In the asymptotic limit, the gravity mode period spacing is
\begin{equation}
\label{eq:dp1_asymp}
\Delta\Pi_\ell=\frac{2\pi^2}{\sqrt{\ell(\ell+1)}} \left[ \int\limits_\text{}^{}{\frac{N}{r}\text{d}r} \right]^{-1},
\end{equation}
where the integral is over the region with $N^2 > 0$, $\ell$ is the spherical harmonic degree,
\begin{equation}
\label{eq:brunt}
N^2 = g \left(\frac{1}{\Gamma_{1}} \frac{\text{d} \ln{p}}{\text{d}r} - \frac{\text{d} \ln{\rho}}{\text{d}r} \right),
\end{equation}
and $g$ is the local gravitational acceleration \citep{1977AcA....27...95D}.  If $ N^2 > 0$ then $N$ is real, which is equivalent to the Ledoux criterion for convective stability
\begin{equation}
\frac{\text{d}\ln{\rho}}{\text{d}\ln{p}} > \frac{1}{\Gamma_{1}},
\end{equation}
where
\begin{equation}
\Gamma_{1} = \left(\frac{\partial \ln{p}}{\partial \ln{\rho}}\right)_\text{ad},
\end{equation}
and the subscript \textit{ad} signifies an adiabatic change.  Another key observable from asteroseismology is the large frequency separation, whose asymptotic value $\Delta\nu$ can be computed as
\begin{equation}
\Delta\nu = \left[ 2 \int_0^R  \frac{\text{d}r}{c} \right]^{-1},
\end{equation}
where $c$ is the adiabatic sound speed \citep{1967AZh....44..786V,1980ApJS...43..469T}.  Under a homologous transformation this scales with the square root of the mean stellar density, which is a finding that is supported by models \citep{1986ApJ...306L..37U}.

If $\Delta\nu$ and $\Delta P$ (or $\Delta\Pi_1$) are both determined for a star, it can be placed on the $\Delta\nu - \Delta P$ (or $\Delta\nu - \Delta\Pi_1$) diagram.  When this procedure is performed for the \textit{Kepler} field stars, two distinct groups are found, comprising the CHeB and RGB stars respectively (e.g., Fig. 3 in \citealt{2012A&A...540A.143M} and Fig. 1 in \citealt{2014A&A...572L...5M}).  The most striking feature of the $\Delta\nu - \Delta\Pi_1$ diagram is how tightly most of the low-mass ($M/\text{M}_\odot \la 1.75$) CHeB stars are grouped, with $250 \la \Delta\Pi_1 (\text{s}) \la 340$ and $\Delta\nu \sim 4 \unit{MHz}$.  To date, however, CHeB models have been unable to properly match the $\Delta\Pi_1$ inferred from the observations (e.g. Figure~\ref{figure_dp1_obs}).

\citet{2013ApJ...766..118M} identified a linear dependence of $\Delta\Pi_1$ on the radius of the convective core for CHeB models.  They also noticed how the $\Delta\Pi_1$ dependence on stellar mass is similar to that for $M_\text{He}$, and emphasized a linear dependence of $\Delta P$ between ``observable'' modes on $M_\text{He}$ for models massive enough to avoid the degenerate ignition of helium ($M \ga 2.2\,\text{M}_\odot$).  Additionally, they suggested that a model with a semiconvection zone will have a lower $\Delta\Pi_1$ than a model with an identically sized convective core but without semiconvection.  Their comparison between $\Delta\Pi_1$ from models (of around solar composition and $0.7 - 3.0$\,$\text{M}_\odot$ computed with the {\sc aton} evolution code; \citealt{2008Ap&SS.316...93V}) and the observations reported by \citet{2012A&A...540A.143M} reveals a general offset, with the theoretical $\Delta\Pi_1$ lower than observed (their Fig. 7).  This offset is also evident from models computed with {\sc mesa} \citep{2012ApJ...744L...6B,2013ApJ...765L..41S} and the Monash University code {\sc monstar} \citep{2014IAUS..301..399C}, making it apparent in at least three independent evolution codes.  It is not as obvious that this offset exists for more massive, higher-$\Delta\nu$ models.  However, the higher-mass models without overshoot by \citet{2013ApJ...766..118M} do not match the whole observed spread of $\Delta\Pi_1$.  Those models have roughly $160 \la \Delta\Pi_1$\,$\text{(s)} \la 230$ compared to $145 \la \Delta\Pi_1$\,$\text{(s)} \la 300$ observed.  Interestingly, for the 1.5\,$\text{M}_\odot$ model at least, it appears that convective overshoot during CHeB considerably increases $\Delta\Pi_1$ (by around 50\,s).  

Recently it has been shown that additional diagnostic information about mixing events may be obtained from the effect that resulting sharp features in the buoyancy frequency have on the observed mode frequencies \citep{2015ApJ...805..127C}.  Such features can arise from composition discontinuities left by first dredge-up during the RGB evolution or, as we discuss in this paper, from those that may arise in the CHeB phase.

\subsection{The core mass at the flash}

\citet{1976ApJS...32..367S} showed that the zero-age horizontal branch (ZAHB) convective core mass depends predominantly on the H-exhausted core mass $M_\text{He}$ (and is insensitive to composition and total mass).  Since there is a close relationship between radius of the convective core (and therefore its mass) and $\Delta\Pi_1$ \citep{2013ApJ...766..118M}, there must also be one between $M_\text{He}$ and $\Delta\Pi_1$.  The mass of the H-exhausted core is therefore a crucial quantity for the seismology of CHeB stars.  

In low-mass CHeB models ($M \la 2.2\,\text{M}_\odot$), helium ignition occurs under degenerate conditions, triggering a thermal runaway known as the core flash.  The minimum $M_\text{He}$ required for this ignition is fairly constant over a wide range of stellar masses and only decreases slightly with increasing metallicity or helium.  \citet{2005essp.book.....S} show, for instance, that there is a 0.03\,$\text{M}_\odot$ difference in $M_\text{He}$ between $Z=0.02$, $Y=0.273$ and $Z=10^{-4}$, $Y=0.245$ solar-mass models.  \citet{1996ApJ...461..231C} explored the uncertainties in the physics in stellar models that could influence $M_\text{He}$, including neutrino losses, rotation, conductive opacity, coulomb effects on the equation of state, reaction rates and screening factors, and element diffusion.  They constrained the possible core mass increase compared with standard models to $\Delta M_\text{He} = 0.01 \pm 0.015$\,$\text{M}_\odot$.  The best observational constraint on $M_\text{He}$ comes from comparisons with globular clusters, but additional factors must be considered there as well, such as the composition, MLT mixing length, bolometric corrections, and distance determination \citep{2013A&A...558A..12V}.
 
\subsection{Clues from subdwarf B stars}

In contrast to the more massive red clump stars, the core mass of sdB stars is less uncertain because of the very thin hydrogen envelope.  The mass distribution of sdB stars, determined from asteroseismology and eclipsing binaries, is peaked at $\sim0.47$ $\text{M}_\odot$ \citep{2013EPJWC..4304007V}.  This corresponds closely to the canonical core mass at the core flash.  What we may be able to learn from this mass distribution though is dependent on understanding the conditions under which helium ignition occurs.  When a sample of sdB stars is plotted in the $\log{g} - T_\text{eff}$ plane there is evidence for two distinct groups \citep{2008ASPC..392...75G}, perhaps suggesting different formation channels \citep{2013EPJWC..4304007V}.  Proposed binary mechanisms include common-envelope ejection, stable Roche lobe overflow and double helium white dwarf mergers \citep{2002MNRAS.336..449H}.  If an sdB star is formed via the stable Roche lobe overflow channel then it is likely that its mass is close to the minimum H-exhausted core mass required for helium ignition.  Even if we cannot reliably deduce the H-exhausted core mass of other CHeB stars from the empirical mass of sdB stars, their observed pulsation properties can still serve as a useful constraint on the physics \textit{during} the CHeB phase. 

Space based observations of g-mode sdB pulsators have proven to be superior to earlier efforts from the ground \citep{ 2010A&A...516L...6C}.   Structural properties, such as core and envelope mass and central helium abundance, of a handful of such stars have now been estimated \citep{2010Ap&SS.329..217V,2010ApJ...718L..97V,2010A&A...524A..63V,2011A&A...530A...3C}.  In their method they determine these quantities by finding a least-squares fit between the observed and the theoretical periods from models with different structural parameters \citep{2008ASPC..392..261B}.  It has been reported, for example, that the sdB star  KPD 1943+4058 has a larger mixed core (defined as the mass interior to the unmixed He-rich radiative zone) than models by \citet{1993ApJ...419..596D} which do not include convective overshooting.  It is unclear, however, whether the determined structure includes a semiconvection region.  

\citet{2011MNRAS.414.2885R} found a period spacing range of $231\leq \Delta\Pi_1$\,$(\text{s}) \leq 271$ with an average of 254\,s for 13 g-mode pulsating sdB stars observed by \textit{Kepler} and another by CoRoT.  This period spacing is clearly lower than the more massive CHeB stars in the \citet{2012A&A...540A.143M} sample, which have reported $\Delta\Pi_1$ typically around 300\,s.

\subsection{The challenges and potential of CHeB asteroseismology}

The pulsations in CHeB models can be far more complex than they are in RGB models.  Importantly for the propagation of g-modes, there is only a single radiative zone in RGB stars, whose structure is well understood, and apart from the discontinuity left by first dredge-up, it contains only smooth variations in chemical composition.  In contrast, the structure of CHeB models is sensitive to the treatment of convective boundaries in the core as well as to prior evolution, especially the core flash.  They may contain multiple convection zones and composition discontinuities.  The deficit in our understanding of the mixing during core helium burning spans a broad mass range, from $M \ga 0.47$\,$\text{M}_\odot$ to at least $M \sim 20$\,$\text{M}_\odot$ \citep{1991A&A...252..669L}.  Part of the reason for this uncertainty is that until now the core mixing has been hidden from view.  

The detailed study of pulsations in models of CHeB stars is imperative if we are to properly interpret asteroseismic observations and gain understanding about the behaviour of convection near the boundary of the convective core.  In this paper we address this need by analysing the non-radial pulsations in a range of CHeB models with disparate internal structures stemming from different treatments of convective boundaries.  Any insights about this mixing will also have implications for the treatment of convective boundaries in stellar models more generally.

\begin{figure}
\includegraphics[width=\linewidth]{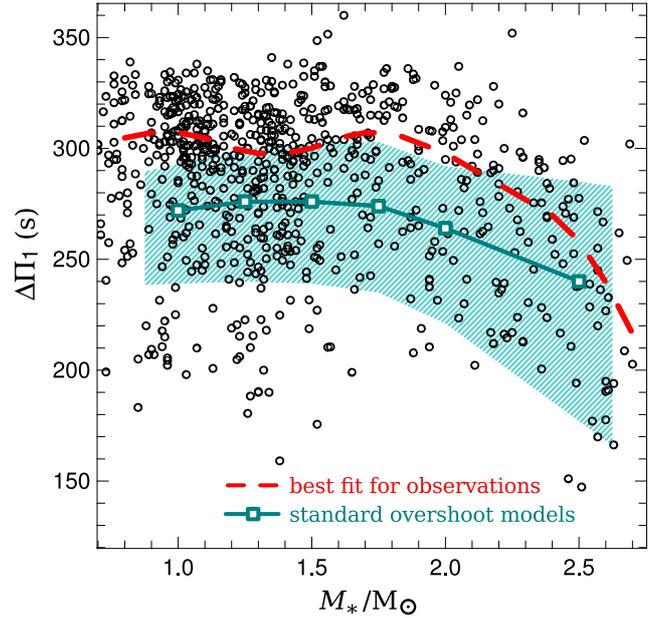}
\caption{Comparison between $\Delta\Pi_1$ inferred from observations of likely CHeB stars with seismic mass determinations \citep[black circles;][]{2014A&A...572L...5M} and the average computed from CHeB models with standard overshoot (cyan line; with markers showing each calculation).  The line of best fit for observations (red dashes) follows the mode of the $\Delta\Pi_1$ distribution.  The shaded area gives the range of $\Delta\Pi_1$ in which the models spend 95 per cent of their CHeB lifetime.}
  \label{figure_dp1_obs}
\end{figure}

\begin{figure}
\includegraphics[width=\linewidth]{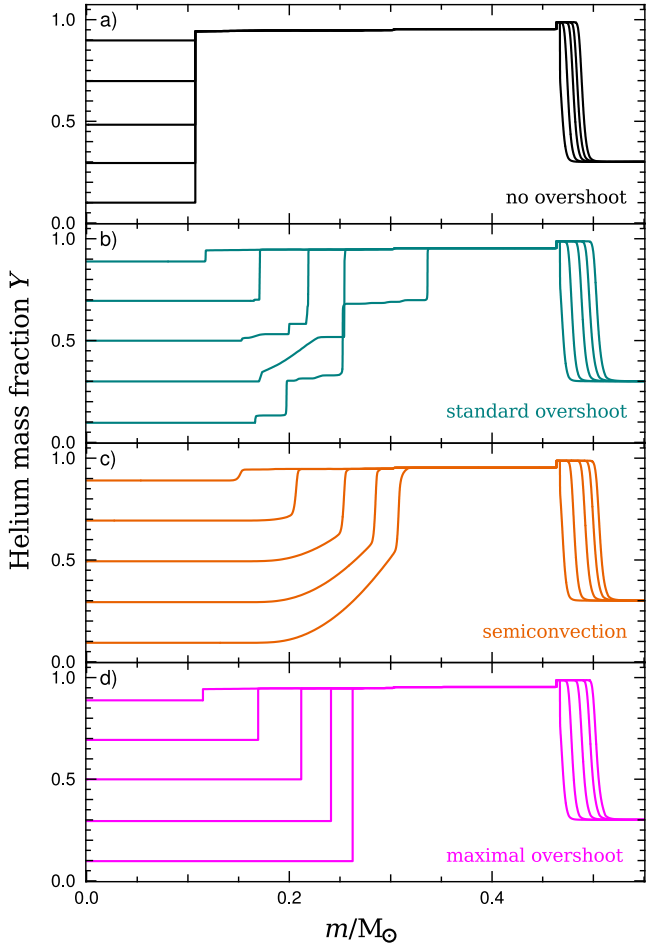}
  \caption{Evolution of internal helium abundance during CHeB with four different mixing prescriptions.  Each panel shows the profile at five different times.  The mixing prescriptions are, from top to bottom, no overshoot (Section~\ref{sec:no_os}; black), standard (pure Schwarzschild) overshoot (Section~\ref{sec:overshoot}; cyan), semiconvection (Section~\ref{sec:sc}; orange), and maximal overshoot (Section~\ref{sec:max_os}; magenta).}
  \label{figure_helium_ev}
\end{figure}

\section{Methods}
\subsection{Evolution code}
In this study our evolution models are computed with  {\sc monstar}, the Monash University stellar structure code, which has been described in detail previously \citep[e.g.,][]{1986ApJ...311..708L,2008A&A...490..769C,2014ApJ...784...56C}.  Unless stated otherwise, our models are 1\,$\text{M}_\odot$, solar metallicity \citep{2009ARA&A..47..481A}, with initial helium mass fraction $Y=0.278$.  

\subsection{Pulsation analysis}
\label{sec:puls_methods}

The models we use for pulsation calculations are usually mid-way through CHeB and have central helium abundance $Y=0.4$ or $Y=0.5$.  The pulsations are computed with the Aarhus adiabatic oscillation package {\sc adipls} \citep{2008Ap&SS.316..113C}.  In this study we restrict our analysis of non-radial modes to the $\ell = 1$ case.  Each structure model we present is converged and in hydrostatic equilibrium, and the Brunt--V{\"a}is{\"a}l{\"a} frequency is calculated directly from $P$, $\rho$, $r$, and $\Gamma_1$ according to Equation~\ref{eq:brunt}. 

The estimates of the frequency of maximum power $\nu_\text{max}$ for the models in this paper use the assumption that it scales with the acoustic cut-off frequency \citep{1991ApJ...368..599B}, and therefore that
\begin{equation}
\label{eq:numax}
\nu_\text{max} = \frac{g}{{g}_\odot} \sqrt{ \frac{\text{T}_{\text{eff},\odot}}{T_\text{eff}} } \nu_{\text{max},\odot},
\end{equation}
where $g$ is the surface gravity, $g_\odot = 2.74\times 10^4\,\text{cm}\,\text{s}^{-2}$, $\nu_{\text{max},\odot}=3.1\,\text{mHz}$, and $\text{T}_{\text{eff},\odot} = 5778\,\text{K}$.

We obtain frequencies of individual modes from the pulsation calculations and hence the actual period spacing $\Delta P$ between modes of adjacent order.   We present these results by showing $\Delta P$ as a function of frequency.  In several examples we use the period \'echelle diagram, where the mode frequency is plotted against the mode period, modulo some period spacing $\Delta P_\text{\'ech}$.  This is used because g-modes tend to be approximately equally spaced in period (the asymptotic limit is given in Equation~\ref{eq:dp1_asymp}).  This is important because in the method developed by \citet{2012A&A...540A.143M}, $\Delta\Pi_1$ corresponds to the $\Delta P_\text{\'ech}$ which produces a regular pattern in the \'{e}chelle diagram.  Thus it allows us to predict the value of $\Delta\Pi_1$ that would be inferred from observations of our theoretical models.

We report the mode inertia $E$ from the pulsation calculations which is defined as 
\begin{equation}
E=\frac{\int_{r_1}^{R_\text{s}} \left [   \xi_\text{r}^2 + \ell(\ell+1)\xi_\text{h}^2 \right ] \rho r^2 \text{d}r}{M \xi_\text{r}(R_\text{s})^2},
\end{equation}
where $R_\text{s}$ is the radius at the outermost point, $r_1$ is the location of the innermost mesh point, and $\xi_r$ and $\xi_h$ are the radial and horizontal displacement eigenfunctions, respectively, which are both functions of $r$.  This is a measure of kinetic energy of a mode relative to the radial displacement at the surface. In the plots of eigenfunctions we show the scaled horizontal displacement
\begin{equation}
\label{eq:y1}
y_2 = \frac{\ell ( \ell +1) }{R}\xi_\text{h},
\end{equation}
where $R$ is the photospheric radius and $\xi_\text{h}$ is scaled so that $\xi_\text{r}/R = 1$ at the surface.

\begin{figure}
\includegraphics[width=\linewidth]{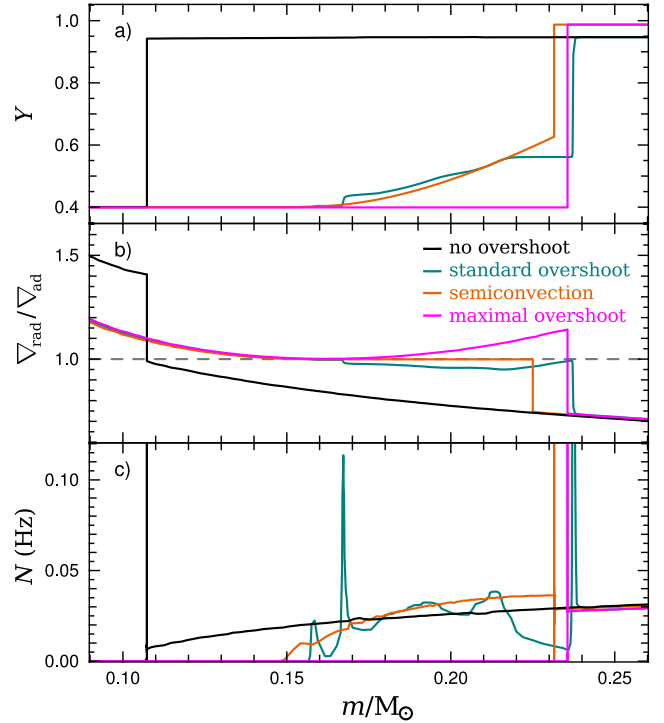}
  \caption{Internal properties of the 1\,$\text{M}_\odot$ models with four different mixing prescriptions when $Y=0.4$ in the centre.  The helium mass fraction $Y$, ratio of temperature gradients $\nabla_\text{rad}/\nabla_\text{ad}$, and Brunt--V{\"a}is{\"a}l{\"a} frequency are shown.  The four mixing prescriptions are no overshoot (Section~\ref{sec:no_os}; black), standard overshoot (Section~\ref{sec:overshoot}; cyan), classical semiconvection (Section~\ref{sec:sc}; orange), and maximal overshoot (Section~\ref{sec:max_os}; magenta).  This colour scheme is used for mixing comparisons throughout this paper.}
  \label{figure_mixing_comparison_zoom}
\end{figure}

\subsection{Core mixing schemes}
\label{sec:mixing_schemes}

\subsubsection{Models without convective overshoot}
\label{sec:no_os}
In the stellar models without convective overshoot the Schwarzschild criterion for convection is strictly applied.  The resulting internal structure is shown in Figure~\ref{figure_helium_ev}a and Figure~\ref{figure_mixing_comparison_zoom}.  In this case a convection zone may only grow (in mass) if the conditions change in a radiative zone so that $\nabla_\text{rad} > \nabla_\text{ad}$.  In the CHeB phase this will give the smallest possible convective core (at least for models with the Schwarzschild criterion) because the region outside the core is close to convective neutrality (see Figure~\ref{figure_mixing_comparison_zoom}).  In this study we do not compute any models using the Ledoux criterion for convection, but we note that we do not expect it to make any difference to our models either with or without convective overshoot, if properly implemented.  This is because in our models without overshoot the convective boundary hardly moves (by less than 0.001\,$\text{M}_\odot$ in the 1\,$\text{M}_\odot$ solar-metallicity run; Figure~\ref{figure_helium_ev}), so any restriction in growth due to the composition gradient would be insignificant.  In the overshooting models, the mixing beyond the boundary tends to erase any stabilizing composition gradients and therefore reduces the Ledoux criterion to the Schwarzschild one.

\subsubsection{Models with convective overshoot}
\label{sec:overshoot}
We have implemented overshooting \citep{2008A&A...490..769C} with an exponential decay in the diffusion coefficient according to the scheme proposed by \citet{1997A&A...324L..81H}.  This is expressed as
\begin{equation}
D_\text{OS}(z)=D_\text{0} e^{\frac{-2z}{H_\text{v}}},
\end{equation}
where $D_\text{OS}(z)$ is the diffusion coefficient at distance $z$ from the convective boundary and $D_\text{0}$ is the diffusion coefficient just inside the boundary.  $H_\text{v}$ is the ``velocity scale height'' defined as
\begin{equation}
H_\text{v}=f_\text{OS} H_\text{p},
\end{equation}
where $H_\text{p}$ is the pressure scale height, and we have chosen $f_\text{OS}=0.001$.  We refer to this as ``standard overshoot'', but the exact value of $f_\text{OS}$ is not important because our models are insensitive to the formulation of convective overshoot.  The resulting internal structure is shown in Figure~\ref{figure_helium_ev}b and Figure~\ref{figure_mixing_comparison_zoom}.

Our models with convective overshoot and time-dependent mixing of chemical species \citep{2008A&A...490..769C} evolve similarly to those with instant mixing and the search for convective neutrality \citep{1986ApJ...311..708L}.  In the latter method, the convective boundary is found by testing whether mixing at the Schwarzschild boundary would cause the next radiative zone to become convective, while the former always mixes beyond the Schwarzschild boundary (i.e. without a test).  The outcome is similar because at these conditions C and O are so much more opaque than He, so this mixing usually results in the radiative zone adjacent to the Schwarzchild boundary at the outer edge of the convective core becoming unstable to convection.  This feedback contrasts with other phases in evolution when the extent of mixing is dependent on the overshooting distance because the resulting mixing does not alter the location of the Schwarzschild boundary.  

In CHeB models, any overshoot tends to grow the core enough so that a minimum in $\nabla_\text{rad}/\nabla_\text{ad}$ appears in the convection zone; see the magenta line in Figure~\ref{figure_mixing_comparison_zoom}b for an example.  Eventually this minimum falls below unity and the convection zone splits.  This process continuously repeats, leaving behind the characteristic stepped abundance profile seen in Figure~\ref{figure_helium_ev}b.  If properly resolved, the partially mixed region created beyond the convective core by overshooting will have temperature gradient $\nabla_\text{rad}/\nabla_\text{ad} \approx 1$, resembling semiconvection (see Section~\ref{sec:sc}). 

\subsubsection{Models with semiconvection}
\label{sec:sc}

We have developed a simple method to mimic the structure that is found using semiconvection routines.  We do this by allowing slow mixing in sub-adiabatic conditions.  Specifically, we set a mixing rate that depends only on how close a zone is to being convective according to the Schwarzschild criterion (which neglects the stabilizing effect of any composition gradients). If $\nabla_\text{rad} < \nabla_\text{ad}$ then we set the diffusion coefficient $D$ according to
\begin{equation}
\log{D}=\log{D'}-c_1 (1-\nabla_\text{rad}/ \nabla_\text{ad}), 
\end{equation}
but also specify a maximum gradient so that
\begin{equation}
\left| \frac{\text{d} \ln{D}}{\text{d} \ln{p}} \right| \leq c_2, 
\end{equation}
where $D'$, $c_1$, and $c_2$ are constants that are chosen at discretion and varied with experience \citep[cf.][]{1967ApJ...147..650I}.  In this study we use $\log{D'}=10$, $c_1=100$, and $c_2=90$ to give $D$ in units of cm$^2$s$^{-1}$.  The structure that is produced by this scheme is shown in Figure~\ref{figure_helium_ev}c.  This differs slightly from previous routines that have a zone with exactly $\nabla_\text{rad} = \nabla_\text{ad}$ and a smooth composition profile which ends with a discontinuity (herein the ``classical'' semiconvection structure).  An example of the classical structure is shown in Figure~\ref{figure_mixing_comparison_zoom}.  The most obvious difference produced by our routine is that the composition is everywhere smoothly varying, i.e. there is no discontinuity at the outer boundary of the partially mixed region. 

We also use a different method to construct classical semiconvection models.  For a given central composition we artificially increase the mass fraction of the convective core until there is a minimum in $\nabla_\text{rad}/\nabla_\text{ad}$ inside the convection zone with $\nabla_\text{rad}/\nabla_\text{ad}=1$ (the structure during this intermediate step is identical to that of the maximal-overshoot models shown in Figure~\ref{figure_mixing_comparison_zoom} and discussed below).  The location of this minimum is the first guess for the boundary between the convection and semiconvection zones, while the edge of the convective region becomes the outer boundary of the semiconvection zone.  We then adjust the composition between these two points until this region has $\nabla_\text{rad}/\nabla_\text{ad}=1$ everywhere (and also make small adjustments to the location of the boundaries if needed).  This contrasts with the method of \citet{1972ApJ...171..309R} where, during each time step, the composition changes due to nuclear burning and then mixing proceeds outwards from the centre, point by point, to give exactly $\nabla_\text{rad} = \nabla_\text{ad}$. 

\subsubsection{Models with maximal overshoot}
\label{sec:max_os}

If the helium-burning convective core is large enough, it will contain within it a minimum $\nabla_\text{rad}/\nabla_\text{ad}$, such as that shown by the magenta line in Figure~\ref{figure_mixing_comparison_zoom}b.  Further core growth will continue to reduce the value of this minimum until it reaches $\nabla_\text{rad}/\nabla_\text{ad} < 1$, which splits the convection zone into two.  This is avoided in our newly developed ``maximal-overshoot'' scheme by making convective overshoot dependent on the value of this minimum, so that the core growth slows (and then can stop) if the convective core is close to splitting.  In its present ad hoc implementation the amount of mass beyond the convective boundary that is mixed each time step is proportional to the minimum of $\nabla_\text{rad}/\nabla_\text{ad}-1$ in the convection zone, and overshoot is stopped if that minimum falls below 0.002.  This ensures that the model attains the largest possible convective core throughout the evolution.  This structure is shown in Figure~\ref{figure_helium_ev}d and Figure~\ref{figure_mixing_comparison_zoom}.  We do not propose a physical justification for achieving this exact structure.  Instead we use it as a comparison to standard models which is interesting because of its extreme core size and the effect on $\Delta\Pi_1$.  Finally, we note that although our maximal-overshoot models are generated by a different mechanism, their structure is similar to some earlier models with large fully mixed cores \citep[e.g.,][]{1986MmSAI..57..411B,2003ApJ...583..878S}.

\subsection{Composition smoothing}

The period spacing pattern in CHeB models taken directly from the evolution code tends to be inconsistent with observations \citep[e.g. Fig. 2 in][]{2012A&A...540A.143M}.  In Section~\ref{sec:postcoreflash} we analyse this in detail and show that it is primarily a relic of the burning during the core flash phase.  Therefore, in a number of our ZAHB models we remove the composition discontinuities between the H-burning shell and the convective core that this burning produces.  We have good reason to do this: the chemical profile that the core flash leaves behind is dependent on unknown factors such as the extent of convective overshoot and mixing at boundaries, the extent of burning during each episode of convection (or ``subflash'') as burning progresses inward, and the number of these subflashes.  We remove this feature either by artificially resetting the composition to the mixture that existed immediately prior to the core flash or by smoothing the composition over a larger interval in mass.  In the latter method we set the mass fraction $X_i$ of species $i$ according to 
\begin{equation}
\label{eq:sine}
X_i = \frac{X_{i,1}+X_{i,2}}{2}+\frac{X_{i,2}-X_{i,1}}{2}\sin \left[ {\frac{m-m_0}{\Delta m}} \pi \right],
\end{equation}
where $\Delta m$ is the mass over which the composition is smoothed, centred at $m_0$, and $X_{i,1}$ and $X_{i,2}$ denote the interior and exterior compositions.  After modifying the composition using either method we then evolve the model in the evolution code to allow it to return to hydrostatic equilibrium before computing the pulsations.

\section{Results}
\subsection{Overview of the CHeB structure}

\begin{figure}
\includegraphics[width=\linewidth]{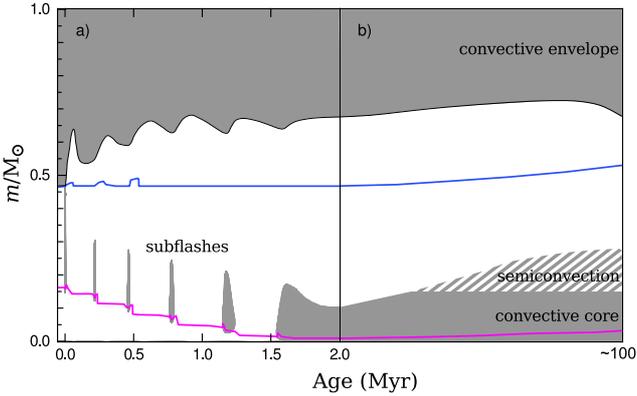}
  \caption{Left panel: Kippenhahn plot of the core flash phase of a 1\,$\text{M}_\odot$ solar-metallicity model.  Right panel: schematic Kippenhahn plot of subsequent quiescent CHeB evolution.  Shaded areas (grey) denote convection zones.  The upper curve (blue) and lower curve (magenta) denote the respective positions of maximum H- and He-burning luminosity.  In the right panel the grey stripes show the region where partial mixing or semiconvection may occur.}
  \label{figure_core_flash_kip}
\end{figure}

We show a schematic of the evolution of the internal structure of a CHeB model in Figure~\ref{figure_core_flash_kip}.  The profile of the Brunt--V{\"a}is{\"a}l{\"a} frequency that develops is crucially dependent on the mixing scheme used at the boundary of the convective core, which is evident from the difference between the models in Figure~\ref{figure_mixing_comparison_zoom}c.  The significant features affecting the Brunt--V{\"a}is{\"a}l{\"a} frequency (proceeding outward from the centre) are as follows.
\begin{enumerate}[(i)]
\item A fully mixed convective core that is Ledoux and Schwarzschild unstable ($N \simeq 0$).
\item A region that may surround the convective core in which material is slowly mixed (see e.g. the grey stripes in Figure~\ref{figure_core_flash_kip}).  Depending on the mixing scheme there can emerge zones with a stabilizing chemical gradient (Schwarzschild marginally stable, Ledoux stable, and $N \gg 0$) or regions which are convective and well-mixed ($N \simeq 0$).  The erratic nature of overshoot can create an irregular $N$ profile in this region that constantly evolves.  In this study, such a region only emerges in the standard-overshoot (Section~\ref{sec:overshoot}) and semiconvection (Section~\ref{sec:sc}) models.
\item A helium-rich radiative region with $N>0$.  In models in which helium ignition begins with the core flash there will be composition gradients between the (fully or partially mixed) core and the H-burning shell (blue line in Figure~\ref{figure_core_flash_kip}).  Only a small fraction of the helium burns in the core flash phase (around 3 per cent in our 1\,$\text{M}_\odot$ models) but this is enough to cause detectable spikes in $N$ from the molecular weight gradients formed at the boundaries of flash and subflash convection zones, e.g. near $r=4.2 \times 10^9\,\text{cm}$ in Figure~\ref{figure_post_core_flash}.  In our models the largest spike is caused by the burning in the initial core flash (closest to the H-burning shell).
\item The H-burning shell, which is strongly stable due to the molecular weight gradient ($N \gg 0$; blue line in Figure~\ref{figure_core_flash_kip}).
\item A radiative zone below the convective envelope (or the surface if the star is not massive enough to have a convective envelope).  $N$ decreases monotonically until the convective envelope, where $N \simeq 0$.  The convective envelope extends all the way to the surface (or close enough for the pulsation calculations we perform in this study).
\end{enumerate}
The radiative region between the convective envelope and the convective core is where g-modes propagate.  Importantly, this includes any partially mixed or semiconvection region surrounding the convective core (e.g., the grey stripes in Figure~\ref{figure_core_flash_kip}).

\begin{figure}
\includegraphics[width=\linewidth]{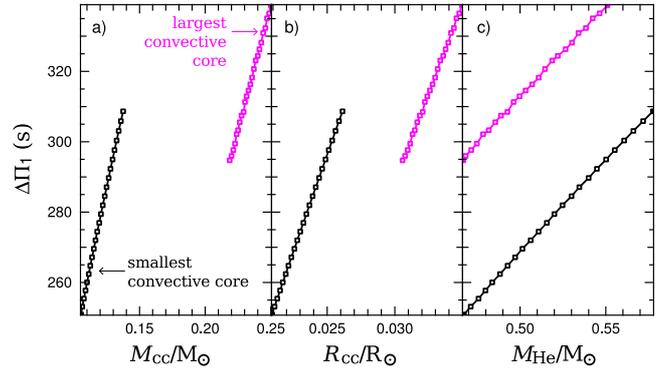}
  \caption{Dependence of the asymptotic g-mode period spacing $\Delta\Pi_1$ on core properties (mass of the convective core $M_\text{cc}$, radius of the convective core $R_\text{cc}$, and mass of the H-exhausted core $M_\text{He}$) for models with the smallest (black) and largest (magenta) possible convective core, artificially constructed by varying $M_\text{He}$.  The models are 1\,M$_\odot$, solar-composition, and consist of a chemically homogeneous convection zone beneath the H-burning shell.  Each model has central helium $Y=0.5$.}
  \label{figure_DP1_dependence}
\end{figure}

\begin{figure}
\includegraphics[width=\linewidth]{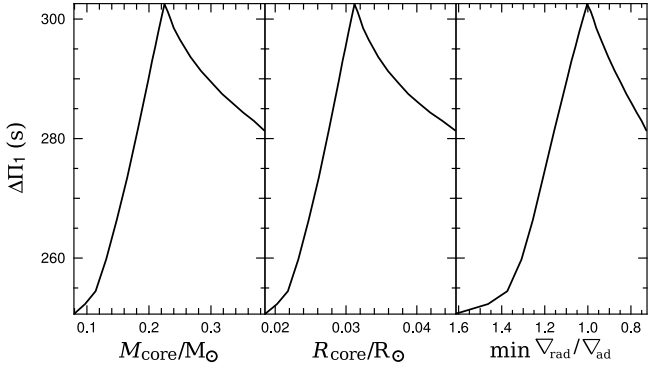}
  \caption{$\Delta \Pi_1$ dependence on the mass, radius and minimum of $\nabla_\text{rad}/\nabla_\text{rad}$ of the fully mixed core $M_\text{core}$.  These models are constructed by varying $M_\text{core}$ while keeping $M_\text{He} = 0.488\,\text{M}_\odot$ and the core convection zone composition constant with $Y=0.5$.  Each model is 1\,M$_\odot$ and solar metallicity.}
  \label{figure_DP1_core_mass}
\end{figure}

\subsection{$\Delta\Pi_1$ dependence on bulk core properties}
\label{sec:bulk_properties}
We have tested the dependence of $\Delta\Pi_1$ on the fundamental properties of the core -- the mass of the H-exhausted core $M_\text{He}$ (Figure~\ref{figure_DP1_dependence}) and the mass of the fully mixed core $M_\text{core}$ (Figure~\ref{figure_DP1_core_mass}).  

In Figure~\ref{figure_DP1_dependence} we increase $M_\text{He}$ through H-burning then construct models with the smallest (black markers) and largest (magenta markers) possible convective cores for a fixed central composition of $Y = 0.5$.  In this test the mass of the convective core $M_\text{cc}$ (Figure~\ref{figure_DP1_dependence}a) and radius of the convective core $R_\text{cc}$ (Figure~\ref{figure_DP1_dependence}b) are dependent variables.  The smallest convective core is found by extending it only until the region adjacent to the convection zone (i.e. the radiative side) is marginally stable to convection, which gives the same structure as the ``no overshoot'' models in this study.  In contrast, the models with the largest core are constructed by extending the convective core as far as possible so that the entire region within it remains convectively unstable, which is the same as for the maximal-overshoot models (Figure~\ref{figure_mixing_comparison_zoom}, Section~\ref{sec:max_os_results}).  We find that in both the smallest and largest convective core cases $M_\text{cc}$, $R_\text{cc}$, and $\Delta\Pi_1$ are linearly dependent on $M_\text{He}$.  \citet{2013ApJ...766..118M} have already highlighted the linear dependence of $\Delta\Pi_1$ on $R_\text{cc}$ in low-mass CHeB models, as well as the importance of $M_\text{He}$, including a linear relationship between $\Delta\Pi_1$ and $M_\text{He}$ for more massive models.  The difference in $\Delta\Pi_1$ between the smallest and largest core cases is 45\,s.  It is evident from the fact that two values of $R_\text{cc}$ can correspond to the same value of $\Delta\Pi_1$ that there is not a single linear dependence of $\Delta\Pi_1$ on $R_\text{cc}$ (Figure~\ref{figure_DP1_dependence}b).  In this case the relationship also depends on how the convective boundary is defined.

In Figure~\ref{figure_DP1_core_mass} we show the effect of artificially changing the mass of the homogeneous (fully mixed) region in the core, $M_\text{core}$, while keeping the central composition and the H-exhausted core mass constant.  Note that $M_\text{core}$ differs from $M_\text{cc}$ in that there is no requirement that the entire region enclosed by $M_\text{core}$ is convective according to the Schwarzschild criterion.  The peak in $\Delta\Pi_1$ occurs exactly when the convective zone is as large as possible ($M_\text{core} = 0.22\,\text{M}_\odot$).  This clearly demonstrates that further extending the fully mixed core (e.g. as a result of overshoot; such as the ``high overshoot'' model from \citealt{2003ApJ...583..878S}) does \textit{not} continue to increase $\Delta\Pi_1$ when part of it becomes stable to convection according to Schwarzschild, allowing g-modes to propagate.  This is of interest because it is not unreasonable to imagine that convective overshoot could allow the composition of two nearby convection zones to remain homogeneous.  In this example, models with $0.22 < M_\text{core}/\text{M}_\odot < 0.33$ have two separate convection zones.  

\begin{figure}
\includegraphics[width=\linewidth]{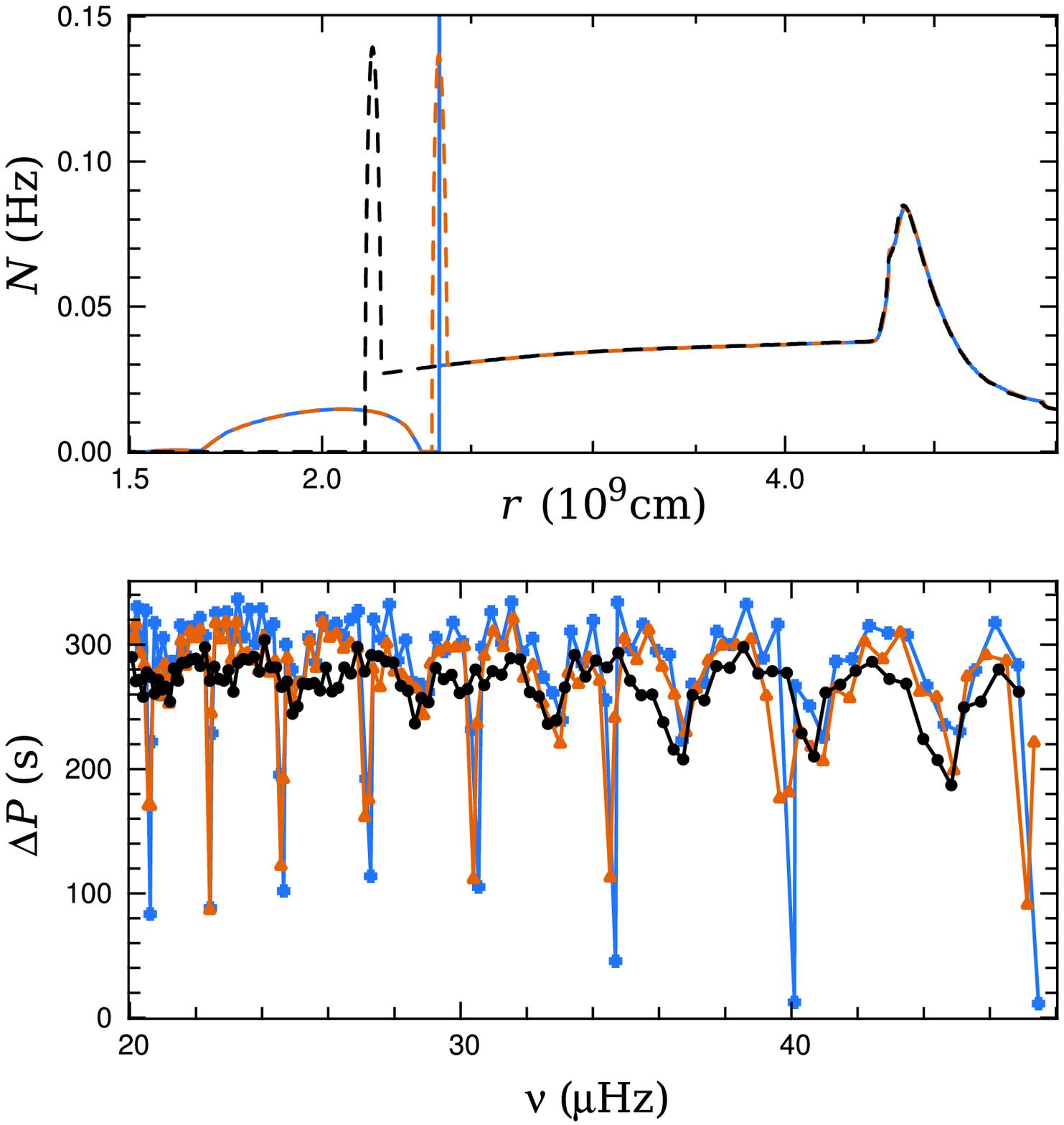}
\par
\vspace{0.35cm}
\includegraphics[width=\linewidth]{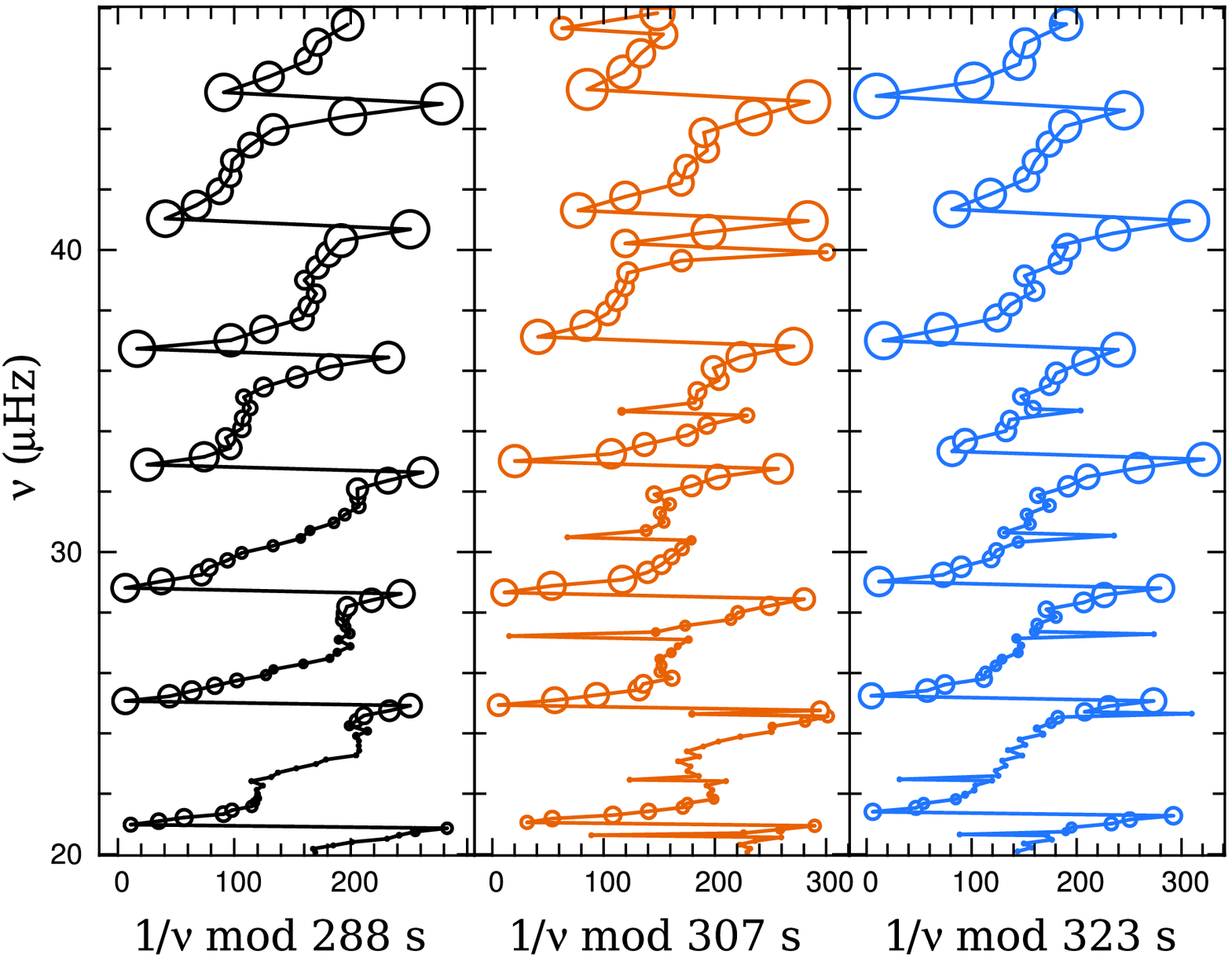}
  \caption{Upper panels: Brunt--V{\"a}is{\"a}l{\"a} frequency and $\ell=1$ period spacing for three models with different sized fully mixed cores and different chemical profiles near the boundary.  Here we have (i) a smooth boundary with a fully mixed core that is convective (black), and (ii) a fully mixed core too large to be convective with a smooth (orange) and (iii) sharp (blue) composition profile at the boundary.  The composition profiles are set according to Equation~\ref{eq:sine} with $\Delta m =  0.01\,\text{M}_\odot$ (smooth profile) and $\Delta m = 10^{-5}\,\text{M}_\odot$ (sharp profile).  Lower panels: period \'echelle diagrams for the three models.  Larger symbols correspond to lower mode inertia.  The best fits for $\Delta P_\text{\'ech}$ are achieved for period modulo 288\,s, 307\,s, and 323\,s.  This compares to the asymptotic g-mode period spacing ($\Delta\Pi_1$) of 287\,s, 284\,s, and 295\,s, respectively.  Each model has approximately $R=10.7\,\text{R}_\odot$, $T_\text{eff}=4760\,\text{K}$, and $\nu_\text{max} = 27\,\mu\text{Hz}$.}
  \label{figure_hos}
\end{figure}

In Figure~\ref{figure_hos} we compare the $\ell = 1$ period spacing for models with fully mixed cores of different sizes: one with a fully mixed core mass of 0.215\,$\text{M}_\odot$ (black dashes) and another otherwise identical model, with a fully mixed core mass of 0.255\,$\text{M}_\odot$ (orange dashes), which is too large to be convective throughout and therefore gives rise to an additional radiative region ($1.7 \times 10^9\,\text{cm} \la r \la 2.3\times 10^9\,\text{cm}$ in Figure~\ref{figure_hos}).  The asymptotic period spacing (from integrating over the entire structure according to Equation~\ref{eq:dp1_asymp}) is slightly lower for the model with the larger core (by 3\,s).  Some modes in this model are very closely spaced in period because they are trapped in the additional radiative region (see also Sections~\ref{sec:overshoot_results} and \ref{sec:sc_results}), whereas most pairs of modes have $\Delta P >\Delta\Pi_1$.  If the mode periods are plotted in the \'echelle diagram (see Section~\ref{sec:puls_methods}), the period spacing required for a regular pattern, $\Delta P_\text{\'ech}$, is 19\,s higher than the asymptotic value.  The reason for this can be understood by considering the local buoyancy radius $\Pi^{-1}(r)$ described by \citet{2008MNRAS.386.1487M}.  At a given point $r$, this is defined by
\begin{equation} 
\label{eq:buoyancy_radius}
\Pi^{-1}(r)= \int_{r_0}^{r} \frac{N}{r'} \text{d}r',
\end{equation}
where $r_0$ is the radius at the edge of the convective core ($r \simeq 1.7 \times 10^9\,\text{cm}$ in Figure~\ref{figure_hos}).  This gives the contribution to the integral in Equation~\ref{eq:dp1_asymp} from the region enclosed by the point at radius $r$.  The total buoyancy radius is the same integral evaluated over the entire g-mode propagation zone.  

In the model with the larger fully mixed core (orange dashes in Figure~\ref{figure_hos}) the interior buoyancy cavity accounts for 7.8 per cent of the total buoyancy radius, which corresponds to the difference between $\Delta\Pi_1$ and $\Delta P_\text{\'ech}$ (which are 284\,s compared to 307\,s respectively).  If we exclude this interior cavity from the calculation of $\Delta\Pi_1$ we get almost exactly $\Delta\Pi_1 = \Delta P_\text{\'ech}$ (308\,s compared to 307\,s).  This divergence between $\Delta\Pi_1$ and $\Delta P_\text{\'ech}$ is also apparent for the model with a sharp composition profile in Figure~\ref{figure_hos} (in blue), demonstrating that it does not depend on the composition profile at the edge of the fully mixed core.  We therefore expect that the existence of a second radiative zone would generally cause the observationally inferred value of $\Delta\Pi_1$ (using the method of \citealt{2012A&A...540A.143M}) to increase above its theoretical value (computed by integrating over the entire structure).  This example highlights the possibility that a difficulty in accurately determining $\Delta\Pi_1$ from observations may contribute to its apparent discrepancy with predictions from models.  This phenomenon is discussed in more detail for a related example in Section~\ref{sec:overshoot_results}.

\begin{figure}
\includegraphics[width=\linewidth]{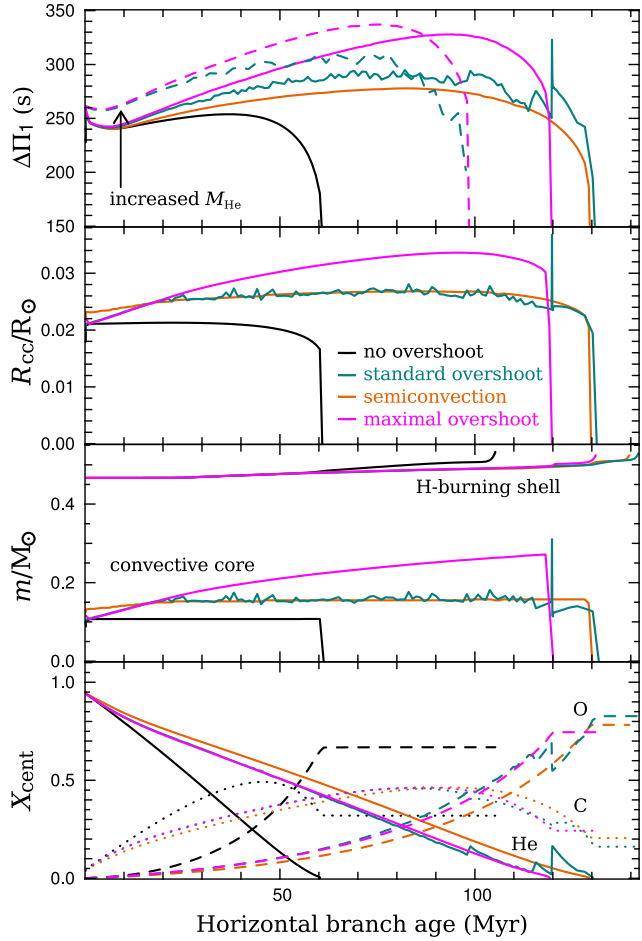}
  \caption{Evolution of the various 1\,$\text{M}_\odot$ models with four different treatments of convective boundaries during the CHeB phase.  Models have the same colours as Figures \ref{figure_helium_ev} and \ref{figure_mixing_comparison_zoom}.  Properties shown, from top to bottom, are the asymptotic g-mode period spacing $\Delta\Pi_1$, the radius of the convective core $R_\text{cc}$, the mass of the H-exhausted and the convective core, and the central helium (solid line), carbon (dots) and oxygen (dashes) mass fractions.  In the top panel dashed lines indicate models with an increased core mass $\Delta M_\text{He}=0.025$\,$\text{M}_\odot$ at the beginning of the CHeB phase.}
  \label{figure_one_msun_core}
\end{figure}

Figure~\ref{figure_one_msun_core} shows the evolution of the internal structure of models with the four different mixing schemes.  The size (in mass and radius) of the convective core in the semiconvection and standard overshoot sequences is similar throughout the evolution, except when overshooting permits core breathing pulses near core helium exhaustion.  This explains the similarity in $\Delta\Pi_1$ evolution.  In both the semiconvection and standard-overshoot sequences almost all of the growth in the mass of the convective core occurs during the first 20\,Myr (Figure~\ref{figure_one_msun_core}c).  Subsequently, helium is transported into the core by the expansion of the partially mixed region.  Interestingly, the rate of depletion of helium in the core is exactly the same for the maximal-overshoot and standard-overshoot runs until the final (and largest) core breathing pulse extends the standard-overshoot model CHeB lifetime. 

At the beginning of CHeB, each of the four standard sequences (solid lines) shows a decrease in $\Delta\Pi_1$ (Figure~\ref{figure_one_msun_core}a).  This can be attributed to the softening of the steep composition gradient at the H-burning shell.  This is further discussed in Section~\ref{sec:postcoreflash} and its effect on $\Delta\Pi_1$ is also explained for an analogous case in Section~\ref{sec:ensemble}.  After hydrogen burning resumes over the entire shell, the evolution of $\Delta\Pi_1$ closely tracks the radius of the convective core, which has been shown by \citet{2013ApJ...766..118M}.  

In dashed lines in Figure~\ref{figure_one_msun_core}a we also show additional sequences with standard and maximal overshoot that result from enlarging $M_\text{He}$ by $\Delta M_\text{He} = 0.025$\,M$_\odot$ at the beginning of CHeB.  This was achieved by delaying helium ignition through an ad hoc increase to the neutrino emission rate during the RGB phase.  This increases the average $\Delta\Pi_1$ during CHeB by 18\,s for the standard-overshoot sequence and 11\,s for the maximal-overshoot case.   Most significantly, it increases $\Delta\Pi_1$ by around 20\,s early in the CHeB phase for both sequences.  The faster rate of helium burning resulting from the larger $M_\text{He}$ exhausts the fuel earlier, shortening the CHeB lifetime by around 25 per cent for both mixing schemes.

As core helium burning progresses, the convective core becomes increasingly C- and O-rich (Figure~\ref{figure_one_msun_core}d) and consequently more dense.  This causes the convective core radius to decrease (Figure~\ref{figure_one_msun_core}b), even when its mass does not and irrespective of the mixing scheme (Figure~\ref{figure_one_msun_core}c).  It is evident from the decrease in $\Delta\Pi_1$ towards the end of CHeB for every sequence shown in Figure~\ref{figure_one_msun_core}a that $\Delta\Pi_1$ is more closely dependent on convective core radius than mass.  The final composition of the degenerate C-O core is another potential diagnostic for mixing, but the range covered by these four different schemes is small (oxygen varies by around 15 per cent) and the situation is further complicated by reaction rate uncertainties \citep{2003ApJ...583..878S}. 

In this section we have explored how differences in the stellar structure affect $\Delta\Pi_1$.  Even among the models without a semiconvection or partially mixed zone, $\Delta\Pi_1$ depends on a number of factors: $R_\text{cc}$, $M_\text{He}$, and core composition.  Additionally, if the chemically homogeneous region in the core is large enough for part of it to become radiative, mode trapping can cause the period spacing to increase above the asymptotic value.  Such an effect would increase the $\Delta\Pi_1$ inferred from observations, and therefore help to explain why standard CHeB models do not match the average $\Delta\Pi_1$ for the \textit{Kepler} field stars.

\subsection{Pulsations in early post core-flash CHeB models}
\label{sec:postcoreflash}

\begin{figure}
\includegraphics[width=\linewidth]{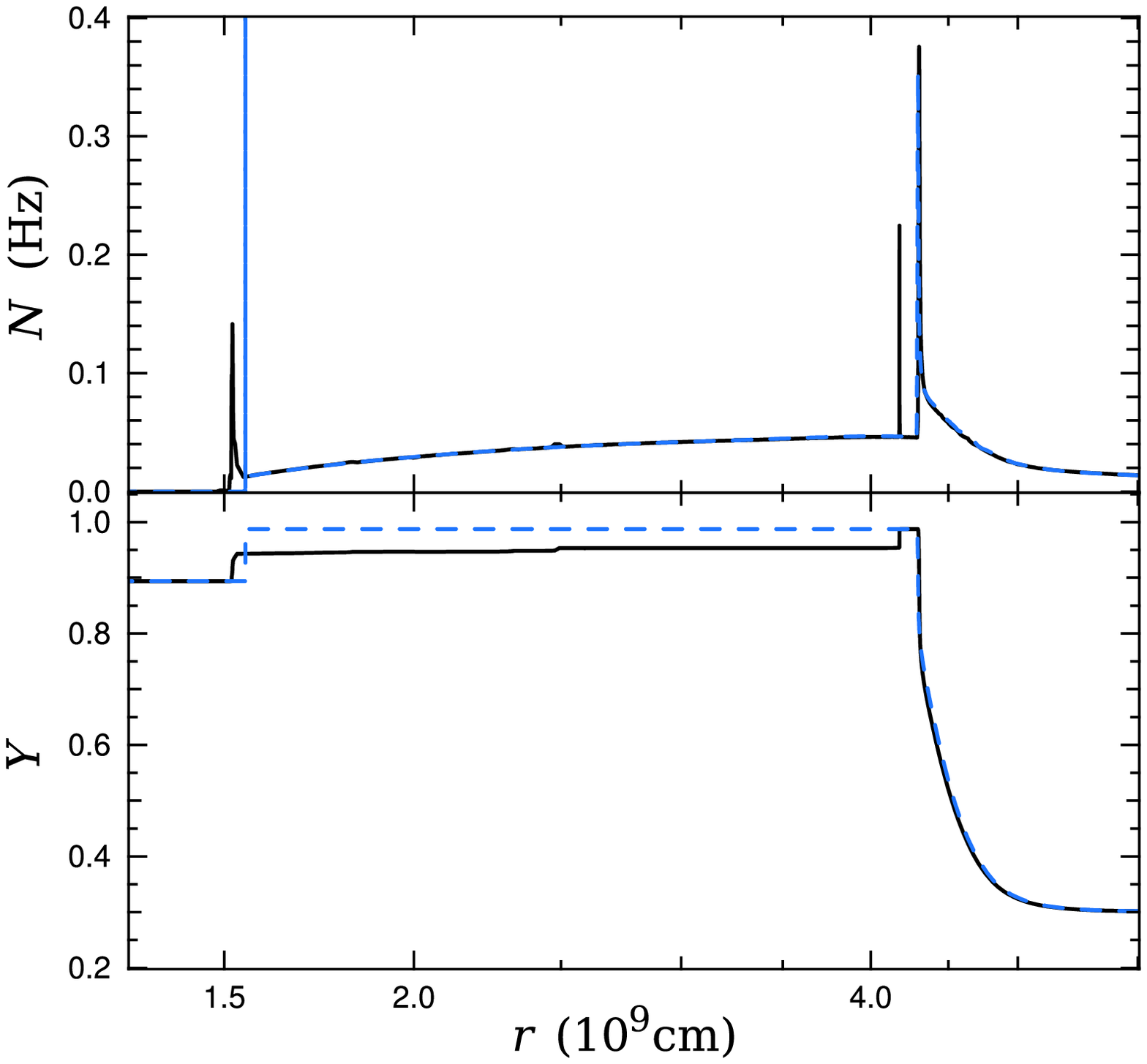}
\par
\vspace{0.35cm}
\includegraphics[width=\linewidth]{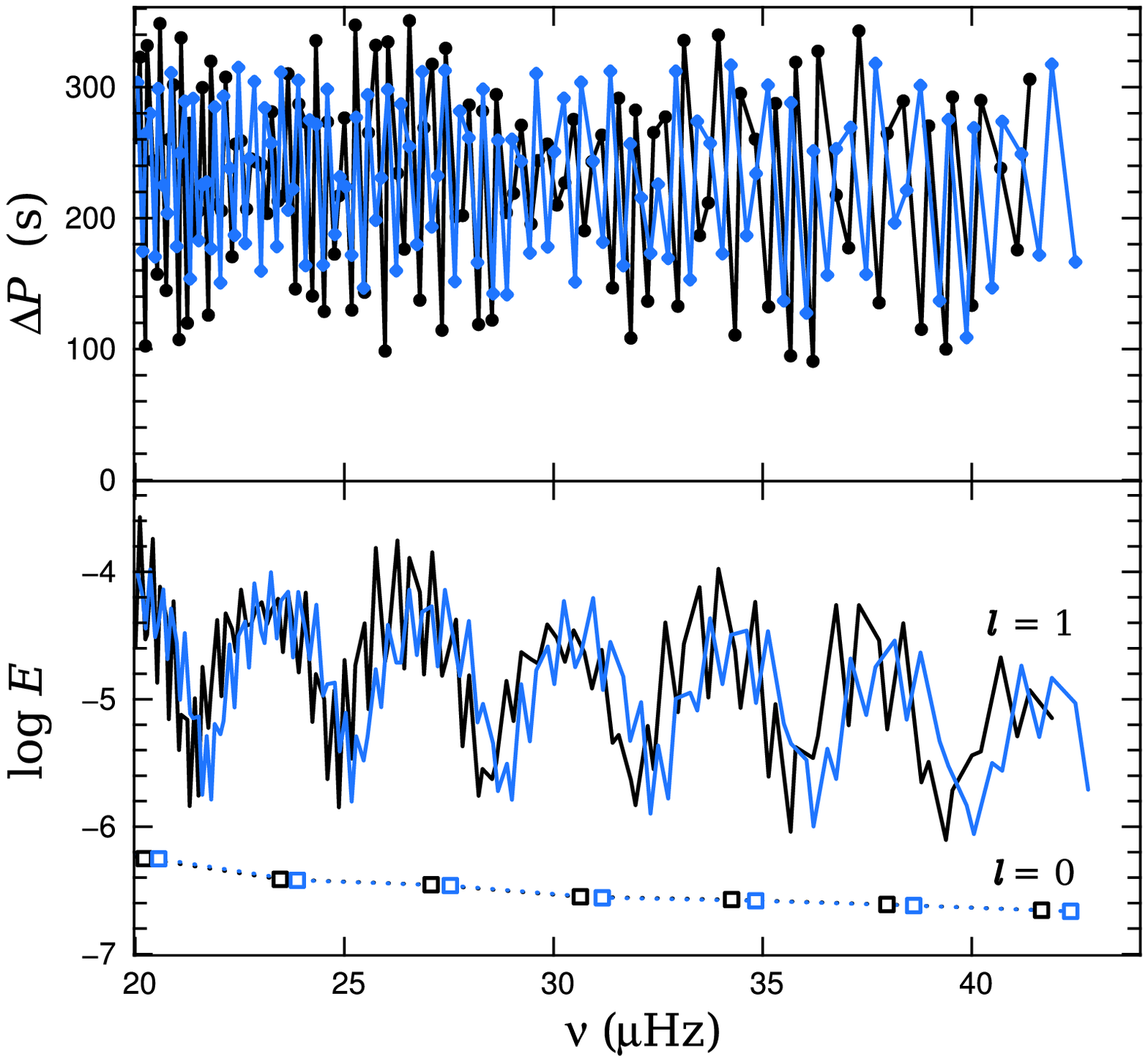}
  \caption{Pulsation properties of 1\,$\text{M}_\odot$ models with standard overshoot 2.7\,Myr after the onset of convective core helium burning taken directly from the evolution code (black) and with the region outside the convective core homogenized (blue).  First panel: Brunt--V{\"a}is{\"a}l{\"a} frequency $N$.  Second panel: helium mass fraction $Y$.  Third panel: $\ell = 1$ period spacing.  Fourth panel: $\ell = 0$ and $\ell = 1$ mode inertia normalized to the surface radial displacement.  The model from the evolution code has $\Delta\Pi_1= 240$\,s, $R=11.6\,\text{R}_\odot$, $T_\text{eff}=4580\,\text{K}$, and $\nu_\text{max} = 23\,\mu\text{Hz}$ whereas the model with the homogenized composition has $\Delta\Pi_1 = 238$\,s, $R=11.5\,\text{R}_\odot$, $T_\text{eff}=4570\,\text{K}$, and $\nu_\text{max} = 24\,\mu\text{Hz}$.}
  \label{figure_post_core_flash}
\end{figure}

Neutrino emission from plasmon decay during the RGB phase is strongest at the centre, where the density is highest.  This energy loss is enough to move the position of maximum temperature, and therefore He-ignition, off-centre.  After helium ignition a sequence of several subflashes move inward until the burning reaches the core and quiescent CHeB begins (Figure~\ref{figure_core_flash_kip}).  In one of the first studies making use of the mixed mode detection in red giants, \citet{2012ApJ...744L...6B} found that it may be possible to identify stars in the core flash phase by using the fact that (between the subflashes) the g-mode period spacing is expected to be much lower than for quiescent CHeB stars, but still higher than for RGB stars.  Their approach of studying the population in the \textit{Kepler} field, if successful, could reveal the lifetime of the core flash phase and the nature of the mixing, and therefore also shed light on the structure before the core flash (which is dependent on neutrino losses for example).  Here we examine the computed pulsation spectra of a model in the early post core-flash phase, and test the effect of the remaining abundance profile.

The inward progression of convection and burning during the core flash phase has a lasting effect on the Brunt--V{\"a}is{\"a}l{\"a} frequency.  Two of these features in our early-CHeB model are visible in Figure~\ref{figure_post_core_flash}: the peak in $N$ at $r = 1.5\times 10^{9}\,\text{cm}$, due to the initial recession of the convective core at the beginning of CHeB; and another at $r = 4.2\times 10^{9}\,\text{cm}$, which is caused by the first episode of core-flash burning.  However, from a seismic perspective, the dominant feature of early-CHeB models is the thinness of the H-burning shell, which is due to the relatively steep temperature gradient in the prior luminous RGB phase.  In Figure~\ref{figure_post_core_flash} we confirm that the composition profile at the H-burning shell causes the irregular period spacing pattern by showing that it still exists for a model with the sharp $N$ feature from the first episode of core-flash burning removed (blue model).

\begin{figure}
\includegraphics[width=\linewidth]{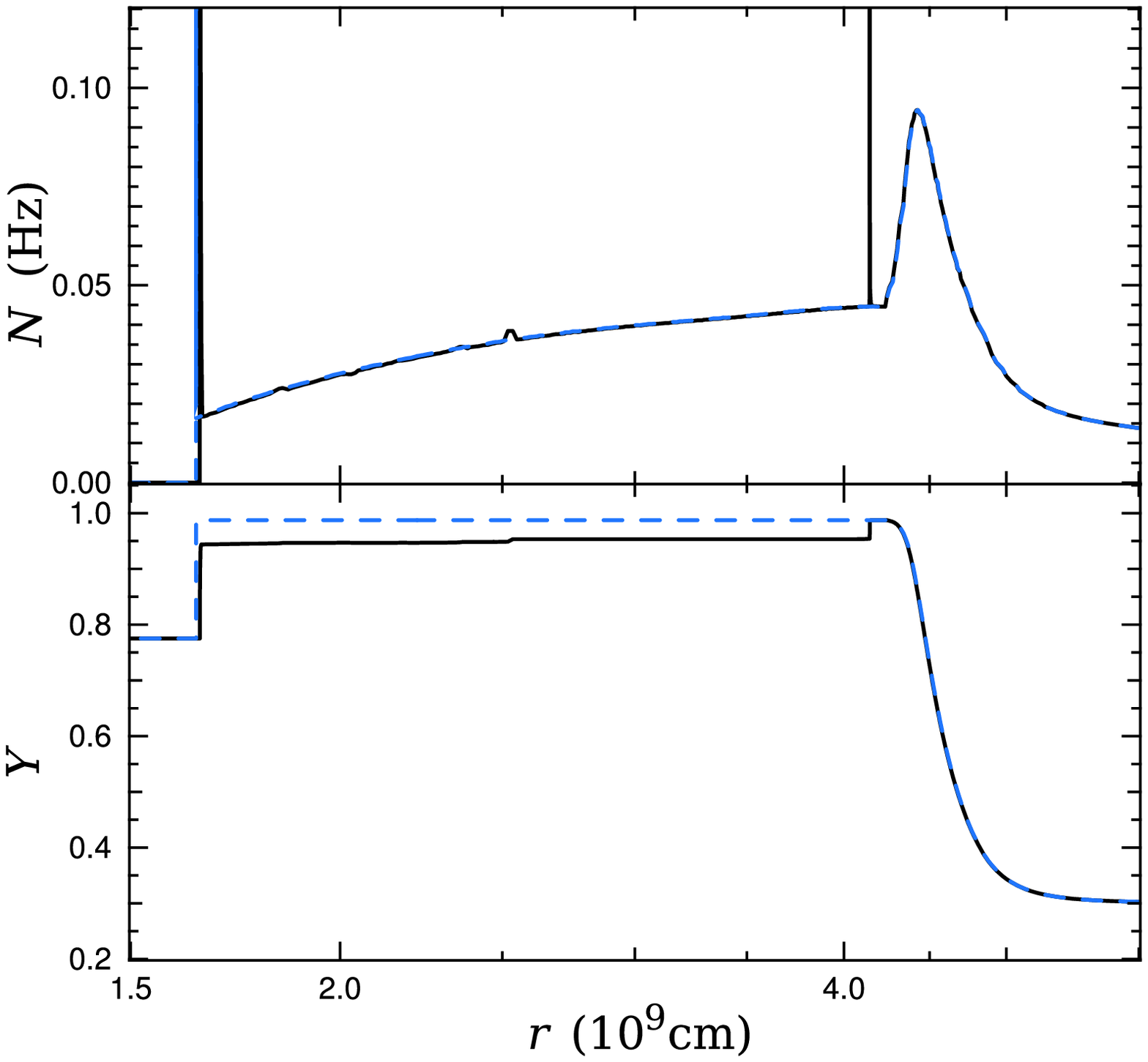}
\par
\vspace{0.35cm}
\includegraphics[width=\linewidth]{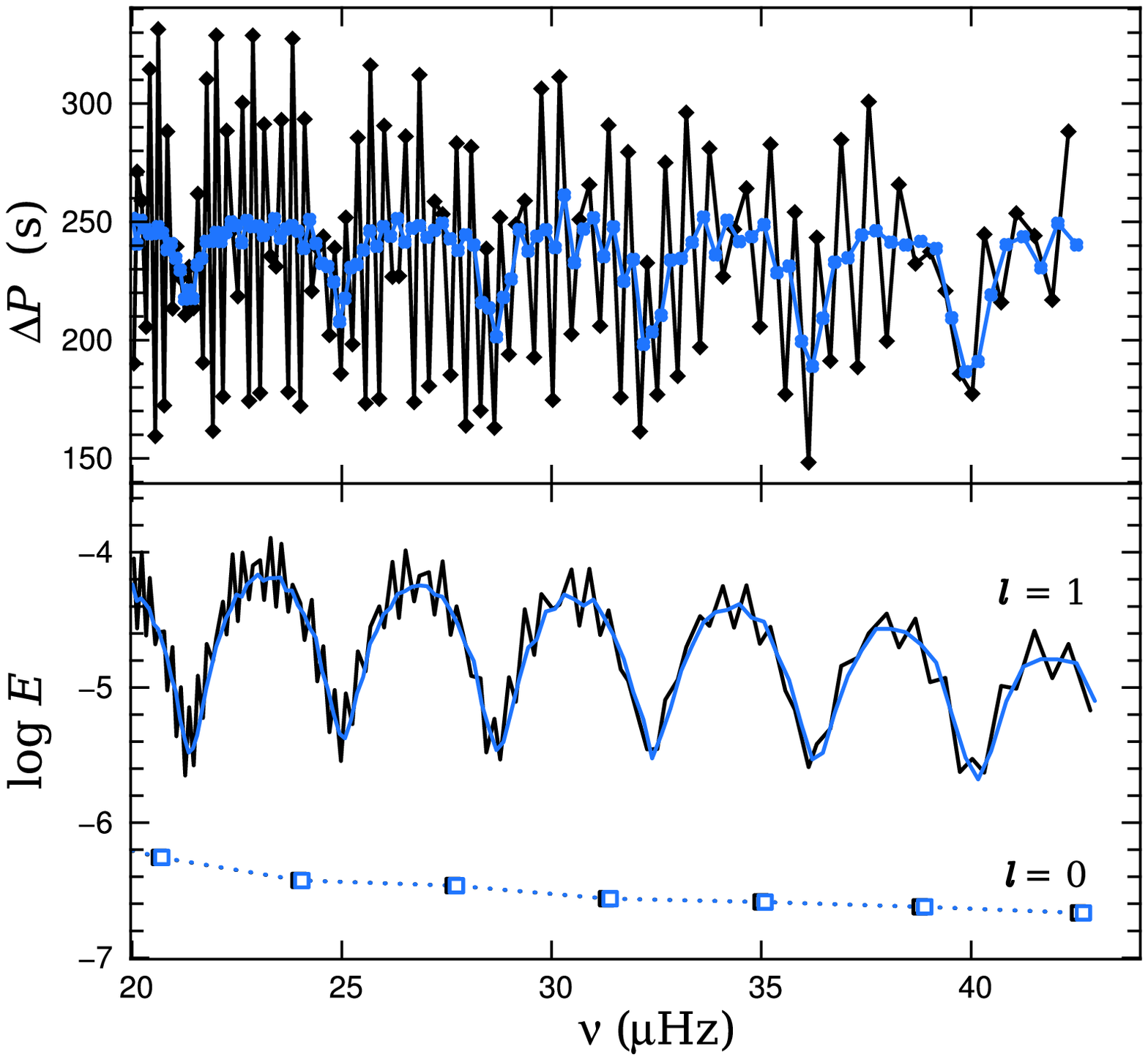}
  \caption{Pulsation properties of 1\,$\text{M}_\odot$ models with standard overshoot 14.8\,Myr after the end of the core flash phase from the evolution code (black) and with the region outside the convective core homogenized (blue).  Both models have $\Delta\Pi_1 = 247$\,s.  These models have $R=11.4\,\text{R}_\odot$, $T_\text{eff}=4590\,\text{K}$, and $\nu_\text{max} = 24\,\mu\text{Hz}$.}
  \label{figure_normal_H_shell}
\end{figure}

\begin{figure}
\includegraphics[width=\linewidth]{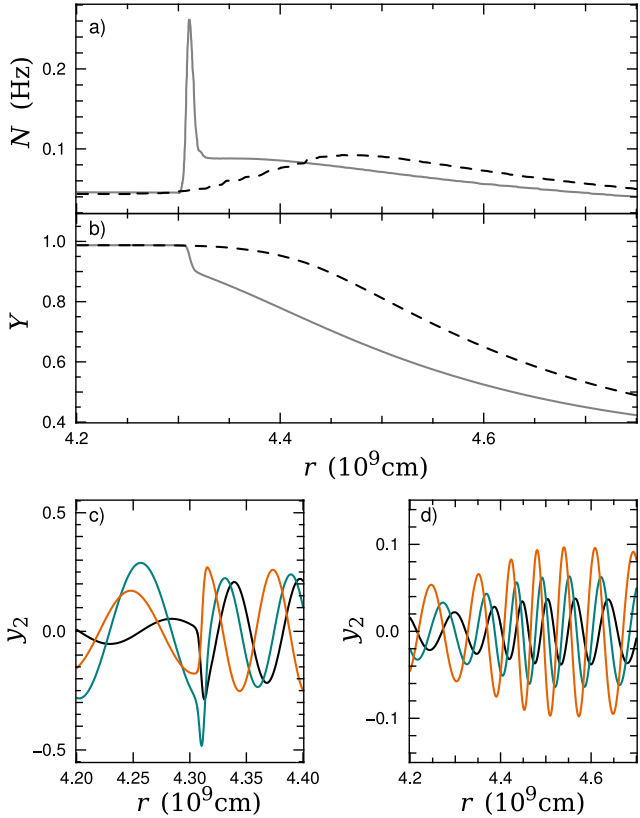}
\caption{Upper panels: comparison of the Brunt--V{\"a}is{\"a}l{\"a} frequency $N$ and helium mass fraction $Y$ near the edge of the H-exhausted core for two early-CHeB models.  The model represented by the solid lines is 6.8\,Myr after the beginning of the CHeB phase and the dashed model has evolved for another 10\,Myr.  Lower panels: scaled horizontal displacement eigenfunctions $y_2$ (defined in Equation~\ref{eq:y1}) for $\ell=1$ modes in c) the earlier, and d) the later model.  The modes are of consecutive radial order $n=-91,-92,-93$ (in orange, cyan, and black, respectively), and have frequency of roughly 42\,$\mu$Hz.  These models have approximately $R=11.3\,\text{R}_\odot$, $T_\text{eff}=4570\,\text{K}$, and $\nu_\text{max} = 24\,\mu\text{Hz}$.}
  \label{figure_post_core_flash_eigen}
\end{figure}

In our solar-mass runs it takes more than 14\,Myr for hydrogen burning to smooth out the composition gradient at the inside of the shell at $r = 4.5\times 10^{9}\,\text{cm}$, as shown in Figure~\ref{figure_normal_H_shell}, and thus for a regular pattern in the period spacing to emerge.  In Figure~\ref{figure_post_core_flash_eigen} we show the effect of this sharp composition gradient on the eigenfunctions.  After the core flash there is a sharp peak in the buoyancy frequency (at $r = 4.3 \times 10^9\,\text{cm}$ in Figure~\ref{figure_post_core_flash_eigen}a) which traps modes of consecutive radial order to very different extents (Figure~\ref{figure_post_core_flash_eigen}c). This buoyancy peak is then slowly eroded by hydrogen burning (Figure~\ref{figure_post_core_flash_eigen}b), and once the hydrogen burning shell completely reactivates the buoyancy peak is broad compared to the characteristic eigenfunction wavelength (Figure~\ref{figure_post_core_flash_eigen}d), leaving the period spacing more regular.  If this picture is true for real stars we anticipate difficulty in determining $\Delta\Pi_1$ for up to 15 per cent of red clump stars (based on a 100\,Myr CHeB lifetime) and also every star in the core flash phase.  

\begin{figure}
\includegraphics[width=\linewidth]{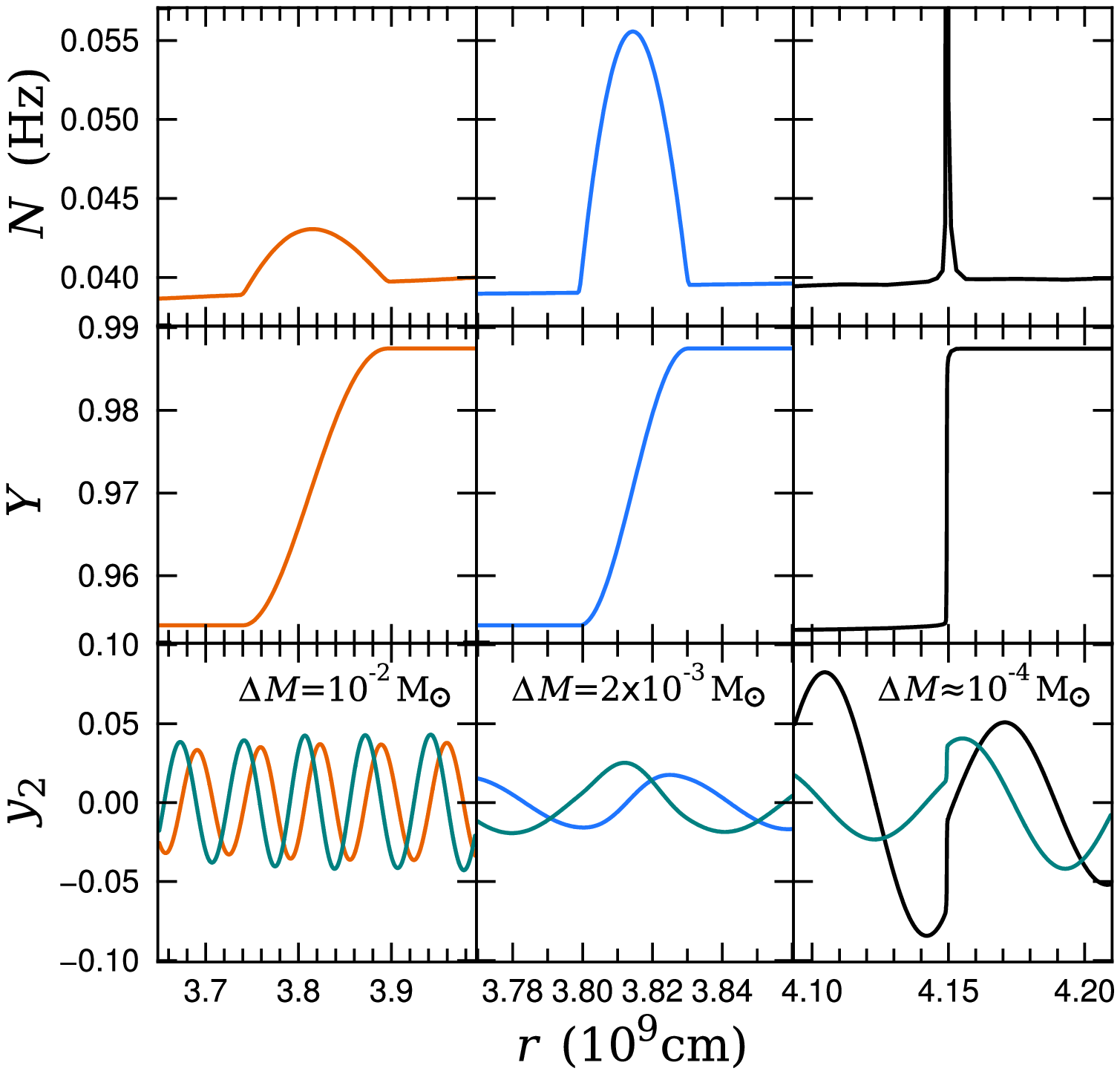}
\par
\vspace{0.35cm}
\includegraphics[width=\linewidth]{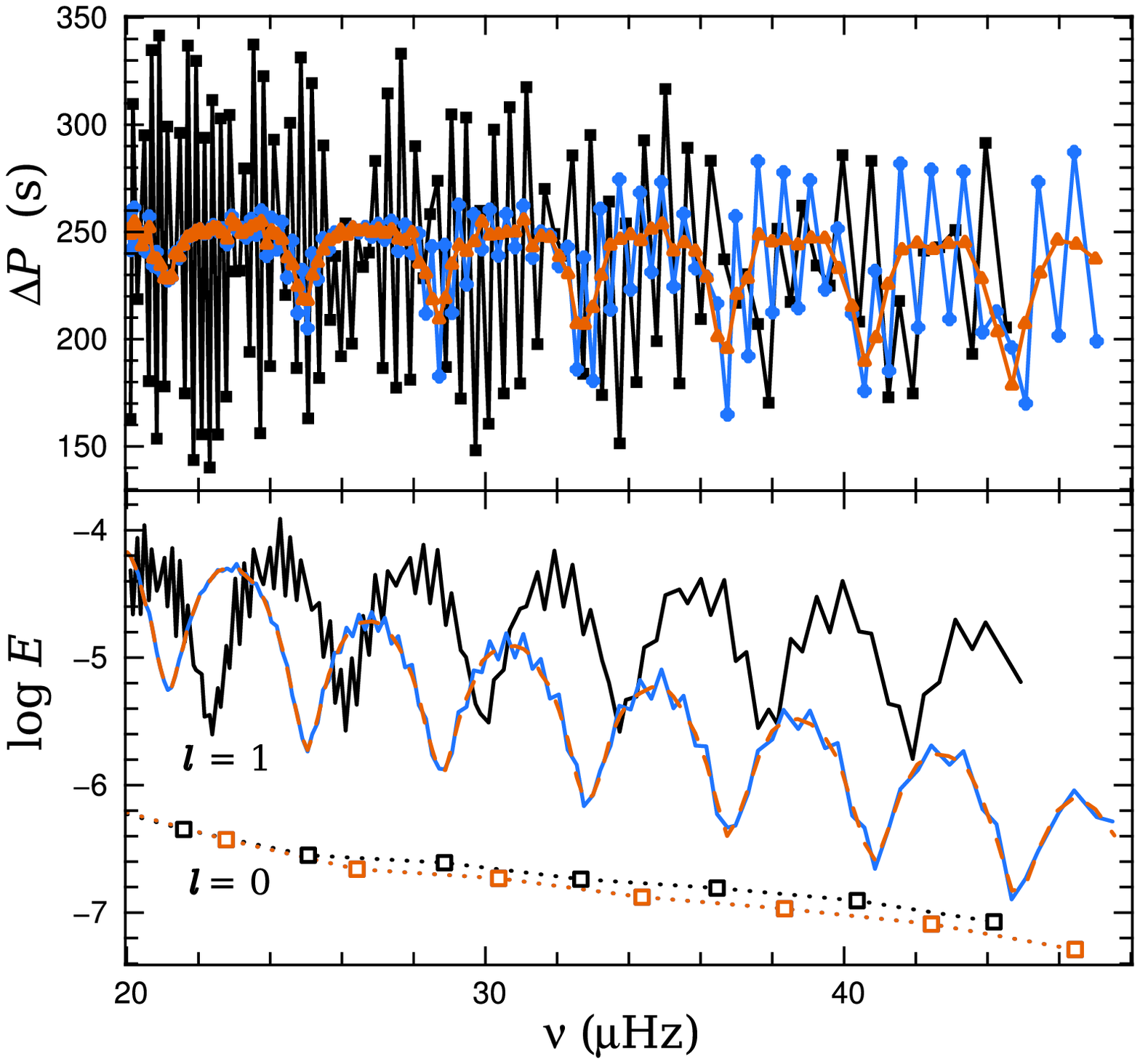}
  \caption{Comparison of pulsation properties of the 1\,$\text{M}_\odot$ models with different composition profiles from the initial core flash.  Upper panels: Brunt--V{\"a}is{\"a}l{\"a} frequency $N$, helium mass fraction $Y$, and scaled horizontal displacement eigenfunctions $y_2$ (defined in Equation~\ref{eq:y1}) for two consecutive $\ell  =1$ modes with $\nu \approx 26\,\mu$Hz.  From left to right the models have chemical profiles artificially smoothed over 0.01\,$\text{M}_\odot$ (orange), 0.002\,$\text{M}_\odot$ (blue), and approximately $10^{-4}$\,$\text{M}_\odot$ (from the evolution code; black).  The former two are of the form described by Equation~\ref{eq:sine} and have approximately $R=11.1\,\text{R}_\odot$, $T_\text{eff}=4750\,\text{K}$, and $\nu_\text{max} = 27\,\mu\text{Hz}$ while the latter model has $R=11.1\,\text{R}_\odot$, $T_\text{eff}=4610\,\text{K}$, and $\nu_\text{max} = 25\,\mu\text{Hz}$.  Lower panels: $\ell = 1$ mode period spacing and normalized inertia for the above models (same colours).}
  \label{figure_core_flash_discontinuity}
\end{figure}

Only after hydrogen burning resumes throughout the shell does the effect of any composition profile left by the core flash become dominant (it is responsible for the difference between the two models in Figure~\ref{figure_normal_H_shell}).  In Figure~\ref{figure_core_flash_discontinuity} we show that the nature of this chemical profile strongly affects the pulsations.  We compare two chemical profiles that vary smoothly according to Equation~\ref{eq:sine} and one that has the composition profile from the evolution code.  It is clear in the case with the composition gradient spread over $\Delta m = 0.002\,\text{M}_\odot$ (central panels in Figure~\ref{figure_core_flash_discontinuity}) that the high-frequency (low radial order) modes are most sensitive to this feature (blue model in Figure~\ref{figure_core_flash_discontinuity}).  This is because adjacent modes are affected differently when the characteristic eigenfunction wavelength is large enough to be comparable to the size of the peak in $N$.  This causes the variation in $\Delta P$ between consecutive pairs of modes.  If the composition profile is smoother, the effect of this feature diminishes (e.g., in the orange model with $\Delta m = 0.01\,\text{M}_\odot$; left panel in Figure~\ref{figure_core_flash_discontinuity}), and the period spacing resembles models where it is absent (e.g. the blue model in Figure~\ref{figure_normal_H_shell}).  Conversely, the sharper composition profile produced by the evolution code (and dependent on the treatment of the convective boundary during the core flash) produces a very obvious effect on the period spacing.  Specifically, it introduces a large mode-to-mode variation in $\Delta P$ throughout the frequency range examined (black model in Figure~\ref{figure_normal_H_shell}).  This behaviour can be explained by analysing the eigenfunctions.  The spike in $N$ is located at about half the buoyancy radius (it is at 57 per cent of the total buoyancy radius in this particular case) where neighbouring eigenfunctions are separated in phase by about $\pi/2$, so it affects consecutive modes differently.  In this case $\Delta m \approx 10^{-4}\,\text{M}_\odot$ which is small compared to the eigenfunction wavelength.

These examples serve as a note of caution when computing pulsations for models with steep composition gradients.  The steepness of those resulting from the core flash phase is particularly dependent on convective overshoot, which is uncertain.  This is the reason we have smoothed the composition profile after the core flash phase for many models shown in this paper.

\begin{figure}
\includegraphics[width=\linewidth]{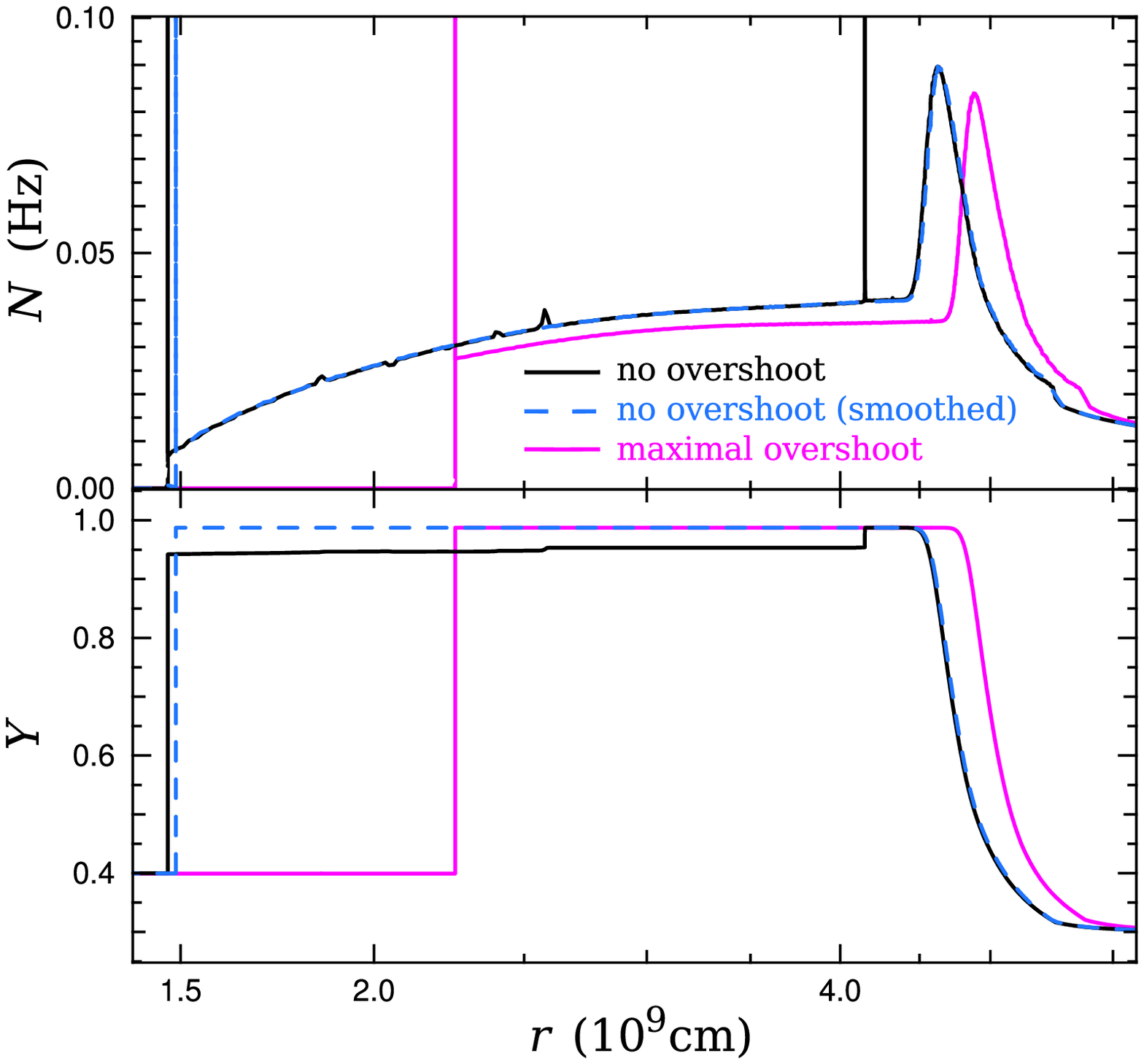}
\par
\vspace{0.35cm}
\includegraphics[width=\linewidth]{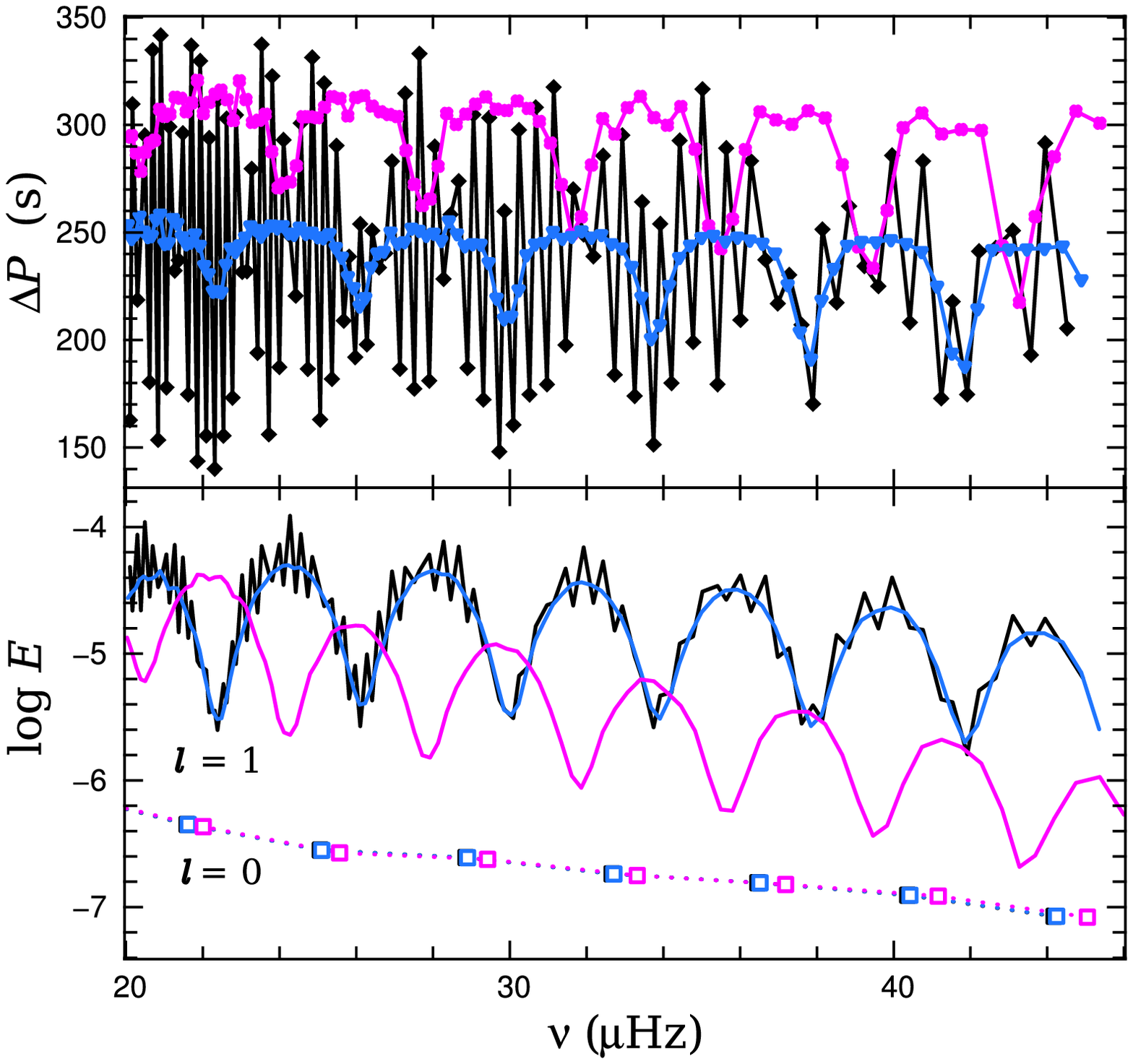}
  \caption{Pulsation properties of 1\,$\text{M}_\odot$ models without convective overshoot from the evolution code (black) and with the region outside the convective core homogenized (blue), and a model with maximal overshoot (magenta). Both models without overshoot have $\Delta\Pi_1=252$\,s while the model with maximal overshoot has $\Delta\Pi_1=314$\,s.  The models have approximately $R=11.1\,\text{R}_\odot$, $T_\text{eff}=4610\,\text{K}$, and $\nu_\text{max} = 25\,\mu\text{Hz}$.}
  \label{figure_no_overshoot}
  \label{figure_max_overshoot}

\end{figure}

\subsection{Pulsation properties for models with different mixing schemes}
\label{sec:pulsations_diff_mixing}

\subsubsection{Models without convective overshoot}

Our models without convective overshoot do not develop a partially mixed region, and experience negligible growth in the mass of the convective core.  When the effects of core flash mixing are excluded, and after H-burning broadens the shell, these models have a simple buoyancy profile, and consequently a simple period spacing pattern (Figure~\ref{figure_no_overshoot}).  This period spacing pattern closely resembles RGB models (e.g. Figure 1b in \citealt{2011Natur.471..608B}) and observations (e.g. KIC 9882316 in Fig. 1 in \citealt{2012A&A...540A.143M}) except that the period spacing is higher.

\subsubsection{Models with standard overshoot}
\label{sec:overshoot_results}

\begin{figure}
\includegraphics[width=\linewidth]{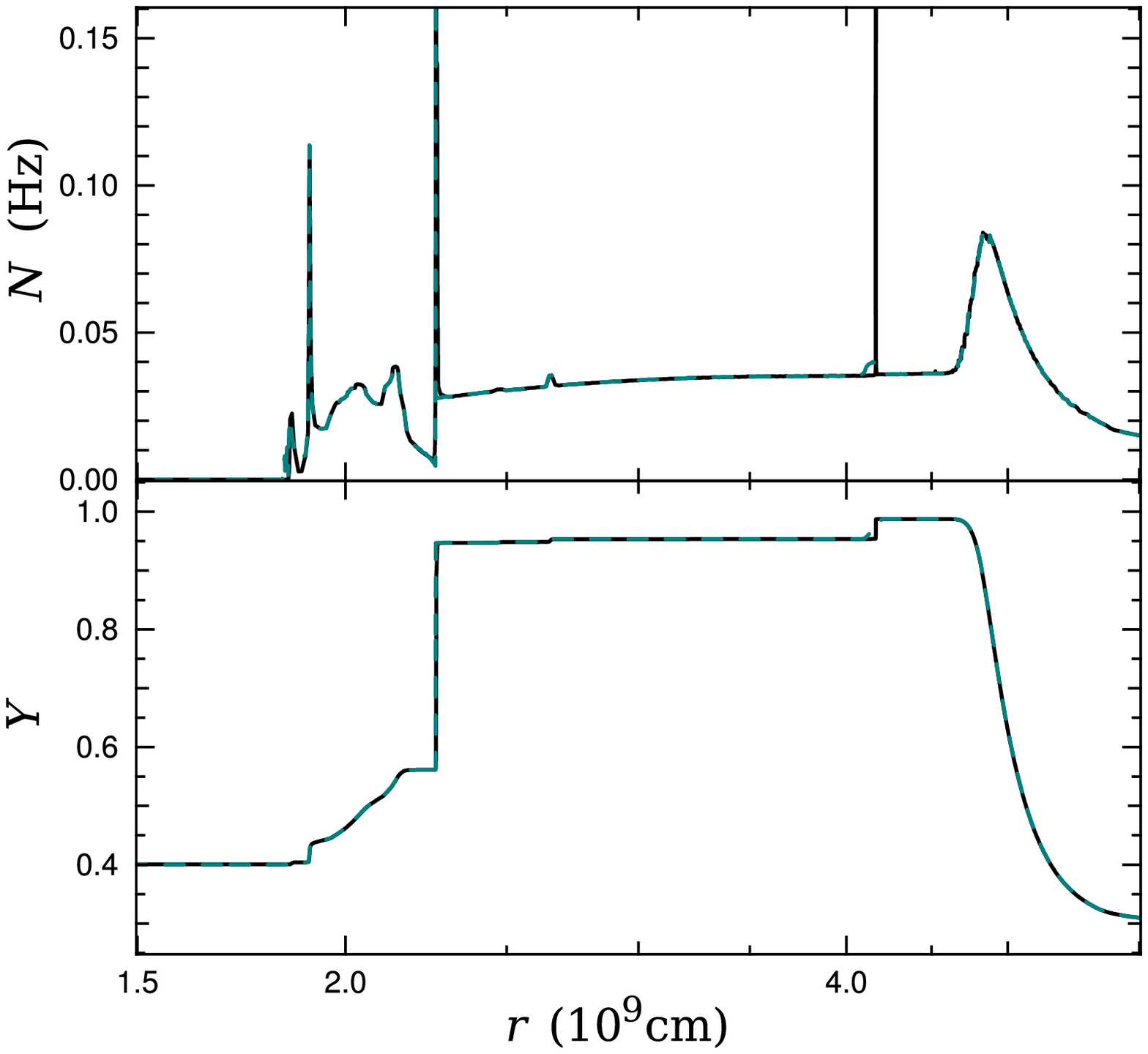}
\par
\vspace{0.35cm}
\includegraphics[width=\linewidth]{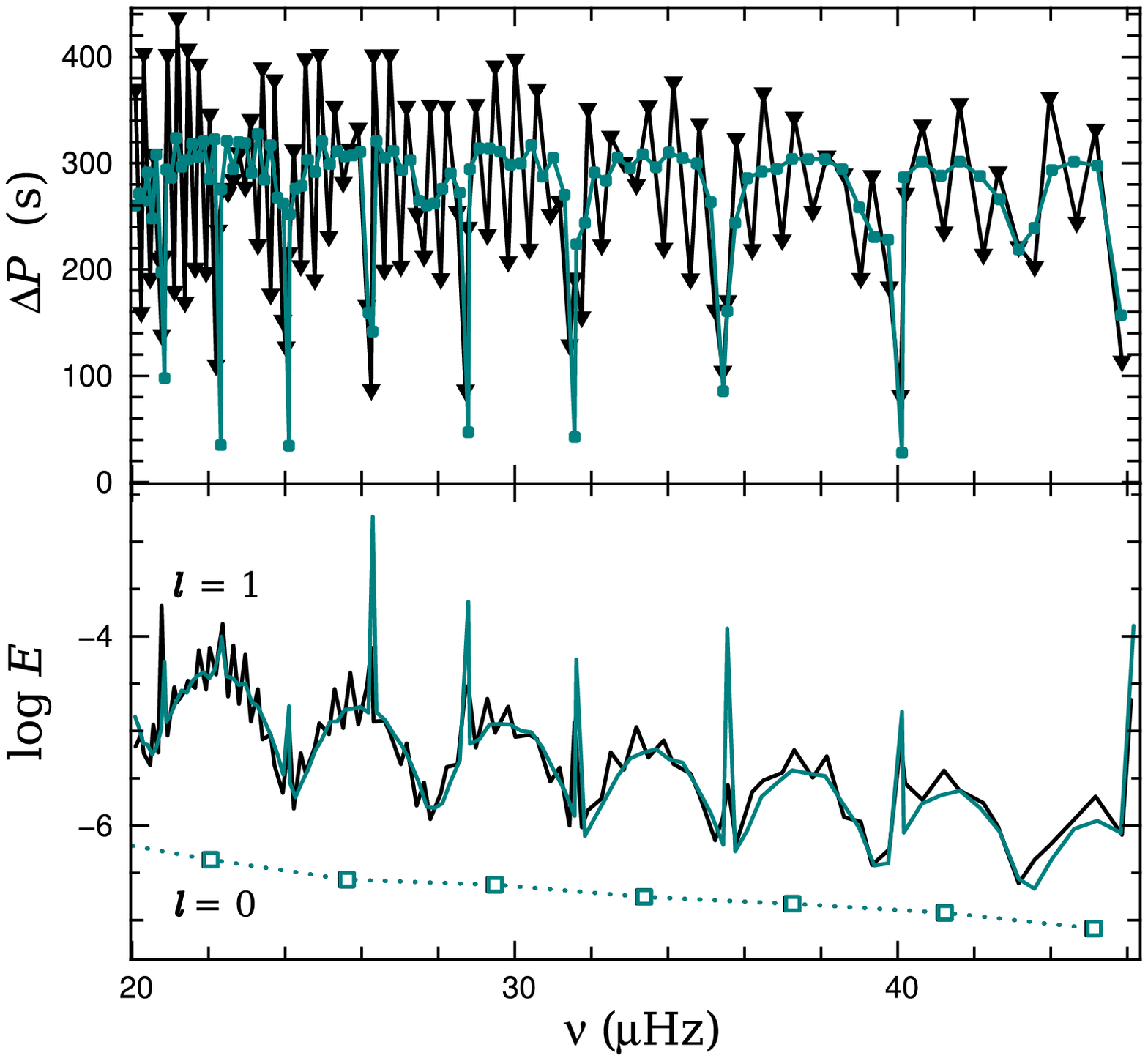}
  \caption{Pulsation properties of 1\,$\text{M}_\odot$ models with standard overshoot from the evolution code (black) and with some composition smoothing (cyan).  The composition profile in the latter model has been smoothed near $r = 4.2 \times 10^9\,\text{cm}$ by using Equation~\ref{eq:sine} with $\Delta m = 0.008\times \text{M}_\odot$, which is sufficient for it to not affect the computed frequencies. The discontinuity at the edge of the partially mixed zone ($2.3 \times 10^9\,\text{cm}$) has been smoothed with $\Delta m = 2\times 10^{-4}\,\text{M}_\odot$.  The model from the evolution code has $\Delta\Pi_1=278$\,s while the model with smoothing has $\Delta\Pi_1=281$\,s.  Both models have $R=11.0\,\text{R}_\odot$, $T_\text{eff}=4600\,\text{K}$, and $\nu_\text{max} = 26\,\mu\text{Hz}$.}
  \label{figure_standard_overshoot}
\end{figure}

The dominant factor in the computed pulsations of our standard-overshoot model shown in black in Figure~\ref{figure_standard_overshoot} is the main composition discontinuity left by the core flash (see Section~\ref{sec:postcoreflash} for the analysis of the effect of this discontinuity).  In order to isolate the effect of the partially mixed region resulting from core helium burning ($r < 2.3 \times 10^9\,\text{cm}$) we have smoothed the composition profile created during the core flash phase.  

\begin{figure}
\includegraphics[width=\linewidth]{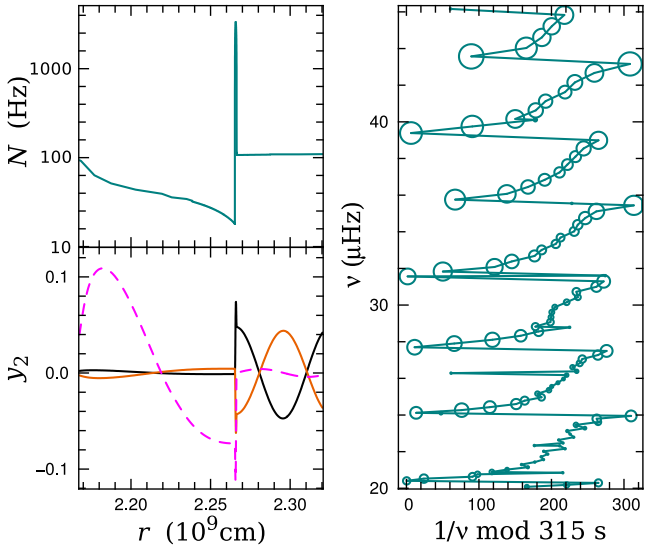}
  \caption{Seismic properties of the standard-overshoot model in Figure~\ref{figure_standard_overshoot} with composition smoothing.  Left panel:   Brunt--v{\"a}is{\"a}l{\"a} frequency and scaled horziontal displacement eigenfunctions (where $y_2$ is defined in Equation~\ref{eq:y1}) with radial order $n=-130,-129,-128$ (black line, orange line, and magenta dashes, respectively) and $\nu \approx 26$\,$\mu$Hz near the abundance discontinuity at the edge of the partially mixed zone ($r=2.3 \times 10^9\,\text{cm}$ in Figure~\ref{figure_standard_overshoot}).  The eigenfunctions of the ``trapped'' mode ($n=-128$; magenta) has been rescaled by a factor of 0.1 for clarity.  Right panel:  \'Echelle diagram with $\Delta P_\text{\'ech} = 315\,\text{s}$. This model has $\Delta\Pi_1=281$\,s, but if the calculation includes only the region exterior to the discontinuity then $\Delta\Pi_1=315$\,s.}
  \label{figure_overshoot_echelle_eigen}
\end{figure}

In addition, we have also smoothed the chemical profile at the edge of the partially mixed zone ($r \simeq 2.3 \times 10^9\,\text{cm}$), in order to make the period spacing slightly more regular.  The resulting period spacing pattern shown in cyan in Figure~\ref{figure_standard_overshoot} differs from the earlier models without a partially mixed region by the appearance of consecutive modes that are very closely spaced in period.  These have a regular dependence on radial order $n$ and are separated by $\Delta n \approx 11$.  These modes are also of very high inertia, and their effect on $\Delta P$ in Figure~\ref{figure_standard_overshoot} appears superimposed on the pattern produced by a structure without a partially mixed region (e.g. the blue model in Figure~\ref{figure_no_overshoot}).  The reason for this is clear from Figure~\ref{figure_overshoot_echelle_eigen}b, where it can be seen that these modes are ``trapped'' in the partially mixed region by the discontinuity at its boundary.  The period spacing between other modes is affected too: the position of nodes in the eigenfunctions of neighbouring modes in the trapping region is nearly identical, so the period spacing between most of them is as if the interior cavity does not exist, i.e. their $\Delta P$ is more consistent with the asymptotic calculation excluding this cavity and is therefore higher.  This is demonstrated in the period \'{e}chelle diagram in Figure~\ref{figure_overshoot_echelle_eigen} (see also the analogous case in Figure~\ref{figure_hos}).  In this example, the $\Delta P_\text{\'{e}ch}$ that produces the best fit for the \'{e}chelle diagram is 315\,s, which is well above the asymptotic value of 281\,s.  This suggests that inferences from asteroseismology about the theoretical $\Delta\Pi_1$ as specified in Equation~\ref{eq:dp1_asymp} may be dubious if the stars have a structure comparable to our standard-overshoot models.  Moreover, the trapped modes that are responsible for this behaviour have a relatively small amplitude at the surface and may therefore be impossible to detect.

\subsubsection{Models with semiconvection}
\label{sec:sc_results}

\begin{figure}
\includegraphics[width=\linewidth]{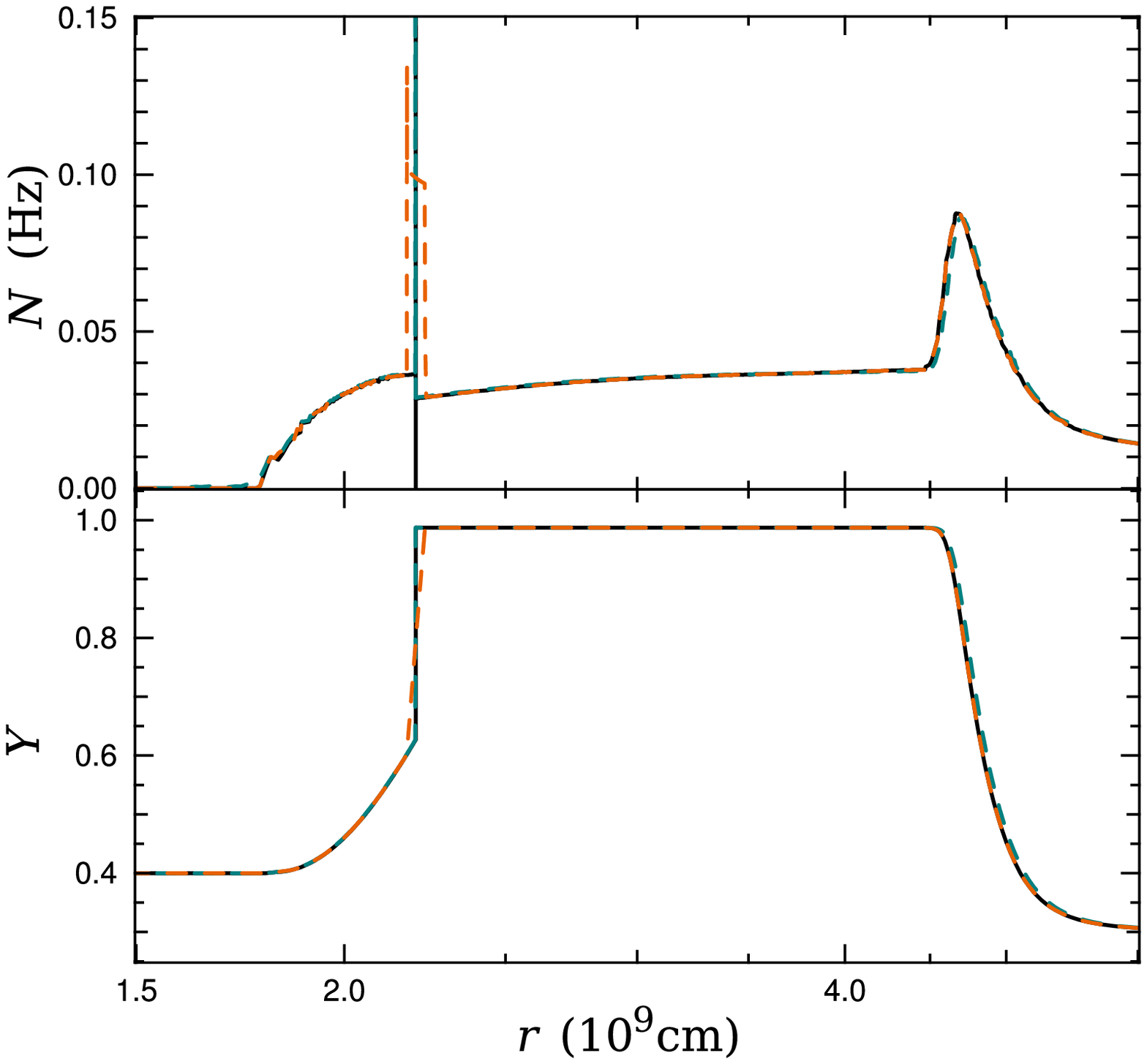}
\par
\vspace{0.35cm}
\includegraphics[width=\linewidth]{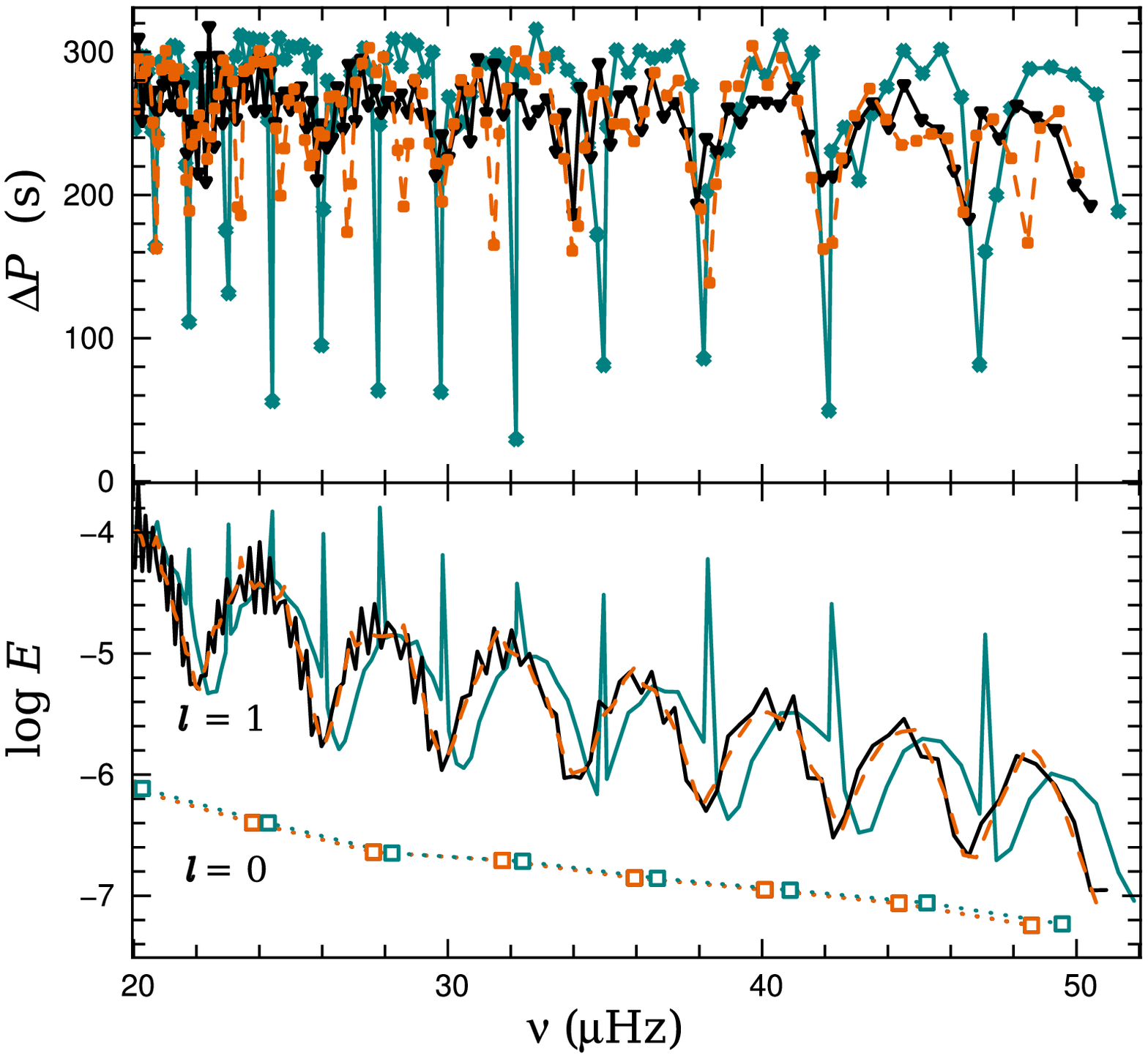}
  \caption{Pulsation properties of a 1\,$\text{M}_\odot$ classical semiconvection model with exactly $\nabla_\text{rad}=\nabla_\text{ad}$ outside the convective core (black) and otherwise identical models where the composition discontinuity at the outer edge of the semiconvection region has been smoothed over 0.01\,$\text{M}_\odot$ (orange dashes) and $2 \times 10^{-4}$\,$\text{M}_\odot$ (cyan).  The region beyond the semiconvection zone has been homogenized in each model (see Section~\ref{sec:puls_methods}). The model without smoothing has $\Delta\Pi_1 = 273$\,s, the model with fine smoothing has $\Delta\Pi_1 = 271$\,s, and the model with broad smoothing has $\Delta\Pi_1 = 264$\,s.  These models have approximately $R=10.4\,\text{R}_\odot$, $T_\text{eff}=4640\,\text{K}$, and $\nu_\text{max} = 28\,\mu\text{Hz}$.}
  \label{figure_semiconvection}
\end{figure}

\begin{figure}
\includegraphics[width=\linewidth]{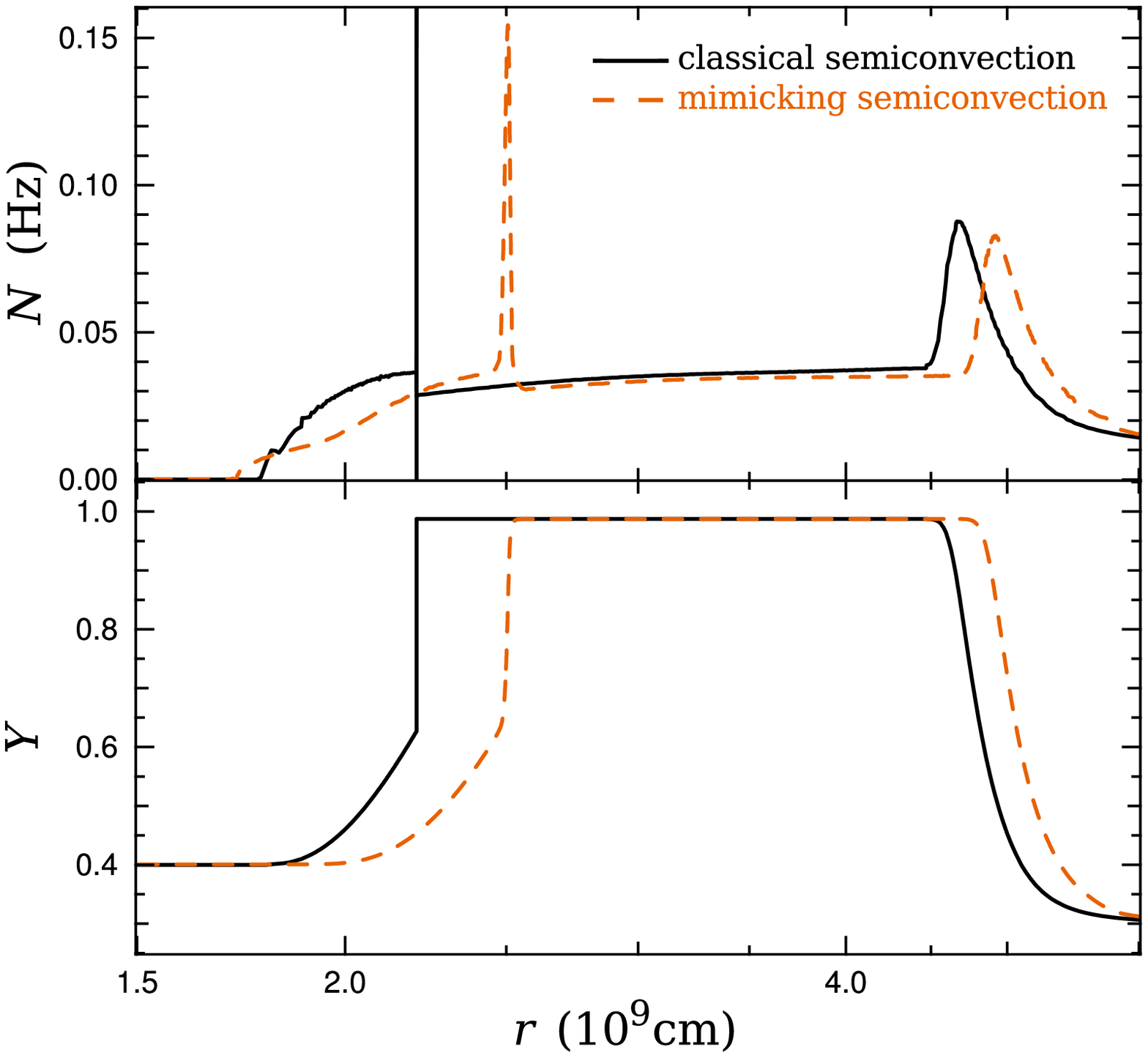}
\par
\vspace{0.35cm}
\includegraphics[width=\linewidth]{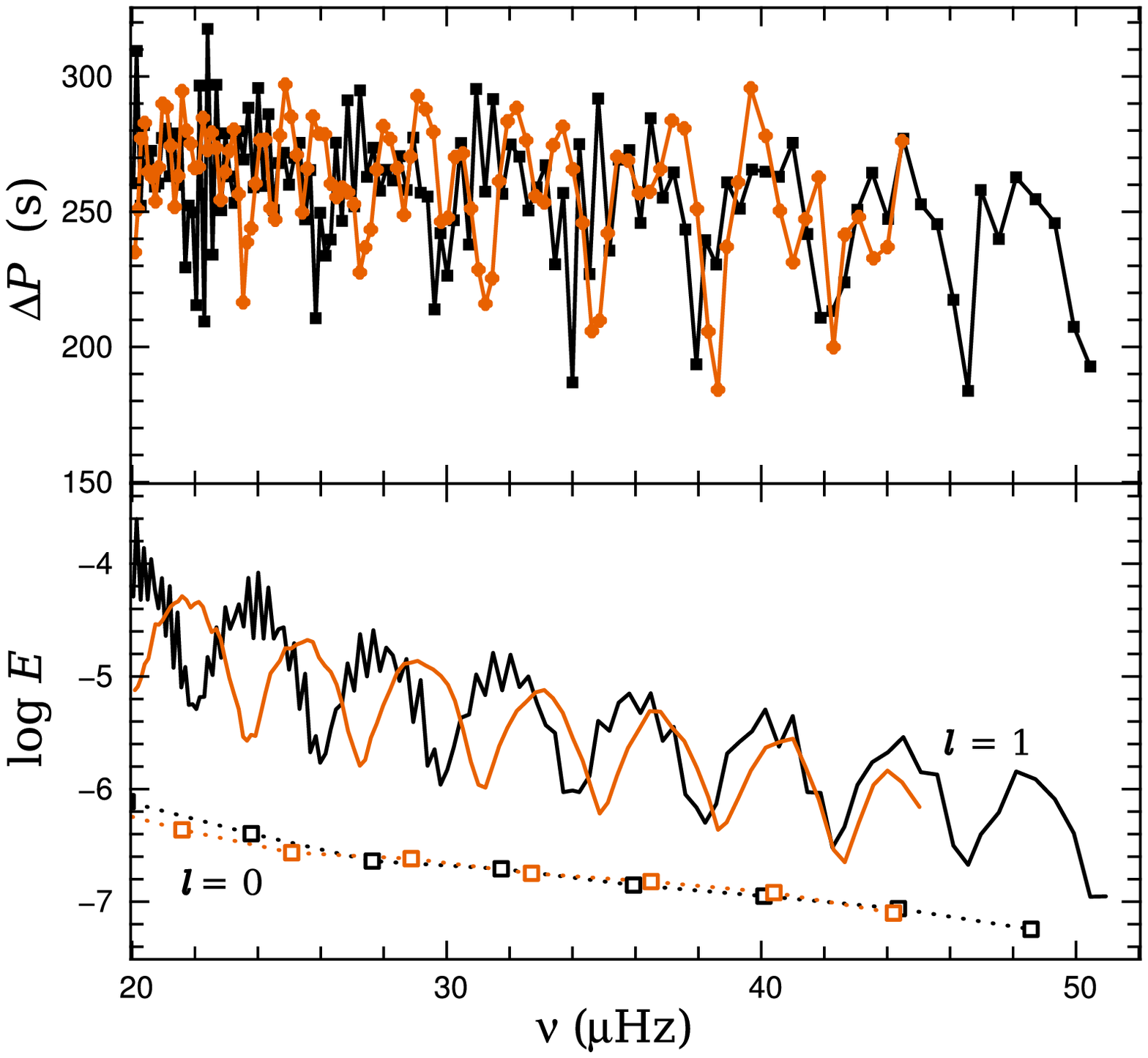}
  \caption{Comparison between the pulsation properties of a 1\,$\text{M}_\odot$ model with the classical semiconvection with exactly $\nabla_\text{rad}=\nabla_\text{ad}$ outside the convective core (black) and a model produced by the evolution code using our new mixing scheme (orange; described in Section~\ref{sec:sc}).  The region beyond the semiconvection zone has been homogenized in both models.  The classical semiconvection model has $\Delta\Pi_1=273$\,s while the model from the evolution code has $\Delta\Pi_1 = 268$\,s. The model from the evolution code has $R=11.1\,\text{R}_\odot$, $T_\text{eff}=4600\,\text{K}$, and $\nu_\text{max} = 31\,\mu\text{Hz}$ (see Figure~\ref{figure_semiconvection} for the properties of the other model).}
  \label{figure_semiconvection_comparison}
\end{figure}

We have computed the pulsation spectra for four models with semiconvection-like structures.  In Figure~\ref{figure_semiconvection} we compare the classical semiconvection structure to those where the abundance discontinuity at the outer boundary of the semiconvection zone has been softened. In Figure~\ref{figure_semiconvection_comparison} we analyse the structure that is produced by our routine that mimics semiconvection. 

In classical semiconvection models the sharp composition gradient between the semiconvection zone (which is relatively C- and O-rich and has stabilizing composition gradient) and the He-rich zone produces a step in the Brunt--V{\"a}is{\"a}l{\"a} frequency (at $r = 2.2 \times10^9\,\text{cm}$ in Figure~\ref{figure_semiconvection}).  This causes variation in $\Delta P$ between consecutive pairs of low-frequency modes and also in the inertia of every second mode (Figure~\ref{figure_semiconvection}).  The replacement of this discontinuity by a linear composition profile spread over 0.01\,$\text{M}_\odot$ introduces a second periodicity in $\Delta P$ of $\Delta n\simeq 7$, where $n$ is the radial order.  This is consistent with the expression derived by \citet{2008MNRAS.386.1487M}:
\begin{equation}
\label{eq:delta_k}
\Delta n \simeq \frac{\Pi_\mu }{\Pi_0},
\end{equation}
where $\Pi_\mu^{-1}$ is the buoyancy radius at the location of the composition gradient and $\Pi_0^{-1}$ is the total buoyancy radius (defined in Equation~\ref{eq:buoyancy_radius}).  This relatively smooth composition profile has only a small effect on period spacing and mode inertia compared with the case where the composition varies over just $2 \times 10^{-4}$\,$\text{M}_\odot$ (cyan dashes in Figure~\ref{figure_semiconvection}).  In the latter case, modes are very strongly trapped in the semiconvection zone (with about the same periodicity), which increases the period spacing between the non-trapped modes to around 300\,s, well above the asymptotic value of 271\,s.  This model has a regular period spacing pattern when plotted in the \'echelle diagram with $\Delta P_\text{\'ech}=306$\,s, which is consistent with the $\Delta\Pi_1$ calculation excluding the region interior to the composition discontinuity.  This is analogous to the model with strong mode trapping in Figures \ref{figure_standard_overshoot} and \ref{figure_overshoot_echelle_eigen}.  In both cases, the low $\Delta P$ between certain pairs of modes provides the only hint that the typical $\Delta P$ is actually above the asymptotic value.  One of the modes in each of these pairs, however, is unlikely to be detected because it has high inertia and is trapped in the semiconvection/partially mixed zone. 

In contrast with the classical semiconvection models, the buoyancy spike produced by the ad hoc semiconvection scheme in the evolution code only weakly traps modes.  This is still enough to clearly add a periodicity to the period spacing (with $\Delta n \simeq 5$; Figure~\ref{figure_semiconvection_comparison}).  This is consistent with Equation~\ref{eq:delta_k} because exactly 20 per cent of the total buoyancy radius is contained within the partially mixed zone.

\subsubsection{Models with maximal overshoot}
\label{sec:max_os_results}

The structure of the maximal-overshoot models is very similar to the no-overshoot models except that the convective core is larger (Figure~\ref{figure_helium_ev}).  There is no mixing beyond the convection zone by design.  One difference is that the growth of the core can eradicate some of the remnants of the previous core-flash burning.  Overall, the period spacing pattern is similar to the no-overshoot models (Figure~\ref{figure_max_overshoot}).  Importantly for the $\Delta\Pi_1$ discrepancy, however, the larger convective core also increases the mean $\Delta P$, while modes of the same radial order have a lower frequency.

\subsection{Matching ensemble $\Delta\Pi_1$ observations}
\label{sec:ensemble}

\begin{figure}
\includegraphics[width=\linewidth]{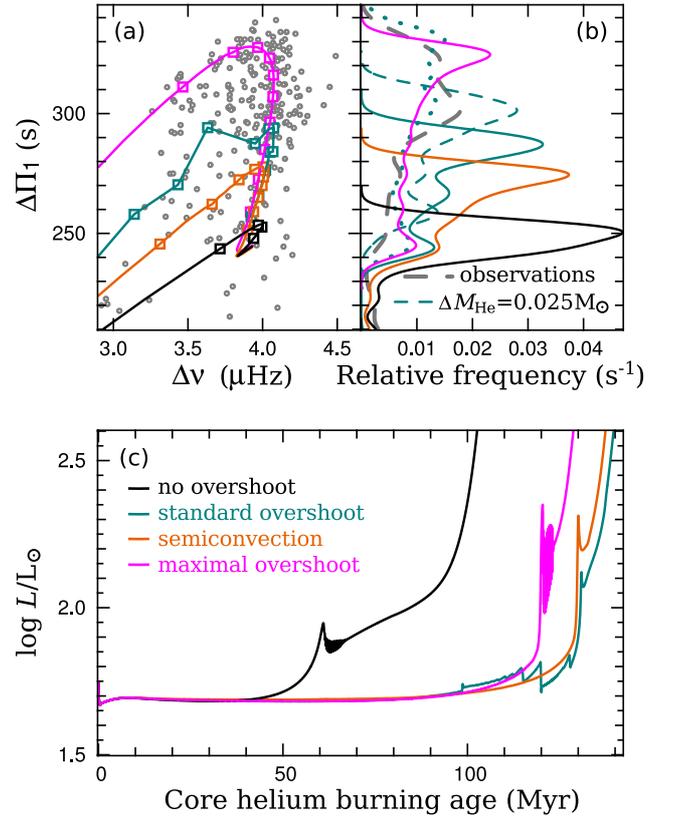}
  \caption{Upper left panel: evolution of 1\,$\text{M}_\odot$ CHeB models with different mixing schemes (no overshoot, standard overshoot, semiconvection, and maximal overshoot; in black, cyan, orange, and magenta respectively) in the $\Delta\nu - \Delta\Pi_1$ plane.  Markers are at 10\,Myr intervals.  Determinations for \textit{Kepler} field stars (grey dots) are from \citet{2014A&A...572L...5M}, and are limited to those with reported mass $0.8<M/\text{M}_\odot<1.25$.  Upper right panel: probability density curves (Equation~\ref{eq:pdf}) for models in the upper left panel (same colours), standard overshoot with $\Delta\Pi_1$ computed using only the region outside the partially mixed zone (cyan dots), standard overshoot with increased $M_\text{He}$ (cyan dashes), and observations (grey dashes).  Lower panel: surface luminosity evolution for the models in the upper left panel.}
  \label{figure_1msun}
\end{figure}

\begin{figure}
\includegraphics[width=\linewidth]{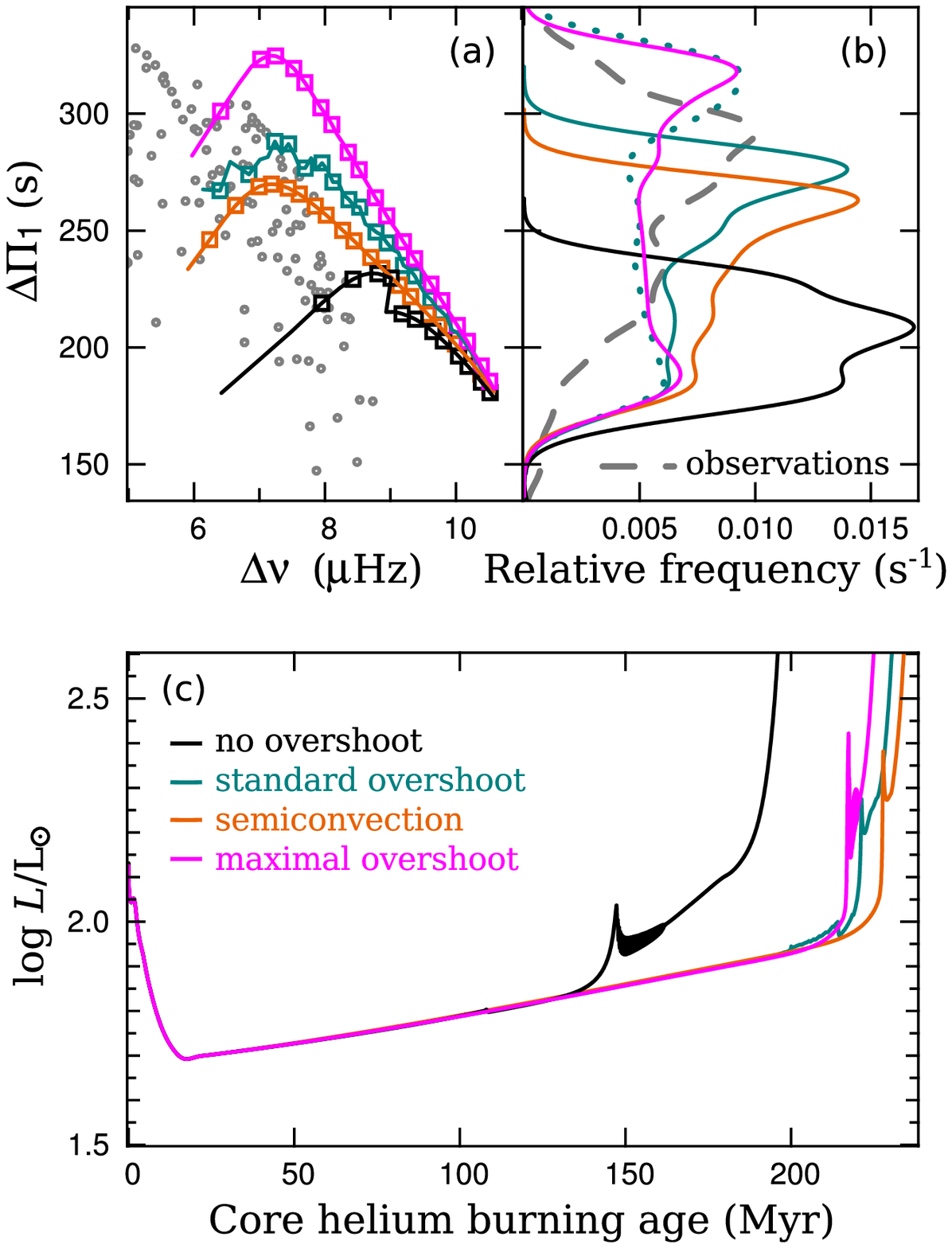}
  \caption{Upper left panel: evolution of 2.5\,$\text{M}_\odot$ CHeB models with different mixing schemes (no overshoot, standard overshoot, semiconvection, and maximal overshoot; in black, cyan, orange, and magenta respectively) in the $\Delta\nu - \Delta\Pi_1$ plane.  Markers are at 10\,Myr intervals.  Determinations for \textit{Kepler} field stars (grey dots) are from \citet{2014A&A...572L...5M}, and are limited to those with reported mass $2<M/\text{M}_\odot<3$.  Upper right panel: probability density curves (Equation~\ref{eq:pdf}) for models in the upper left panel (same colours), observations (grey dashes), and standard-overshoot model with $\Delta\Pi_1$ computed using only the region outside the partially mixed zone (cyan dots).  Lower panel: surface luminosity evolution for the models in the upper left panel.}
  \label{figure_2.5msun}
\end{figure}

In this section we compare the inferred $\Delta\Pi_1$ from the population of \textit{Kepler} field stars with predictions from evolution models.  We have chosen two representative masses: 1\,$\text{M}_\odot$ which experiences typical evolution for a red clump star (due to the uniformity of H-exhausted core mass at the flash), and 2.5\,$\text{M}_\odot$ which is massive enough to undergo core He-ignition in non-degenerate conditions (i.e. avoid the core flash) and then move to the so-called secondary clump in the HR diagram.  We compare the models and observations with probability density functions $P(\Delta\Pi_1)$ constructed by the addition of Gaussian functions according to
\begin{equation}
\label{eq:pdf}
P(\Delta\Pi_1) = \frac{1}{N} \sum\limits_{i=1}^N  \frac{1}{\sigma \sqrt{2\pi}}\exp{\left[-\frac{\left(\Delta\Pi_1-\Delta\Pi_{1,i}\right)^2}{2\sigma^2}\right]},
\end{equation}
where $\Delta\Pi_{1,i}$ represent each value from observations, or in the case of models, calculations at 1\,Myr intervals, and $N$ is the total of number of observations or calculated values.  We use a standard deviation of $\sigma = 4\,\text{s}$ and $\sigma = 8\,\text{s}$ for the 1\,$\text{M}_\odot$ and 2.5\,$\text{M}_\odot$ cases, respectively (Figures~\ref{figure_1msun} and~\ref{figure_2.5msun}).

In the 1\,$\text{M}_\odot$ case the maximal-overshoot models have the highest $\Delta\Pi_1$, followed by the standard-overshoot, semiconvection, and no-overshoot models.  The spreads of the $\Delta\Pi_1$ probability density functions for the semiconvection and overshoot cases are smaller than is observed, and offset to lower values, as shown in Figure~\ref{figure_1msun}.  In contrast, the spread for the maximal-overshoot 1\,$\text{M}_\odot$ model appears too broad, especially considering that we have computed single evolution sequences rather than a population which would widen the distribution.  

Every one of the four low-mass models appears to spend too much time with a low $\Delta\Pi_1$.  Two possible resolutions are i) an increased H-exhausted core mass at the flash, which increases $\Delta\Pi_1$ (dashed lines in Figures~\ref{figure_one_msun_core} and~\ref{figure_1msun}), or ii) that there is a difficulty in observationally determining $\Delta\Pi_1$ for stars that have recently begun core helium burning (discussed in Section \ref{sec:postcoreflash}).  Both of these affect the beginning of the CHeB, when $\Delta\Pi_1$ is lowest.  The fact that this discrepancy exists even for the maximal-overshoot run, when the convective core is the largest possible, suggests that the treatment of convective boundaries cannot be the sole reason for it.  In addition, Figure~\ref{figure_2.5msun} shows that there is no evidence that this problem exists for any of the higher-mass models.  These more massive models do not experience the core flash, do not ascend the RGB to as high luminosity, and have more luminous hydrogen burning at the beginning of the CHeB phase, and thus would be unaffected by the proposed resolutions.  In Figure~\ref{figure_1msun} the appearance of the discrepancy at low $\Delta\Pi_1$ is worsened for the semiconvection and standard-overshoot runs by the slow decrease in $\Delta\Pi_1$ towards the end of CHeB.  This alone cannot explain the discrepancy, however, because it is still present for sequences that do not undergo this slow drop in $\Delta\Pi_1$ late in CHeB (e.g. the dotted curve in Figure~\ref{figure_1msun}; discussed later in this section).

The maximal-overshoot model is the only one of the four with different mixing prescriptions that can reach $\Delta\Pi_1$ values consistent with the bulk of the low-mass observations.  Among the remaining cases, the standard-overshoot model is closest to the observations.  The shape of its $\Delta\Pi_1$ probability density function also looks reasonable, except that it is offset by at least 25\,s.  Even a substantial increase in the H-exhausted core mass $\Delta M_\text{He} = 0.025\,\text{M}_\odot$ (the most permitted by \citealt{1996ApJ...461..231C}) at the start of core helium burning is not enough to match the entire observed $\Delta\Pi_1$ range.  In that case it shifts the $\Delta\Pi_1$ probability density function higher by around 20\,s.  

The 1\,$\text{M}_\odot$ semiconvection sequence has a lower $\Delta\Pi_1$ than our standard-overshoot case, by around 10\,s.  This is despite the similar evolution of $R_\text{cc}$ and $M_\text{He}$ which strongly influence $\Delta\Pi_1$ (see Section~\ref{sec:bulk_properties}).  It is also evident from Figure~\ref{figure_helium_ev} that the evolution of the size of the partially mixed region is similar for both sequences.  We therefore attribute the difference in $\Delta\Pi_1$ to the way the composition always varies smoothly in the semiconvection case, increasing $N$ over a large interval in radius instead of over sharp spikes.  

We have performed an explicit test of the effect of the steepness of composition profiles on $\Delta\Pi_1$.  The three models in Figure~\ref{figure_semiconvection} are identical except for the composition near the edge of the semiconvection zone at $r \simeq 2.2 \times 10^9$\,cm.  The buoyancy frequency is nearly identical elsewhere in the structure (Figure~\ref{figure_semiconvection}b) so any difference in $\Delta\Pi_1$ must be due to the composition smoothing.  In this case, smoothing the discontinuity over $\Delta m = 0.01\,\text{M}_\odot$ decreases $\Delta\Pi_1$ by 9\,s.  

This effect is also apparent in Figure~\ref{figure_max_overshoot_smooth}, where smoothing the edge of the fully mixed core increases $\Delta\Pi_1$.  It can be seen in panel (a) that this smoothing increases the width of the peak in $N$ (in the log scale) by more than it reduces its height.  This increases the area under the curve, which reduces $\Delta\Pi_1$.  This is evident when we rewrite the integral in the asymptotic solution for $\Delta\Pi_\ell$ in terms of $\log{r}$ to get
\begin{equation}
\resizebox{.93\hsize}{!}{$
\Delta\Pi_\ell=\frac{2\pi^2}{\sqrt{\ell(\ell+1)}} \left[ \int\limits_\text{}^{}{\frac{N}{r}\text{d}r} \right]^{-1} = \frac{2\pi^2}{\sqrt{\ell(\ell+1)}} \left[ \int\limits_{}^{}N\text{d}\ln{r} \right]^{-1}.
$}
\end{equation}

The 1\,$\text{M}_\odot$ sequence without overshoot has the lowest $\Delta\Pi_1$.  $\Delta\Pi_1$ stays around 250\,s, around 50\,s below the bulk of the observations, for the entire CHeB phase. This can be attributed to the lack of growth of the convective core (Figure~\ref{figure_one_msun_core}c).  The evolution of $\Delta\Pi_1$ in our 1\,$\text{M}_\odot$ and 2.5\,$\text{M}_\odot$ sequences without overshoot is almost identical to the corresponding models (also without overshoot) from {\sc mesa} \citep{2013ApJ...765L..41S}. 

We have emulated the effect of mode trapping on the more easily observable (non-trapped) modes in the standard-overshoot models (discussed in Section~\ref{sec:overshoot_results}) by excluding the partially mixed region from the calculation of $\Delta\Pi_1$ (dotted curve in Figure~\ref{figure_1msun}b).  The impact of this is increasingly significant as core helium burning progresses and the partially mixed region grows.  This makes the $\Delta\Pi_1$ evolution very similar to that resulting from the maximal-overshoot scheme, except that it slightly exceeds the observed values (by less than 10\,s) near the end of core helium burning.  However, this is late in CHeB when this crude approximation of the effects of mode trapping is least valid, because the mode trapping cavity, and consequently the fraction of modes that become trapped, is large (making a neat fit in the period \'echelle diagram difficult; see Section~\ref{sec:late_rc}).  At the other extreme, $\Delta\Pi_1$ is still too low in the early stages of CHeB compared to the observations.

The CHeB lifetime of the 1\,$\text{M}_\odot$ no-overshoot model is by far the shortest, followed by the maximal-overshoot case (Figure~\ref{figure_1msun}c).  The semiconvection and overshoot sequences have nearly identical lifetimes.  The surface luminosity of the models is independent of the mixing scheme when they are still burning helium in the core (the variation in $\log{L/\text{L}_\odot}$ is less than 0.01).  The relative energy generation rates from hydrogen and helium burning differ by a little more.  The semiconvection model has more luminous H burning than the other sequences, while the no-overshoot model has the most luminous He burning, and the standard-overshoot and maximal-overshoot sequences are almost identical until the occurrence of a core breathing pulse after 98\,Myr.

Star counts in globular clusters can be used as a constraint on the mixing scheme.  This is because the CHeB lifetime is dependent on the amount of helium that is transported into the core.  Models that consume less helium during CHeB have more helium that must be burnt during subsequent shell helium burning, and consequently have a longer early-AGB lifetime (compare the swift exit from the red clump of the no-overshoot sequence with its sluggish ascent of the early-AGB in Figure~\ref{figure_1msun}c).  The parameter $R_2=n_\text{AGB}/n_\text{CHeB}$ (the number ratio of observed AGB to CHeB stars) for globular clusters is thought to correspond to the ratio of the respective phase lifetimes.  \citet{1989ApJ...340..241C} argued that models with semiconvection, but without breathing pulses, give the best fit to observations of the globular cluster M\,5.  In their models, the suppression of breathing pulses (by not allowing the growth in the convection zone if it would increase the central helium abundance) increased $R_2$ from $0.10$ to $0.14$ or $0.15$ (depending on the extent of core-flash burning), matching observations.  

We have computed $R_2$ for our models by considering luminosity bins comparable to the observed range in metal-rich globular clusters.  We have defined the CHeB lifetime to be when $\log{L/\text{L}_\odot}$ is within 0.1 of its mean value before core helium depletion, and the AGB to be when $\log{L/\text{L}_\odot}$ is no more than 1.0 higher than the CHeB range.  We find values of $R_2=t_\text{AGB}/t_\text{CHeB}$ of 0.110, 0.113, and 0.117 for the standard-overshoot, semiconvection, and maximal-overshoot schemes respectively.  This would make them practically indistinguishable from one another by observations of star clusters.  In contrast, the no-overshoot model has $R_2=0.743$, which is a difference that could easily be detected.  We will address constraints from star counts in the next paper in this series (by computing less massive and more metal-poor models relevant to Galactic globular clusters; \citealt{1996AJ....112.1487H}).

There are a number of common trends between the 1\,$\text{M}_\odot$ (Figure~\ref{figure_1msun}) and 2.5\,$\text{M}_\odot$ (Figure~\ref{figure_2.5msun}) models.  In the 2.5\,$\text{M}_\odot$ runs the mixing scheme has a very similar effect on mean $\Delta\Pi_1$, CHeB lifetime, and the H- and He-burning luminosity.  We also find a very similar effect from our emulation of mode trapping in the standard-overshoot model (dotted curve Figure~\ref{figure_2.5msun}).  The probability density functions for the 2.5\,$\text{M}_\odot$ models are very similar in shape to the those for the 1\,$\text{M}_\odot$ models, except that they cover a larger range of $\Delta\Pi_1$.  The more substantial increase in $\Delta\Pi_1$ during their evolution can be explained by the greater extent of the growth of the H-exhausted core (roughly 0.2\,$\text{M}_\odot$ compared with 0.05\,$\text{M}_\odot$ for the 1\,M$_\odot$ runs), the importance of which was shown in Section~\ref{sec:bulk_properties}.

Compared to the lower-mass case, the agreement with observations is markedly better for the 2.5\,$\text{M}_\odot$ sequences, with the exception of the no-overshoot model.  The semiconvection and standard-overshoot models, however, still do not reach the highest $\Delta\Pi_1$ observations.  In contrast, the $\Delta\Pi_1$ evolution for the mode trapping and maximal-overshoot sequences match each other even more closely, and both exceed the highest observed values by considerably more than does the 1\,$\text{M}_\odot$ mode trapping case.  The comparison between observations and models, however, is more complex than for the low-mass case.  More of the increase in $\Delta\Pi_1$ is due to the growth of the H-exhausted core, and we are comparing the models to a population more diverse in mass and smaller in number.  Therefore it would be imprudent to draw strong conclusions about the mixing from this sample.  We note that our models do not match the observed $\Delta\nu$ (but do match the shape of the population's distribution in $\Delta\nu-\Delta\Pi_1$ space).  This is not problematic because $\Delta\nu$ can easily be decreased by adjusting (in this case reducing) the MLT mixing length parameter, without affecting $\Delta\Pi_1$.  Finally, we note that by the end of core helium burning, the 2.5\,$\text{M}_\odot$ models are considerably more luminous than at the beginning (by around a factor of 2, apart from the shorter-lived no-overshoot model; Figure~\ref{figure_2.5msun}).  This could introduce an observational bias for the secondary clump towards more luminous evolved stars, which have higher $\Delta\Pi_1$ (for all of the mixing schemes we have examined).  Accounting for such a bias would help to resolve the excess of predicted low-$\Delta\Pi_1$ stars that is apparent in Figure~\ref{figure_2.5msun}b.

In Section~\ref{sec:pulsations_diff_mixing} we demonstrated how mode trapping may lead to an overestimation of $\Delta\Pi_1$.  This is made possible because only a subset of mixed modes can be detected.  Here we briefly consider how mode trapping affects the period spacing between the pairs of modes that are most likely to be detected, i.e. those with low inertia.  For this, we compare a standard-overshoot model with mode trapping (Figure~\ref{figure_standard_overshoot}) to a maximal-overshoot model without mode trapping (Figure~\ref{figure_max_overshoot}).  Although these two models have different $\Delta\Pi_1$ (281\,s and 314\,s, respectively), the respective values determined from the period \'{e}chelle diagram, $\Delta P_\text{\'ech} = 315$\,s (Figure~\ref{figure_overshoot_echelle_eigen}) and $\Delta P_\text{\'ech} = 316$\,s, are nearly identical.  The average $\Delta P$ between all modes with $20\,\mu\text{Hz} < \nu < 40\,\mu\text{Hz}$ for the standard-overshoot model is 270\,s, which increases to 293\,s if all of the (presumably undetectable) trapped modes are excluded, compared with 295\,s for the maximal-overshoot model.  When this calculation is restricted to the six pairs of modes closest to each low inertia trough (e.g., near $\nu = 24\,\mu\text{Hz}$ in Figure~\ref{figure_standard_overshoot}) we find $\Delta P = 275\,\text{s}$ and $\Delta P = 277\,\text{s}$ for the standard-overshoot and maximal-overshoot models, respectively.  If we restrict the count to sets of four pairs of low-inertia modes instead of six we again find that the two models have a similar average $\Delta P$, except that it is reduced further, by 9\,s in both cases.  Moreover, the average frequency spacing between these troughs is the same for both models.  This indicates that knowing the typical observed $\Delta P$ would not assist with the detection of mode trapping.  It also supports our suggestion in Section~\ref{sec:overshoot_results} that modes that are not trapped behave as though the buoyancy cavity is smaller than its true size, i.e. it excludes the semiconvection/partially mixed region with $N^2 > 0$ that is surrounded by a steep composition gradient that can trap modes.  Because the observationally determined $\Delta P$ depends on how many modes are detected it is difficult to compare these results to the average or median $\Delta P$ found in populations of CHeB stars \citep[e.g.][]{2011A&A...532A..86M,2013ApJ...765L..41S}.

\subsection{The effect of the boundary of the convective core on pulsations}

\begin{figure}
\includegraphics[width=\linewidth]{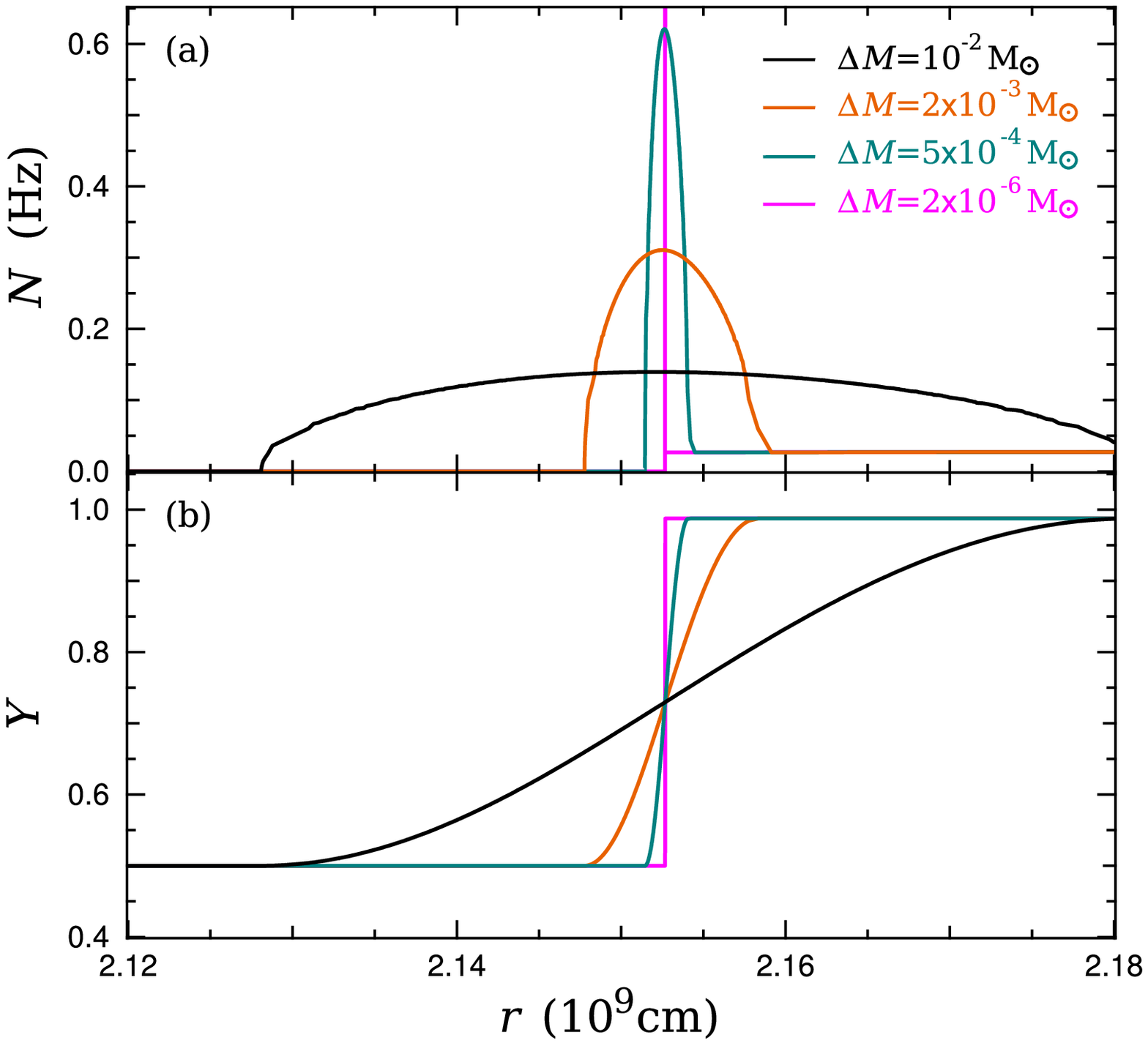}
\par
\vspace{0.35cm}
\includegraphics[width=\linewidth]{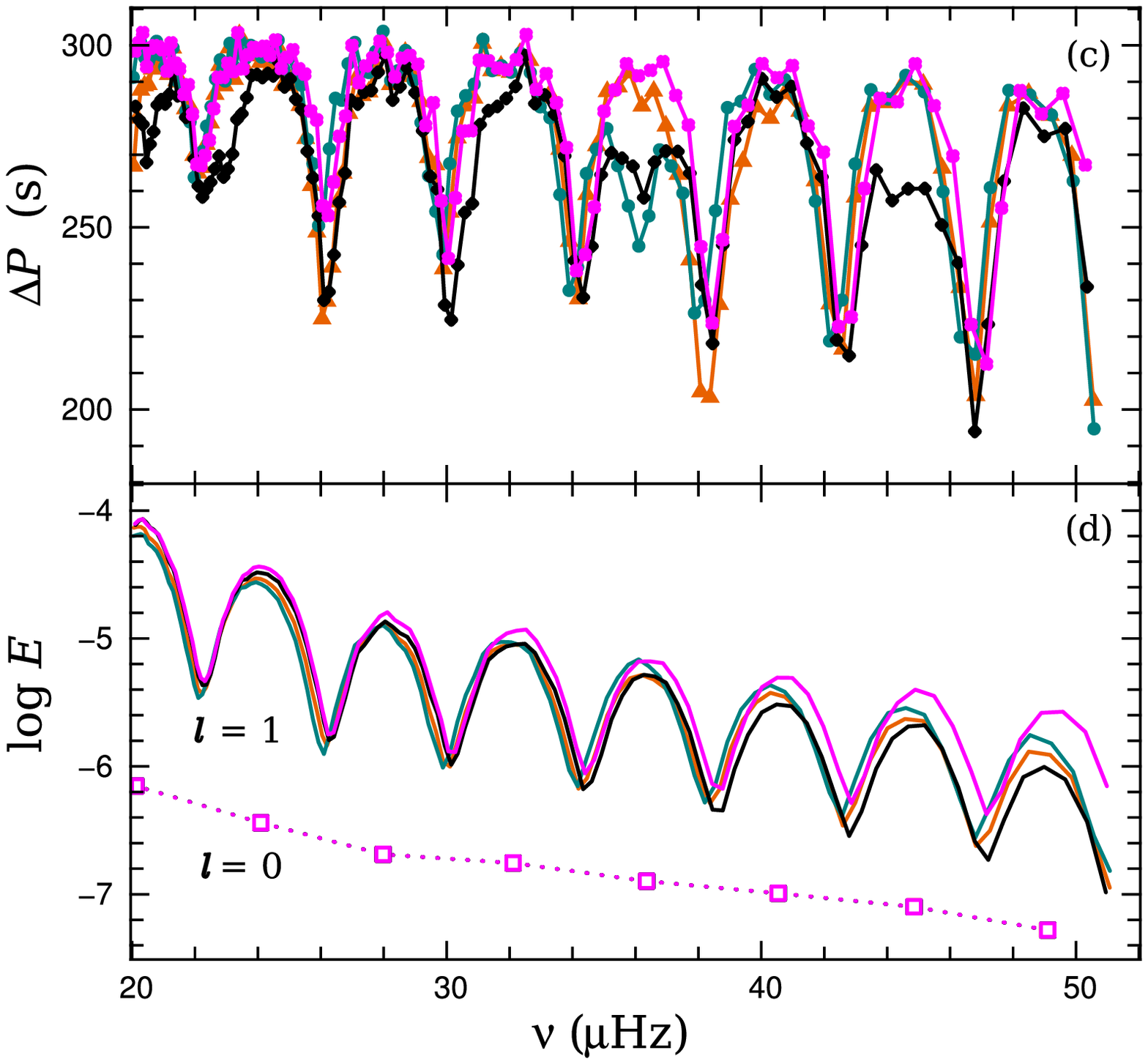}
  \caption{Pulsation properties of 1\,$\text{M}_\odot$ models with a small partially mixed regions outside the convective core.  In all four cases the composition profile in the overshoot region has been set according to Equation~\ref{eq:sine} with $\Delta m/\text{M}_\odot$ values of $2 \times 10^{-6}$, $5 \times 10^{-4}$, 0.002, and 0.01 in magenta, cyan, orange, and black, respectively.  In the same order $\Delta\Pi_1$ is 287\,s, 294\,s, 297\,s, and 300\,s.  The composition between the convective core and H-burning shell has been homogenized.  These models have approximately $R=10.4\,\text{R}_\odot$, $T_\text{eff}=4770\,\text{K}$, and $\nu_\text{max} = 29\,\mu\text{Hz}$.}
  \label{figure_max_overshoot_smooth}
\end{figure}

In Figure~\ref{figure_max_overshoot_smooth} we examine the pulsations resulting from a structure with a fully mixed convective core but with a smooth composition profile at its boundary.  This structure (which was produced by artificially smoothing according to Equation~\ref{eq:sine}) is interesting because of the physical implausibility of the core boundary in the maximal- and no-overshoot models (e.g. the magenta and black lines in Figure~\ref{figure_mixing_comparison_zoom}b).  In these models there is a true composition discontinuity where material that is strongly convectively unstable ($\nabla_\text{rad} \gg \nabla_\text{ad}$) does not partially mix with the material directly adjacent to it.  

All four models in Figure~\ref{figure_max_overshoot_smooth} with different composition profiles have a comparable period spacing over much of the frequency range shown.  The model in black with the largest partially mixed region ($\Delta m = 0.01\,\text{M}_\odot$), however, has several frequency ranges where the modes are more closely spaced (e.g., at around 35 and 45\,$\mu$Hz).  This behaviour is also seen near $\nu = 38\,\mu$Hz and $\nu = 36\,\mu$Hz, for the models with $\Delta m = 0.002\,\text{M}_\odot$ and $\Delta m = 5 \times 10^{-4}\,\text{M}_\odot$, in orange and cyan, respectively.  

These interruptions to the regular $\Delta P$ pattern shown in Figure~\ref{figure_max_overshoot_smooth}c are more prevalent when the composition is smoother.  This can be explained by the increasing buoyancy radius (see Section~\ref{sec:bulk_properties}) of the mode trapping region enclosed by smoother composition gradients (detailed in Section~\ref{sec:ensemble}).  This trapping region is adjacent to the convective core, however, so it always has a small buoyancy radius and therefore has little effect on mode inertia (Figure~\ref{figure_max_overshoot_smooth}d).  This small buoyancy radius of the trapping region also explains the long periodicity (in radial order) in its effect on $\Delta P$, because this gives a large $\Delta n$ according to Equation~\ref{eq:delta_k}.  This contrasts with standard-overshoot and semiconvection models (e.g. Figures \ref{figure_standard_overshoot} and \ref{figure_semiconvection}), where a similar buoyancy feature is surrounded on both sides by a g-mode cavity which triggers mode trapping at regular intervals in radial order $n$ with smaller $\Delta n$.  Finally, we note that the disruption to the regular period spacing caused by a composition gradient at the edge of the convective core appears most obvious for gravity-dominated modes, which are the most difficult to detect.

\subsection{Subdwarf B models}
\label{sec:sdb}

We have also tested the effect of the core mixing scheme in subdwarf B models.  At the beginning of core helium burning in these runs we homogenized the composition between the H-shell and the convective core to remove traces of core-flash burning.  In each model we also set the helium mass fraction $Y$, at the H-exhausted core boundary according to 
\begin{equation}
Y(m)=Y_\text{surf}+\frac{\Delta Y}{2}\left \{ 1+\cos{\left[ \left( \frac{m-M_\text{He}}{\Delta m} \right)^2 \pi \right ]  }\right \},
\end{equation}
where $\Delta Y$ is the difference between the surface and interior helium abundance, $M_\text{He}$ is the mass of the H-exhausted core, $m$ is the mass coordinate, and we have chosen $\Delta m = 0.002\,\text{M}_\odot$.  We chose this smooth profile because we are only interested in the effect of the composition profile at the boundary of the convective core.  We set the total mass and mass of the shell to match the mass, gravity, and effective temperature typical of the stars in the \citet{2011MNRAS.414.2885R} sample.  This ad hoc approach is obviously inadequate for precision studies of particular stars \citep[e.g., those found in][]{2011A&A...530A...3C,2013A&A...553A..97V}, but suits our purpose here.

\begin{figure}
\includegraphics[width=\linewidth]{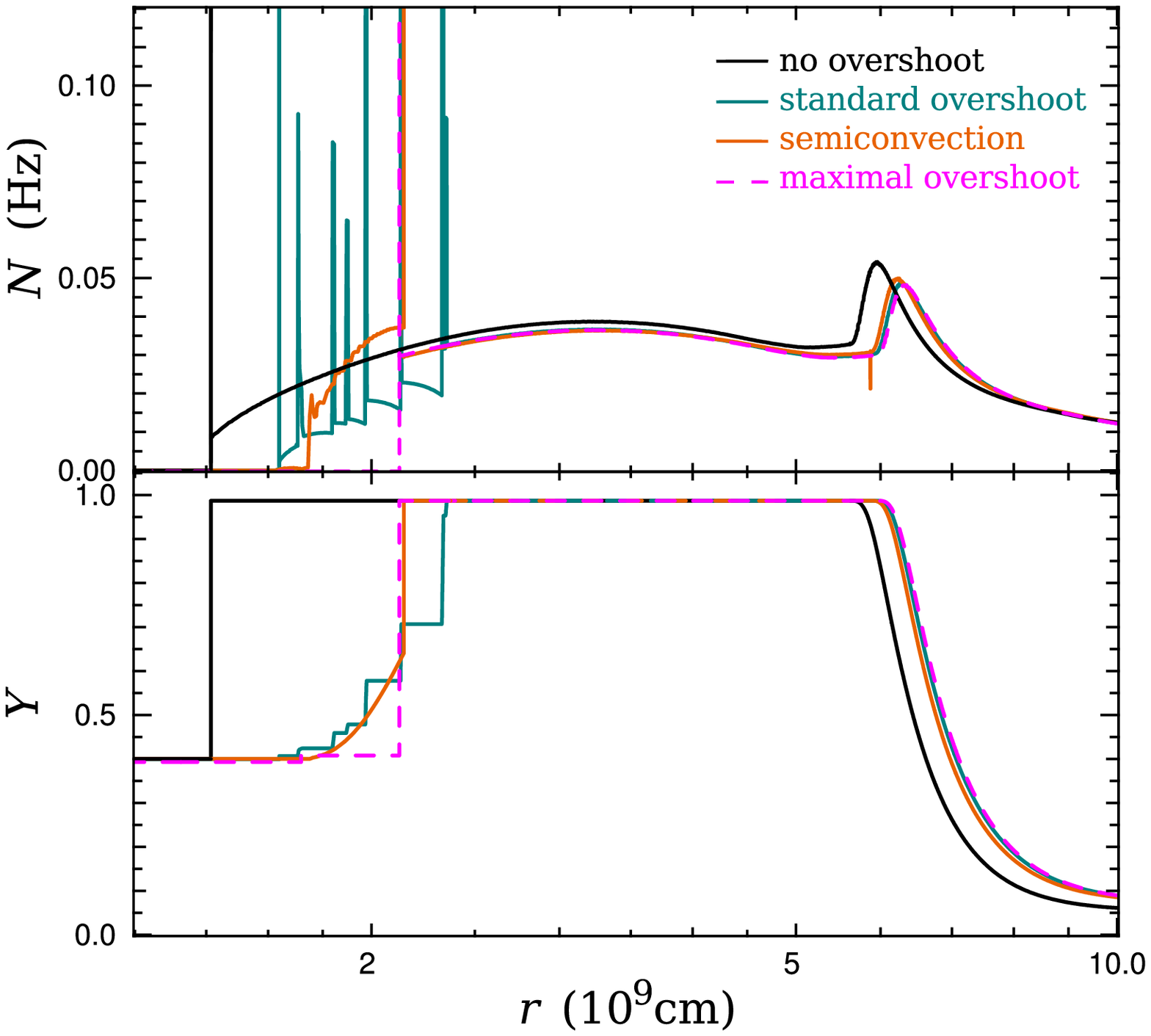}
\par
\vspace{0.35cm}
\includegraphics[width=\linewidth]{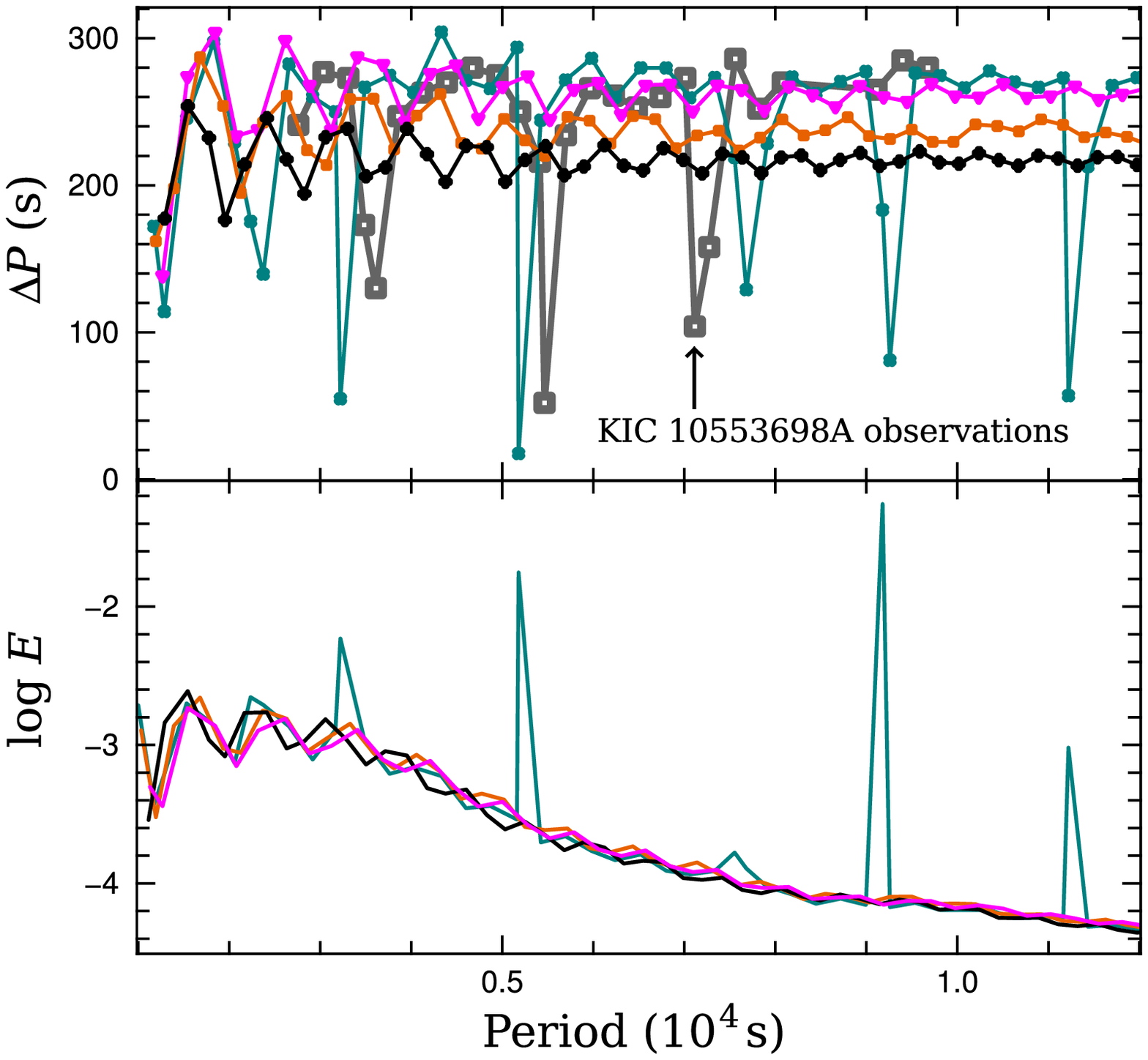}
  \caption{Seismic properties of synthetic subdwarf B models and an observed pulsator in the \textit{Kepler} field.  Upper panels: Brunt--V{\"a}is{\"a}l{\"a} frequency $N$ and helium mass fraction $Y$ for models with different mixing prescriptions.  Lower panels: $\ell =1$ mode spacing $\Delta P$ and inertia.  The observations are of $\ell =1$ modes classified by \citet{2014A&A...569A..15O} for KIC 10553698A (thick grey lines and squares).  The models were generated according to the method outlined in Section~\ref{sec:sdb}.  They have $M= 0.475\,\text{M}_\odot$, solar metallicity, and $Y_\text{cent}=0.4$.  The models have no overshoot (black), standard overshoot (cyan), semiconvection (orange), and maximal overshoot (magenta).  These models have $\Delta\Pi_1$ of 222\,s, 245\,s, 238\,s, and 269\,s, respectively, and approximately $R=0.20\,\text{R}_\odot$ and $T_\text{eff}=27000\,\text{K}$.}
  \label{figure_sdb}
\end{figure}

The results of the pulsation calculations for the four models with different mixing schemes are presented in Figure~\ref{figure_sdb}.  The appearance of the $\Delta P$ pattern for each model is broadly similar, except for a few trapped modes in the standard-overshoot model (Figure~\ref{figure_sdb}).  These modes have much higher inertia than their neighbours and are more closely spaced in period.  This behaviour is similar to our red clump standard-overshoot model (Figure~\ref{figure_standard_overshoot}) and the semiconvection model that includes a region with a stabilizing molecular weight gradient at the edge of the semiconvection zone (Figure~\ref{figure_semiconvection}).  

There is a substantial difference in the mean $\Delta P$ between the four different mixing cases.  $\Delta P$ spans a range of around 60\,s, with the no-overshoot sequence having the lowest average value, followed by the semiconvection case, then the maximal-overshoot and standard-overshoot cases (if the trapped-modes are excluded).  Like its more massive counterparts, the sequence without overshooting has a lower $\Delta \Pi_1$ than is observed.  In this case it is more than 10\,s too low to match any of the observations reported by \citet{2011MNRAS.414.2885R}, which is an especially strong constraint because the model in Figure~\ref{figure_sdb} is from the stage of CHeB when $\Delta \Pi_1$ is near its maximum.  The range of $\Delta P$ between pairs of modes found for KIC 5807616 spans less than 30\,s \citep{2011MNRAS.414.2885R} which is consistent with the range for our models (except near the high-inertia modes in the standard-overshoot case).

Recently, \citet{2014A&A...569A..15O} found evidence for mode trapping in KIC 10553698A, an sdB star in the \textit{Kepler} field, by classifying $\ell =1$ and $\ell =2$ modes.  They identified both the C-O/He and He/H transition zones as possible origins of the mode trapping and highlighted the resemblance of the period spacing pattern to existing theoretical calculations, e.g. Fig. 3 in \citet{2002ApJS..139..487C}.  Due to the use of $q = \log \left[ 1- m/M \right] $ for the horizontal axis of that figure, the structure near the core is difficult to discern, but it appears that in their ``evolutionary model'' there is a relatively smooth buoyancy peak near where partial mixing can occur in our models.  In our standard-overshoot model the mode trapping is certainly a result of the sharp composition gradient at the edge of the partially mixed zone outside the convective core.  Moreover, that model's pulsations bear perhaps an even more remarkable similarity to the observations shown in grey in Figure~\ref{figure_sdb} (keeping in mind we made no attempt to match the frequencies).  The trapped modes in our model, however, reside deep within the core, so their observability is uncertain.  Finally, the theoretical $\Delta P$ for our non-trapped modes is nearly an exact match for the $\Delta P$ between most observed modes in KIC 10553698A, which suggests that the size of the convective core in the standard-overshoot model is reasonable.

\begin{figure}
\includegraphics[width=\linewidth]{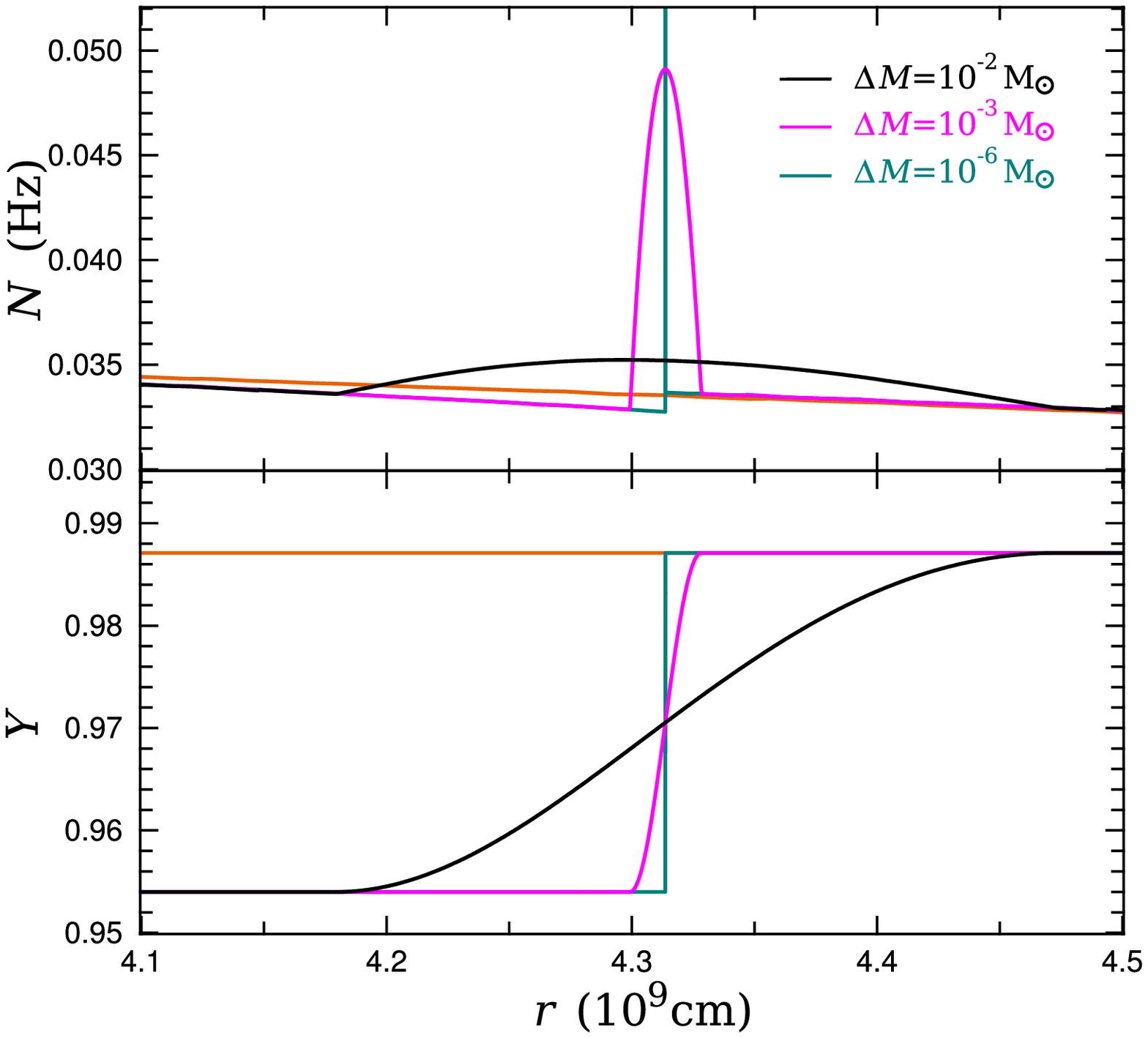}
\par
\vspace{0.35cm}
\includegraphics[width=\linewidth]{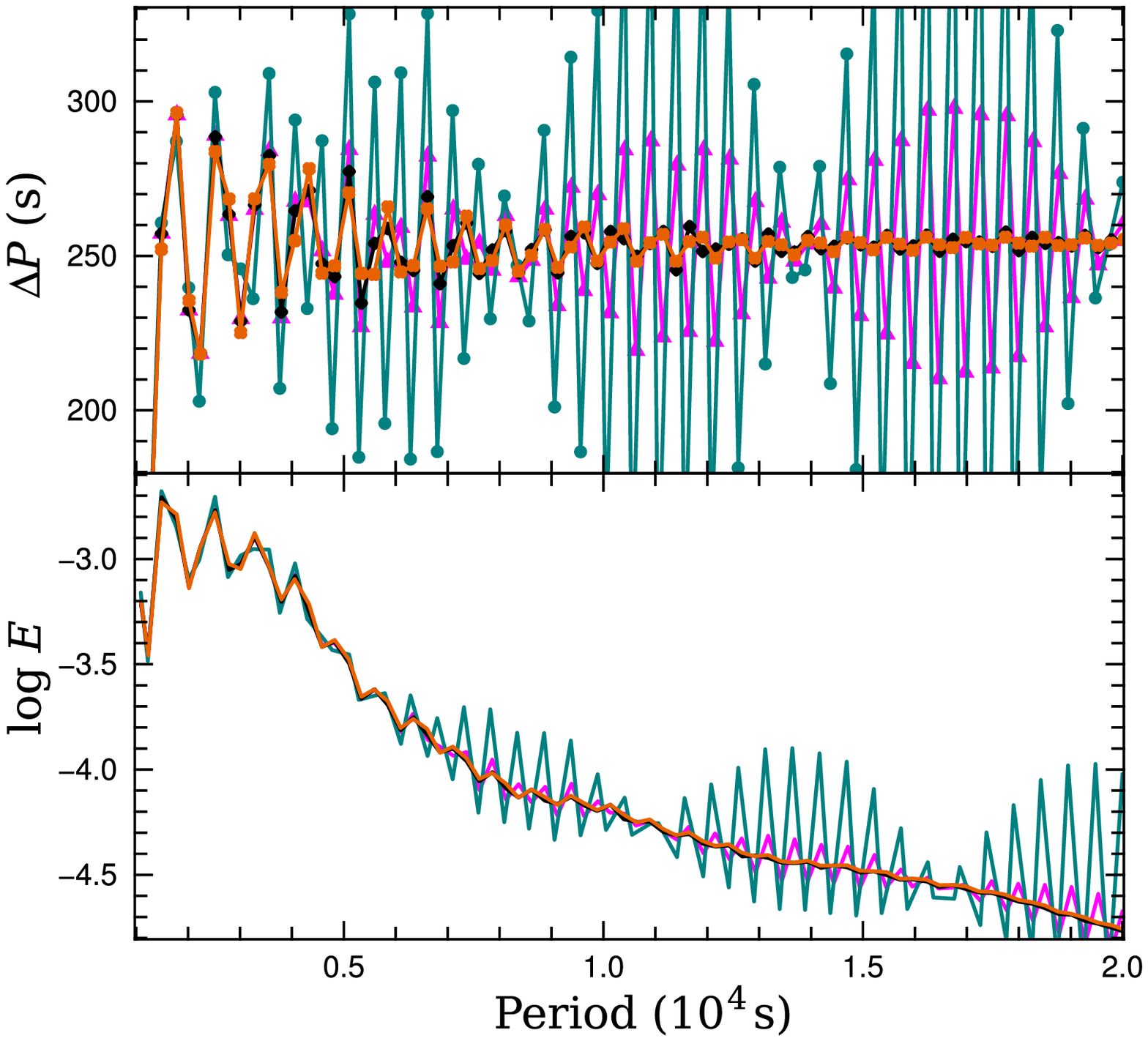}
  \caption{Comparison of the pulsation properties of sdB models with different composition discontinuities left behind by the core flash.  One model (orange) has a constant composition between the convective core and the H-burning shell.  The models in black, magenta, and cyan have chemical gradients over 0.01\,$\text{M}_\odot$, 0.001\,$\text{M}_\odot$, and $10^{-6}$\,$\text{M}_\odot$, respectively.  Each composition profile is set according to Equation~\ref{eq:sine}.  These models have $R=0.20\,\text{R}_\odot$ and $T_\text{eff}=27400\,\text{K}$.}
  \label{figure_sdb_core_flash_comparison}
\end{figure}

We have also examined the effect of core flash phase burning on the pulsations in our sdB models.  In Figure~\ref{figure_sdb_core_flash_comparison} we demonstrate that this effect is strongly dependent on the smoothness of the remaining composition profile.  The four models that we use to test this include a model without the discontinuity from the core flash (constant composition), and others with sine wave composition profiles (Equation~\ref{eq:sine}), with $\Delta m$ set as 0.01\,$\text{M}_\odot$, 0.001\,$\text{M}_\odot$, and $10^{-6}$\,$\text{M}_\odot$.  The consequences for the computed frequencies are increasingly apparent for models with sharper composition profiles.  The mode period spacing and inertia for the model with the smoothest composition profile (spread over 0.01\,$\text{M}_\odot$) is nearly identical to the model with a constant composition. By comparison, the model with the chemical profile spread over 0.001\,$\text{M}_\odot$ shows up to four times the period spacing variation for high frequency modes ($P<1.3\times 10^4$\,s) and mode to mode variation of almost 100\,s at lower frequency (where the two smoother models show almost constant $\Delta P$).  The model with the sharpest composition profile shows a pattern similar to the $\Delta m = 0.001\,\text{M}_\odot$ case, except with more extreme variation in $\Delta P$.  In both cases the amplitude of this variation oscillates, with a period of around $1.8 \times 10^{4}$\,s.  We also note the similarity between the period spacing pattern of these two models and the model by \citet{2014IAUS..301..397C}.  

Overall, these results suggest that the possibility of using pulsations to determine whether a low-mass CHeB star has experienced the core flash depends principally on how discontinuous is the composition profile it has left behind.

\subsection{Dependence on input physics}
\label{sec:input_physics}

\setlength{\tabcolsep}{2pt}

\begin{table}

  \caption{Properties of the evolution sequences.  The models have standard overshoot (SO), no overshoot (NO), semiconvection (SC), and maximal overshoot (MO).  The triple--$\alpha$ and ${}^{12}\text{C}(\alpha,\gamma)^{16}\text{O}$ reaction rates are denoted by $r_{\text{C}\alpha}$ and $r_{3\alpha}$, respectively.  $\Delta\Pi_{1,\text{mean}}$ is the average $\Delta\Pi_1$ value in the CHeB phase which has duration $\tau_\text{CHeB}$.  The initial and final H-exhausted core masses for the CHeB phase are denoted by $M_\text{He,i}$ and $M_\text{He,f}$ respectively.\label{table_dp1}}
\begin{tabular}{ccccccc}
\hline
{Model notes} &{$M$} &{Mixing} & {$\Delta\Pi_{1,\text{mean}}$} & {$\tau_\text{CHeB}$} & {$M_\text{He,i}$} & {$M_\text{He,f}$} \\ 
& {(M$_\odot$)} &  & {(s)} & {(Myr)} & {($\text{M}_\odot$)} & {($\text{M}_\odot$)}  \\
\hline
. . . &  1 &  SO  & 267  &  135.0   &    0.466   &    0.503 \\
. . . &  1 &  SC  & 258  &  129.6   &    0.467   &    0.503 \\
. . . &  1 &  NO  & 204  &  79.5    &    0.467   &    0.499 \\
. . . &  1 &  MO  & 293  &  119.7   &    0.467   &    0.499 \\
. . . &  2.5 & SO & 232  &  221.1   &    0.331   &    0.509 \\
. . . &  2.5 & SC & 225  &  227.5   &    0.331   &    0.514 \\
. . . &  2.5 & NO & 182  &  170.0   &    0.331   &    0.475 \\
. . . &  2.5 & MO & 251  &  216.8   &    0.331   &    0.502 \\
$\Delta M_\text{He} = 0.025\,\text{M}_\odot$ &  1 &  SO  & 285  &  99.8   &    0.493   &    0.520 \\
$\Delta M_\text{He} = 0.025\,\text{M}_\odot$ &  1 &  MO  & 304  &  100.4    &    0.493   &    0.519 \\
$r_{\text{C}\alpha}\text{, }r_{3\alpha}  \times 2$  &  1 &  SO  & 276  &  137.5   &    0.462   &    0.494 \\
$r_{\text{C}\alpha} \times 2$ &  1 &  SO  & 275  &  132.8   &    0.466   &    0.503 \\
$r_{3\alpha} \times 2$ &  1 &  SO  & 270  &  119.9   &    0.462   &    0.491 \\
$Y+0.1$ &  1 &  SO  & 272  &  112.0   &    0.448   &    0.548 \\
$\text{[Fe/H]}-1.0$ &  1 &  SO  & 285  &  112.4   &    0.467   &    0.523 \\

\hline
\end{tabular}

\end{table}

In Section~\ref{sec:bulk_properties} we showed that $1\,\text{M}_\odot$ standard-overshoot models need an increase in the H-exhausted core mass at the flash of more than $\Delta M_\text{He} > 0.025\,\text{M}_\odot$ to match the range of $\Delta\Pi_1$ reported for low-mass CHeB stars by \citet{2014A&A...572L...5M}.  The effect of uncertainties in the input physics on $M_\text{He}$ at the core flash has been examined in detail previously \citep[e.g.][]{1996ApJ...461..231C}.  Some of these uncertainties are not important to subsequent CHeB evolution.  For instance, the expansion of the core during the flash phase decreases both its rotation rate and neutrino emission so these effects need only be considered in light of how they affect the core mass at the flash.  In contrast, helium burning reaction rates and initial composition also affect the later evolution, including $\Delta\Pi_1$ (Table~\ref{table_dp1}).

Doubling the $^{12}\text{C}(\alpha,\gamma)^{16}\text{O}$ reaction rate increases the average $\Delta\Pi_1$ during CHeB by 8\,s.  Once there is enough carbon in the core ($X_\text{C} \ga 0.1$) the $^{12}\text{C}(\alpha,\gamma)^{16}\text{O}$ reaction proceeds more efficiently, slowing the rate of increase of the central temperature and density (and therefore also the triple--$\alpha$ rate).  This increases the convective core radius and consequently $\Delta\Pi_1$, but does not significantly affect the CHeB lifetime.  In contrast, increasing the triple-$\alpha$ rate reduces the core temperature and density from the beginning of CHeB.  Although this tends to increase $\Delta\Pi_1$, it is offset by the lower $M_\text{He}$, which starts smaller and grows more slowly due to the consequently reduced hydrogen-burning luminosity.  This results in only a 3\,s increase in the average $\Delta\Pi_1$ during the CHeB phase.  In the relevant conditions the uncertainty in the triple-$\alpha$ rate is less than 15 per cent while for $^{12}\text{C}(\alpha,\gamma)^{16}\text{O}$ rate it is around 40 per cent \citep{1999NuPhA.656....3A}, and more recent data favours the lower limit \citep{2013NuPhA.918...61X}.  Taking both of these uncertainties into account, they could together only account for around a 5\,s change in $\Delta\Pi_1$, considerably less than the size of the disparity between standard models and observations (of around 30\,s).

We have also tested the consequences of varying the initial composition.  Increasing helium raises the average $\Delta\Pi_1$, but the dependence is weak: a large increase of $\Delta Y = 0.1$ only increases the average $\Delta\Pi_1$ by 5\,s.  This may be attributed to the more rapid growth of the H-exhausted core during CHeB compared to the standard case, making it 0.04\,$\text{M}_\odot$ larger at core helium exhaustion.  This is partly offset, however, by the lower H-exhausted core mass at helium ignition, limiting the increase in average $\Delta\Pi_1$.  Reducing the metallicity by a factor of 10 increases the average $\Delta\Pi_1$ during the CHeB phase by 18\,s.  This is due to a reduction in the heavy element opacity (which we confirmed by evolving an $\text{[Fe/H]}=-1$ model but with solar heavy element opacity; which had a negligible effect on $\Delta\Pi_1$).  This initially increases the helium burning rate and consequently $\Delta\Pi_1$.  The hydrogen burning rate increases even more substantially, which further increases $\Delta\Pi_1$ by accelerating the growth of the H-exhausted core.  Composition, however, is not likely to be the cause of the $\Delta\Pi_1$ discrepancy because the stars in the \citet{2014A&A...572L...5M} sample are typically around solar metallicity \citep{2014ApJS..215...19P}, consistent with the models in Section~\ref{sec:ensemble}.  Indeed, none of these factors, nor any reasonable combination of them, can explain why the 1\,M$_\odot$ standard-overshoot run fails to match the observations.

\subsection{Late-CHeB and early-AGB models}
\label{sec:late_rc}

A number of authors have identified possible late-CHeB and AGB stars in the \textit{Kepler} field through seismology.  \citet{2012A&A...540A.143M} found five stars with the same $\Delta\nu$ as the low-mass CHeB group but with lower $\Delta\Pi_1$ (around 250\,s) and posited that these stars have exhausted helium in their cores.  \citet{2012ApJ...757..190C} identified several members of the open clusters in the \textit{Kepler} field (NGC 6811, NGC 6819 and NGC 6791) that are likely to be evolved red clump stars because they have similar $\ell = 1$ $\Delta P$ to the majority of clump stars, but have lower $\Delta \nu$.  In their examination of field stars, \cite{2012A&A...541A..51K} suggested that lower-$\Delta\nu$ stars belong to the early-AGB.  These stars also have a distinct central radial ($\ell = 0$) mode phase shift, which can be attributed to a difference in the structure of the convective envelope \citep{2014MNRAS.445.3685C}.  Although it is not examined in this paper, it would be interesting to determine if and how this phase shift depends on the CHeB mixing scheme.

Our models disagree with the earlier suggestion by \citet{2012A&A...540A.143M} that red clump stars with a low $\Delta\Pi_1$ but typical $\Delta \nu$ can be explained as being post-CHeB (they are now classified with the other red clump stars in \citealt{2014A&A...572L...5M}).  These are unlikely to be post-CHeB because every one of our low-mass models -- irrespective of mixing scheme -- shows a decrease in $\Delta\nu$ when $\Delta\Pi_1$ begins to decrease, which occurs prior to central helium exhaustion (Figure~\ref{figure_one_msun_core}).  This causes them to move away from the location of the suspected post-CHeB stars in $\Delta\nu-\Delta\Pi_1$ space (Figure~\ref{figure_1msun}a), which is in agreement with the {\sc mesa} models without overshoot shown in Figure 4b in \citet{2013ApJ...765L..41S}.  

\begin{figure}
\includegraphics[width=\linewidth]{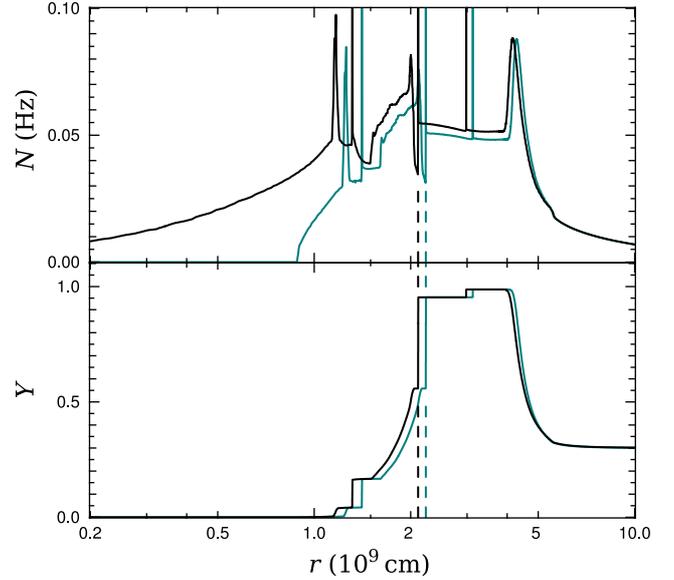}
  \caption{The Brunt--V{\"a}is{\"a}l{\"a} frequency $N$ (upper panel) and helium abundance Y (lower panel) for a 1\,M$_\odot$ solar-metallicity run with standard overshoot before (black) and after (cyan) core helium exhaustion.  The first model has central helium abundance $Y_\text{cent}=4 \times 10^{-4}$.  Between the two models 160\,kyr elapses.  In this time $\Delta\Pi_1$ decreases from 153\,s to 99\,s, $\Delta \nu$ decreases from 1.80\,$\mu$Hz to 1.47\,$\mu$Hz, radius increases from $18.9\,\text{R}_\odot$ to $21.0\,\text{R}_\odot$, $T_\text{eff}$ decreases from 4300\,K to 4250\,K, and $\nu_\text{max}$ decreases from $9\,\mu\text{Hz}$, to $7\,\mu\text{Hz}$.}
  \label{figure_1msun_core_exhaustion}
\end{figure}

In Figure~\ref{figure_1msun_core_exhaustion} we show a standard-overshoot model before and after core helium exhaustion, separated by 160\,kyr.  During this period there is a rapid increase in luminosity ($\log{L/\text{L}_\odot}$ increases from 2.029 to 2.117), a decrease in $\Delta\nu$ (from 1.80\,$\mu$Hz to 1.47\,$\mu$Hz), and a decrease in $\Delta\Pi_1$ (from 153\,s to 99\,s; it then drops to 65\,s after a further 1\,Myr).  This sudden decrease in period spacing has also been shown for models computed with {\sc mesa} \citep{2013ApJ...765L..41S}.   If the region enclosed by the outer edge of the partially mixed zone (dashed lines in Figure~\ref{figure_1msun_core_exhaustion}) is excluded from the calculation of $\Delta\Pi_1$ (to emulate the effect of mode trapping) the drop in period spacing is less severe (from 251\,s to 234\,s).  This, however, still suggests that the high $\Delta P$ ($\sim 250$\,s) stars, identified as possible members of the AGB by \citet{2012A&A...541A..51K}, are in fact still CHeB stars.  Similarly, the relatively low $\Delta\nu$ open cluster stars identified as evolved red clump stars in \citet{2012ApJ...757..190C} appear to be correctly classified, while the one suggested early-AGB star in NGC 6811 is probably also in the late-CHeB phase.  Lastly, the position in $\Delta\nu-\Delta\Pi_1$ space of our late-CHeB models, and those from \citet{2013ApJ...765L..41S}, is generally consistent with the observed group marked by `A' in Fig. 1 in \citet{2014A&A...572L...5M}.  We suggest that care should be taken when describing these stars, because ``red clump'' and ``core helium burning'' are not interchangeable terms.  According to models, stars leave the red clump when they are still burning helium in the core.

Determinations of $\Delta\Pi_1$ from observations of stars near core helium exhaustion could be very uncertain if there is mode trapping in the partially mixed region.  This is because late in CHeB the buoyancy radius of the partially mixed region, where the modes are trapped, becomes large compared to the total buoyancy radius.  In the pre- and post-core helium exhaustion models in Figure~\ref{figure_1msun_core_exhaustion} the partially mixed regions account for 41 per cent and 68 per cent of the total buoyancy radius respectively.  There are thus few modes of consecutive radial order that are both not trapped, unlike the model in Figure~\ref{figure_overshoot_echelle_eigen} for instance.  The extensive mode trapping in these models would make it difficult to accurately determine $\Delta\Pi_1$ from observations, but also make it unlikely they could be interpreted as having an erroneously high $\Delta\Pi_1$ from the period \'{e}chelle diagram. 

\begin{figure}
\includegraphics[width=\linewidth]{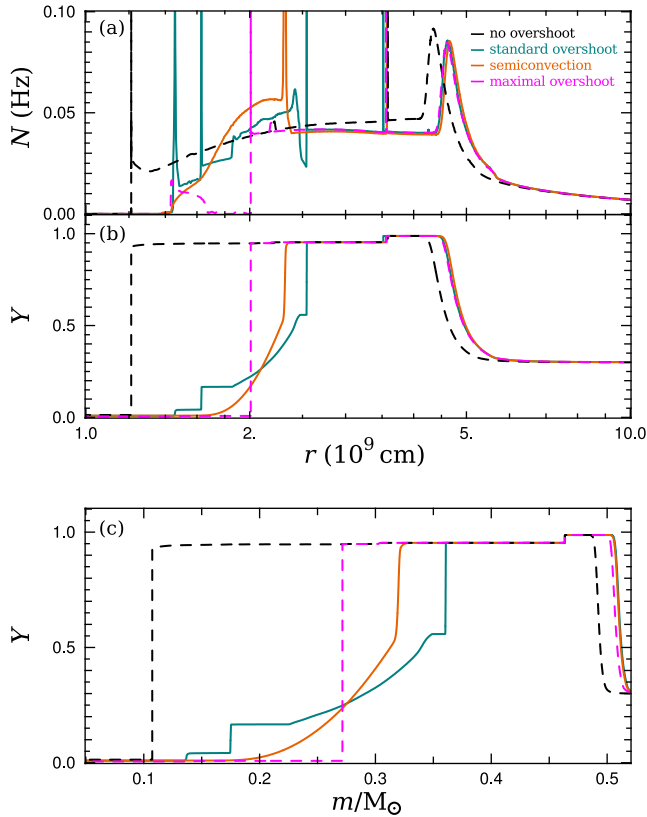}
  \caption{Comparison of the structure of four late-CHeB models with different treatments of convective boundaries, each with $Y_\text{cent} \approx 0.01$.   The mass range in the lower panel approximately corresponds to the radius range in the upper two panels.  The colours are the same as in Figure~\ref{figure_mixing_comparison_zoom}.  The standard-overshoot, semiconvection, no-overshoot, and maximal-overshoot models have $\Delta\Pi_1$ of 208\,s, 298\,s, 195\,s, and 247\,s, respectively.  These models have approximately $R=15\,\text{R}_\odot$, $T_\text{eff}=4440\,\text{K}$, and $\nu_\text{max} = 14\,\mu\text{Hz}$.}
  \label{figure_late_RC_comparison}
\end{figure}

Core breathing pulses (CBP) only occur in the standard-overshoot model (note the rapid increases in central helium abundance that begin after 98\,Myr in Figure~\ref{figure_one_msun_core}).  CBP do, however, occur in each of the remaining models if the mixing scheme is changed to standard overshoot late in CHeB (when the central helium abundance is $Y=0.1$).  This demonstrates that CBP are prevented by the mechanics of each mixing prescription rather than by the very different late-CHeB structures they eventually produce.  An example of the divergence of the internal composition is shown in Figure~\ref{figure_late_RC_comparison}c and is discussed below.  Although these structural differences do not prevent CBP, they do affect the magnitude of them: a larger convective core, or the existence of a partially mixed region outside it, reduces the amount of helium transported into the core by the breathing pulses.

Finally, in Figure~\ref{figure_late_RC_comparison} we show four examples of late-CHeB models that were evolved with different mixing schemes until they have central helium abundance of around $Y=0.01$.  The contraction of the fully convective core is evident in all but the no-overshoot model (note the convectively stable region near $r=1.6\times 10^9\,\text{cm}$ in the maximal-overshoot model in magenta).  By this stage the partially mixed regions in the semiconvection and standard-overshoot models extend well beyond the edge of the maximal-overshoot core, whereas earlier in the evolution their sizes are comparable (e.g. Figure~\ref{figure_mixing_comparison_zoom}a).  It is also clear that by the end of core helium burning those models have burned more helium than the maximal-overshoot case.  By the end of CHeB the internal structures have diverged significantly enough to suggest that i) the mixing scheme could affect the early-AGB evolution, perhaps to an extent that is detectable in a large enough homogeneous population (e.g. globular clusters; which will be explored in a forthcoming paper) and ii) asteroseismic studies of the population of late-CHeB (such as that found in \citealt{2014A&A...572L...5M}) and early-AGB stars may provide vital clues about CHeB evolution.

\section{Summary and conclusions}
\label{sec:conclusions}

The asteroseismic detection of mixed modes in core helium burning stars in the \textit{Kepler} field offers an unprecedented insight into the internal structure of these stars. With the aim of  better constraining the models, we have investigated two discrepancies between the predicted asymptotic g-mode $\ell = 1$ period spacing $\Delta\Pi_{1}$ from standard low-mass ($1\,\text{M}_{\odot}$) stellar models and those reported for the \textit{Kepler} field stars \citep{2012A&A...540A.143M,2014A&A...572L...5M}:

\begin{itemize}
   \item The average value of $\Delta\Pi_{1}$ predicted by the models is significantly below the average inferred from observations (by more than 25\,s) and the models never reach the highest observed values of $\Delta\Pi_{1}$. 
   \item The models spend more time with low values of $\Delta\Pi_1$ during core helium burning than is implied by the observed population (Figure~\ref{figure_dp1_obs}).
\end{itemize}

One possible source of these discrepancies could be that there are systematic problems with the internal stellar structure of standard models. Indeed, it is well known, yet often ignored, that the models of this phase are uncertain (e.g. Figure~\ref{figure_helium_ev}). To explore these uncertainties in the light of the new asteroseismic observations we computed non-radial adiabatic pulsations and $\Delta\Pi_{1}$ for a suite of core helium burning models with varying physical inputs and mixing algorithms. The stellar models were calculated with:
\begin{enumerate}[(i)]
   \item Four different mixing schemes (Section~\ref{sec:mixing_schemes}).
   \item Three different initial chemical compositions (Section~\ref{sec:input_physics}).
   \item Altered He-burning reaction rates (Table~\ref{table_dp1}).
\end{enumerate}

We found that varying the stellar composition or altering the He-burning reaction rates cannot reconcile the models and observations (Section~\ref{sec:input_physics}; Table~\ref{table_dp1}). Three of the four mixing schemes also failed to increase $\Delta\Pi_{1}$ by the magnitude required. The only models that can match the average observed $\Delta\Pi_{1}$ values reported are those with large convective cores, such as those calculated with our newly proposed ``maximal-overshoot'' scheme (Section~\ref{sec:max_os}). In this scheme the extent of convective overshoot is adjusted so that it produces the most massive convective core possible.  This treatment was implemented, however, only as a demonstration of the effect of a large convective core: we have not proposed any physical basis for it. In the case of more massive stars ($M > 2\,\text{M}_{\odot}$), the smaller number of observations, and the fact that their H-exhausted cores grow substantially during the core helium burning phase, allowed us to only rule out the no-overshoot model (Figure~\ref{figure_2.5msun}).

Another possible source of the $\Delta\Pi_{1}$ discrepancies is that the observations may be biased in some way. By comparing our non-radial adiabatic pulsation calculations against $\Delta\Pi_{1}$ across the suite of models, we identified a potential difficulty in inferring $\Delta\Pi_{1}$ from observations: any mode trapping that results from a convective region between two radiative zones (e.g. Section~\ref{sec:bulk_properties}; Figure~\ref{figure_hos}), or a steep composition gradient at the outer boundary of a semiconvection or partially mixed region (e.g. Section~\ref{sec:overshoot_results}; Figure~\ref{figure_overshoot_echelle_eigen}), increases the period spacing between most pairs of modes of consecutive radial order, and therefore the observationally inferred value of $\Delta\Pi_{1}$. The difference between these values could explain much of the disagreement between standard models and observations (dotted curve in Figure~\ref{figure_1msun}).

However, even after accounting for these two proposed resolutions to the discrepancy in average $\Delta\Pi_{1}$ values, the models still predict more core helium burning stars with low $\Delta\Pi_{1}$ ($< 270$\,s) than observed. We suggested two possible remedies for this problem: i) there may be a difficulty in observationally determining $\Delta\Pi_{1}$ for early core helium burning stars (when $\Delta\Pi_{1}$ is lowest) because the sharp composition profile at the hydrogen burning shell causes the period spacing pattern to be highly irregular compared to more evolved models (Section~\ref{sec:postcoreflash}), or ii) the mass of the helium core at the flash may be higher than predicted by standard models, thereby raising the initial $\Delta\Pi_{1}$ (Figure~\ref{figure_1msun}).  Further information about the selection effects in asteroseismic population studies, which are alluded to by \citet{2014A&A...572L...5M}, would help to establish the validity of the first point.  At present, the possibility of unknown systematic biases in the observations limits our ability to use them to make firm conclusions about stellar evolution theory.

We also investigated the dependence of $\Delta\Pi_{1}$ on the radius of the convective core, as shown by \citet{2013ApJ...766..118M}. We found that the relationship to the observed period spacing is more complicated than a simple linear relationship in a number of respects (Section~\ref{sec:bulk_properties}).  Furthermore, $\Delta\Pi_1$ is also dependent on the steepness of any chemical profiles outside the convective core, such as those found in the semiconvection zone (Section~\ref{sec:ensemble}).

The structure of low-mass CHeB stars is further complicated by the stabilizing chemical gradient left behind by helium burning during the core-flash phase. This can have a significant effect on the period spacing pattern, depending on the steepness of the gradient (Figure~\ref{figure_core_flash_discontinuity}). In fact, in models taken directly from our evolution code the mode trapping from the main discontinuity produced in the core-flash phase is the most important feature in the period spacing (e.g.  Figure~\ref{figure_no_overshoot}). This contrasts to composition gradients that may be created by overshooting (a small distance) from the convective core which have a more subtle effect on the period spacing (Figure~\ref{figure_max_overshoot_smooth}).

We also tested low-mass models that imitate sdB stars (Section~\ref{sec:sdb}).  In these models we find the same dependence of $\Delta\Pi_{1}$ and $\Delta P$ from pulsation calculations on mixing scheme as for our solar-mass models.  We also found that it may be difficult to use asteroseismology to constrain sdB formation channels.  This is because the effect of the composition discontinuity resulting from core-flash burning is smallest for the low radial-order modes (Figure~\ref{figure_sdb_core_flash_comparison}) that are typically detected.  We noted, however, that there may be other evidence from asteroseismology such as differences in the H--He transition region \citep{2008A&A...490..243H}.

In Section~\ref{sec:late_rc} we showed that our models of core helium exhaustion suggest that early-AGB stars will not be found near the bulk of core helium burning stars in the $\Delta\nu$ - $\Delta\Pi_{1}$ diagram, independent of mixing scheme. This is because both $\Delta\Pi_{1}$ and $\Delta\nu$ have decreased by the time core helium burning ceases.  This expands on the earlier finding by \citet{2013ApJ...765L..41S} for models without convective overshoot.  

Finally, although we have highlighted some possible explanations for the discrepancies in $\Delta\Pi_{1}$, further work is needed to pinpoint the cause(s).  In order to better gauge the extent of the problem, and therefore the merit of our proposed solutions, it is necessary to account for any selection bias in the observations.  Our possible solution involving the mode trapping phenomenon affecting the observationally inferred $\Delta\Pi_{1}$ values (e.g. Section~\ref{sec:overshoot_results}) could be investigated by comparing models to specific frequency patterns observed.  Constraints on the core mass at the helium flash and the mixing during the CHeB phase could be investigated by using the latest photometry of globular clusters -- this is the subject of the next paper in this series.

\section*{Acknowledgments}
This research was supported under Australian Research Council’s Discovery Projects funding scheme (project numbers DP1095368
and DP120101815).  Funding for the Stellar Astrophysics Centre is provided by The Danish National Research Foundation (Grant DNRF106). The research is supported by the ASTERISK project (ASTERoseismic Investigations with SONG and Kepler) funded by the European Research Council (Grant agreement no.: 267864).  This work was supported in part by computational resources provided by the Australian Government through the National Computational Infrastructure under the National Computational Merit Allocation Scheme (projects g61 and ew6).

\clearpage
\footnotesize{
 \bibliographystyle{mnras}
  \bibliography{paper1}
}

\end{document}